\documentclass[twocolumn]{aastex631}

\usepackage{amsmath}
\usepackage{subfigure}
\usepackage{float}

\shorttitle{$pp$ interactions in AGN jets as the origin of the DNB}
\shortauthors{Xue et al.}

\begin{document}

\title{Hadronuclear interactions in AGN jets as the origin of the diffuse high-energy neutrino background}

\author[0000-0003-1721-151X]{Rui Xue}
\affiliation{Department of Physics, Zhejiang Normal University, Jinhua 321004, China}

\author[0000-0002-3883-6669]{Ze-Rui Wang}
\affiliation{College of Physics and Electronic Engineering, Qilu Normal University, Jinan 250200, China}

\author{Jagdish C. Joshi}
\affiliation{Aryabhatta Research Institute of Observational Sciences (ARIES), Manora Peak, Nainital 263001, India}
\affiliation{Centre for Astro-Particle Physics (CAPP) and Department of Physics, University of Johannesburg, PO Box 524, Auckland Park 2006, South Africa}

\author{Wei-Jian Li}
\affiliation{Department of Physics, Zhejiang Normal University, Jinhua 321004, China}

\correspondingauthor{Rui Xue}
\email{ruixue@zjnu.edu.cn}

\begin{abstract}
The origin of diffuse high-energy neutrinos from TeV to PeV energies detected by IceCube Observatory remains a mystery. In our previous work, we have shown that hadronuclear ($p-p$) interactions in AGN jets could be important and generate detectable very-high-energy emissions. Here, we further explore these interactions in the AGN jets based on their luminosity function. The diffuse neutrino flux and corresponding $\gamma$-ray flux have been calculated and compared with observational data. In our modeling, two beaming patterns are considered separately. To make sure that the corresponding $\gamma$-ray flux does not overshoot the diffuse $\gamma$-ray background, we find that if the neutrino production region in jet is opaque to $\gamma$ rays, $p-p$ interactions in AGN jets with a small viewing angle (the blazar case) are able to interpret the PeV neutrino background. Similarly, AGN jets with a large viewing angle (the radio galaxy case) may interpret the TeV neutrino background. While, if the neutrino production region is transparent to $\gamma$ rays, only blazars have the potential to interpret the DNB around PeV band. Some caveats are also discussed.
\end{abstract}


\section{Introduction}\label{intro}
The origin of the diffuse neutrino background (DNB) with energies ranging from TeV to PeV is not yet clear since its discovery by the IceCube Observatory \citep{2013Sci...342E...1I, Aartsen2013PhR103A, Aartsen2017ApJ.51A}. The DNB is distributed isotropically across the sky and does not align with the Galactic plane, suggesting an extragalactic origin \citep{Aartsen2017ApJ..67A}. Several potential sources, including active galactic nuclei (AGNs), gamma-ray bursts, starburst galaxy, and tidal disruption events, are being considered as promising candidates of the DNB \citep[see][for reviews]{2008PhR...458..173B,2021PhyU...64.1261T, 2021NatAs...5..436H, 2022NatRP...4..697G, 2022ARNPS..72..365K, 2022arXiv220203381M}. At present, only two AGNs, namely TXS 0506+056 \citep{2018Sci...361.1378I, 2018Sci...361..147I} and NGC 1068 \citep{2022Sci...378..538I}, have been confirmed as neutrino emitters. In addition, possible associations between neutrino events and individual sources are being thoroughly investigated \citep{moharana2015JCAP, 2015EPJC...75..273S, moharana2016JCAP,2016NatPh..12..807K, 2019ApJ...880..103G, 2020A&A...640L...4G, 2020ApJ...902...29P, 2020ApJ...899..113P, 2021JCAP...10..082O, 2021ApJ...912...54R, 2021NatAs...5..510S, 2022ApJ...932L..25L, 2022PhRvL.128v1101R, 2023MNRAS.519.1396S, 2023ApJ...958L...2F, 2024ApJ...965L...2J, 2024MNRAS.527.8746P}. In the list of potential neutrino sources \citep{2017PhRvD..96h2001A, 2020PhRvL.124e1103A, 2020ApJ...892...92A}, AGNs are the predominant constituents.

Regarded as the most powerful persistent sources of electromagnetic (EM) radiation in the Universe, radio-loud AGNs with relativistic jets are believed to be accelerating particles beyond $1~\rm EeV$ \citep[e.g.,][]{1987ApJ...322..643B,eich2018JCAP36E}. This is also supported by their contribution towards diffuse gamma-ray background \citep{pp2014ApJ_1D,2016JCAP019H}. For jetted AGNs with large viewing angles, also known as radio galaxies, hadronuclear ($p-p$) interactions within the jets are considered as a possible origin for the DNB \citep{2014PhRvD..89l3005B}, although a super-Eddington jet power is introduced. Blazars are the subclass of radio-loud AGNs with jets pointing to observers. Owing to the relativistic beaming effect, the EM and neutrino emissions from blazars' jets are amplified significantly, which has attracted widespread study and discussion on the extent of the contribution that blazars can make to the DNB. Since various soft photon fields exist in the blazar's environment, photohadronic ($p-\gamma$) interactions are naturally considered as the most likely hadronic process. In the framework of conventional one-zone models, powerful blazars (i.e., flat spectrum radio quasars; FSRQs) with strong external photon fields are suggested to be effective neutrino emitters \citep{2014JHEAp...3...29D, 2014PhRvD..90b3007M, 2018ApJ...854...54R}. However, they are most likely the sources for (sub-)EeV neutrinos \citep{2021PhRvL.126s1101R, 2020A&A...642A..92R}, not $<10\rm~PeV$ of IceCube. On the other hand, weak blazars (i.e., BL Lac objects; BL Lacs) are suggested to contribute marginally to the DNB \citep{2014PhRvD..90b3007M, 2018ApJ...854...54R, 2021PhRvL.126s1101R}. Despite the prevailing view that BL Lacs are not efficient neutrino emitters, the potential for these objects to account for the DNB may not be dismissed. This is particularly true when considering more complex jet models, such as the spine-sheath model \citep{2014ApJ...793L..18T, 2015MNRAS.451.1502T} and the inner-outer blob model \citep{2019ApJ...886...23X, 2021ApJ...906...51X}. Although blazars have been considered as possible neutrino emitters, stacking analysis suggested that 2LAC blazars may not be able to dominate the DNB \citep{2017ApJ...835...45A}. On the other hand, \cite{2016PhRvL.116g1101M} argued that if the diffuse neutrinos originate from $p-\gamma$ interactions, these emitters should be $\gamma$-ray hidden objects. It may exclude the $p-\gamma$ origin of blazars jets, since the majority of blazar jets are transparent to $\gamma$ rays \citep{2015ApJ...800L..27A}. Furthermore, it has been proposed that the $\gamma$-ray opaque AGN core, i.e., the accretion disk, could account for the DNB ($p-\gamma$ and $p-p$ interactions are both considered) in theory \citep{2019ApJ...880...40I, 2020PhRvL.125a1101M}. The merit of this idea is that it significantly increases the number of potential neutrino emitters, given that all AGNs have the disk structure. 

Our recent preliminary work \citep{2022A&A...659A.184L} proposed an analytical method, indicating that $p-p$ interactions could play an important role in jets of blazars and radio galaxies. Numerical modeling \citep{2022PhRvD.106j3021X, 2024ApJS..271...10W} has shown that $p-p$ interactions can generate detectable TeV spectra without invoking extreme physical parameters. In this work, we investigate if $p-p$ interactions in AGN jets can explain the DNB, as $p-p$ interactions will not be constrained by the hidden source argument that arises in the case of $p-\gamma$ interactions \citep{2016PhRvL.116g1101M}. On the other hand, besides blazars and radio galaxies, other types of AGNs, such as FR0 radio galaxies \citep{2016MNRAS.457....2G} and narrow-line Seyfert 1 galaxies \citep{2023MNRAS.523..404L, 2023MNRAS.523..441Y}, may also have jets that can accelerate particles. Here we boldly hypothesize that AGNs' radio and $\gamma$-ray emissions originate from jets \citep[cf., see][]{2019NatAs...3..387P}. By utilizing the luminosity function of AGNs, all jetted AGNs could be collected together, which may help increase potential contribution to the DNB in theory. This paper is organized as follows. We describe our $p-p$ jet model and the cosmic evolution of jetted AGNs in Section~\ref{method}. The corresponding results are shown in Section~\ref{result}. In Section~\ref{sum}, we end with discussions. Throughout the paper, the cosmological parameters $H_{0}=69.6\ \rm km\ s^{-1}Mpc^{-1}$, $\Omega_{0}=0.29$, and $\Omega_{\Lambda}$= 0.71 \cite{2014ApJ...794..135B} are adopted in this work.

\section{Method}\label{method}
\subsection{The $p-p$ interactions in the jet}\label{framework}
Here we briefly recall the basic assumptions about $p-p$ interactions in AGNs' jet from our preliminary work \citep{2022A&A...659A.184L, 2022PhRvD.106j3021X}. Various charged and neutral particles exist in the relativistic jet, the composition of which remains uncertain at present. In our model, we do not specify the jet composition. Instead, we assume that the absolute jet power $P_{\rm jet}$ is predominantly governed by the injection power of relativistic protons and kinetic power in cold protons. Herein, cold protons are construed as the part of protons that are not accelerated within the jet. In the following, all quantities are considered in the comoving frame, unless specified otherwise.

\begin{figure}[htbp]
\includegraphics[width=0.5\textwidth]{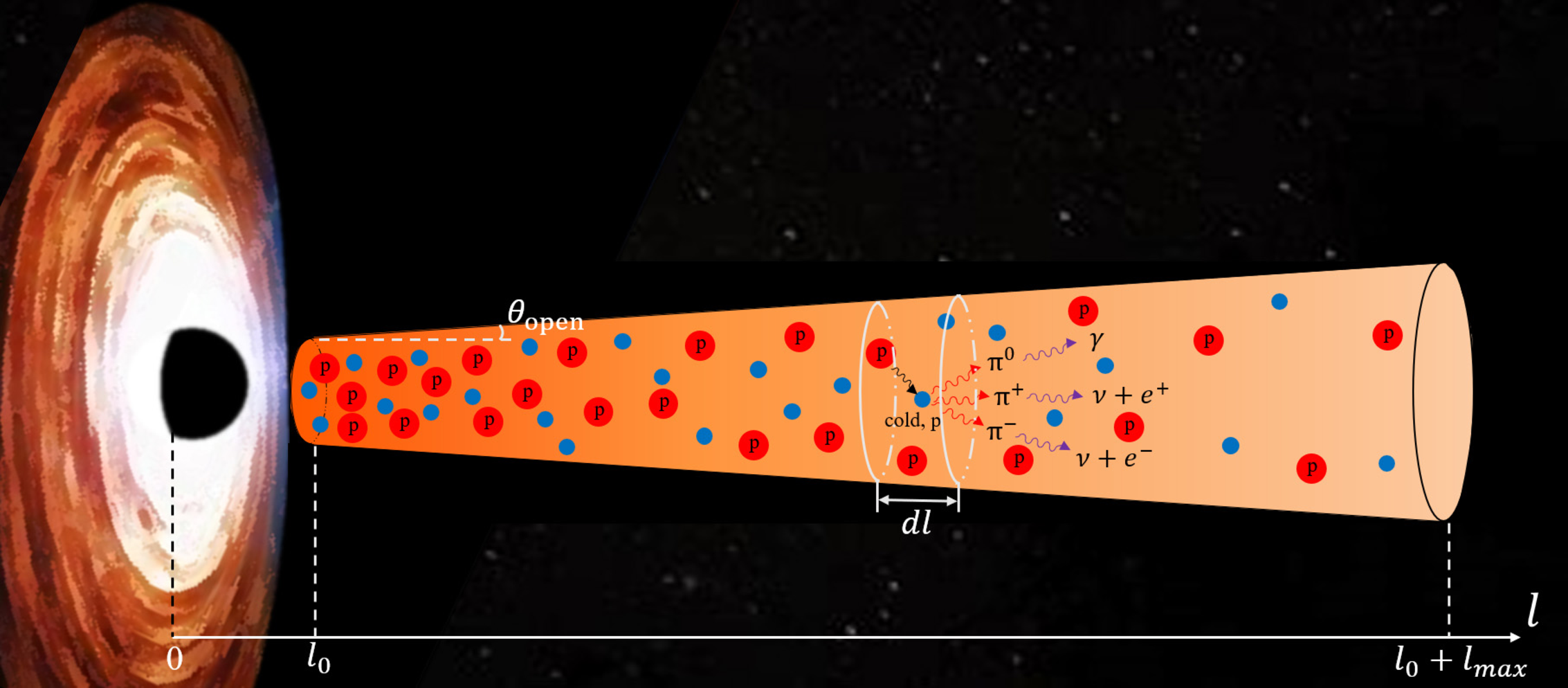}
\caption{A schematic illustration (not to scale) of the conical jet model having $p-p$ interactions. Relativistic (red circles) and cold protons (blue circles) are injected at the jet base, interacting with each other as they transport along the jet.
\label{sketch}}
\end{figure}

A sketch of our model is shown in Fig.~\ref{sketch}. We assume that the jet has a conical structure with a constant half-opening angle $\theta_{\rm open}$, length $l_{\rm max}$, distance between the supermassive black hole (SMBH) and the jet base is $l_0$, and jet's bulk Lorentz factor is $\Gamma$. The radius of jet's cross section in the comoving frame is given by $R_l=l \times  \textrm{tan}\theta_{\rm open}$. Relativistic and cold protons are injected at the jet base, interacting with each other as they propagate along the jet. Here $P_{\rm jet}$ is considered as a fraction of the Eddington luminosity $L_{\rm Edd}$ of the SMBH, which is usually seen as the upper limit of the jet power \citep{2015MNRAS.450L..21Z}, i.e., $P_{\rm jet}=\xi L_{\rm Edd}, \xi \leqslant 1$. As the power of electrons is normally negligible compared to that of protons, the number density of cold protons in the jet $n_{\rm H}$ can be estimated by \citep{2022A&A...659A.184L}
\begin{equation}\label{n_H}
    n_{\rm H}=\frac{(1-\chi_{\rm p})\xi L_{\rm Edd}}{\pi R_l^2\Gamma^2m_{\rm p}c^3},
\end{equation}
where $\chi_{\rm p}$ represents the ratio of the injection power of relativistic protons to $P_{\rm jet}$, $m_{\rm p}$ is the proton rest mass, and $c$ is the speed of light. 

For the relativistic protons, they are assumed to be injected with a power-law energy distribution at a constant rate,
\begin{equation}
    \dot{Q}_{\rm inj}(\gamma)=\dot{Q}_{\rm 0}\gamma^{-\alpha}, \gamma_{\rm min}<\gamma<\gamma_{\rm max},
\end{equation}
where $\dot{Q}_{\rm 0}$ is the normalization in units of $\rm s^{-1}$, $\gamma_{\rm min/max}$ are the minimum, and maximum proton Lorentz factors, and $\alpha$ is the spectral index. By giving the proton
injection luminosity $L_{\rm rel, p}\approx \chi_{\rm p}P_{\rm jet}/\Gamma^2$ in the jet, $\dot{Q}_{\rm 0}$ can be determined by $\int \dot{Q}_{\rm inj}\gamma m_{\rm p}c^2 d\gamma=L_{\rm rel, p}$. Please note that $L_{\rm rel, p}$ is a Lorentz invariant, and $\Gamma^2L_{\rm rel,p}$ represents the absolute power of relativistic protons that can be compared with $L_{\rm Edd}$ and $P_{\rm jet}$ in the AGN frame \citep{1993MNRAS.264..228C, 2015ApJ...809..174D, 2016ApJ...825L..11P}.

From the view of analytical estimation, the observed EM and neutrino luminosities from a AGN's jet can be calculated by \citep{2013PhRvD..88l1301M, 2022A&A...659A.184L}
\begin{equation}\label{analytical}
\begin{split}
L_{\gamma/\nu}^{\rm obs}\approx& \delta^{n} \int_{l_0}^{l_0+l_{\rm max}}L_{\rm rel, p}f_{\rm pp}(l)dl\\
\approx&L_{\rm Edd}\frac{2K\kappa_{\rm pp}\chi_{\rm p}(1-\chi_{\rm p})\xi^2}{\textrm{tan}^2\theta_{\rm open}}\big(\frac{\sigma_{\rm pp}}{\sigma_{\rm T}}\big)\big(\frac{R_{\rm S}}{l_0}\big)\frac{\delta^{n}}{\Gamma^4},
\end{split}
\end{equation}
where $\delta=[\Gamma(1-\beta \rm cos\theta_{\rm obs})]^{-1}$ is the Doppler factor with $\theta_{\rm obs}$ being the viewing angle and $\beta c$ being the speed of jet, $f_{\rm pp}(l)dl \approx \frac{t_{\rm ad}}{t_{\rm pp}}\approx K\kappa_{\rm pp}\sigma_{\rm pp}n_{\rm H}dl$ is the $p-p$ interaction efficiency with $t_{\rm pp}$ being the cooling timescale of $p-p$ interactions, $t_{\rm ad}$ being the adiabatic timescale \citep{2022A&A...658A.173T}, $\kappa_{\rm pp}$ being the inelasticity coefficient, $\sigma_{\rm pp}$ being the cross section for the $p-p$ interactions, $K$ being the branching ratio into $\gamma$ rays or neutrinos, and $R_{\rm S}\approx 1.5\times10^{14}(\frac{M_{\rm BH}}{10^9 M_{\odot}})~\rm cm$ is the Schwarzschild radius of the SMBH with $M_{\rm BH}$ being the mass of SMBH in units of the solar mass, $M_{\odot}$ and the specific value of $n$ represent the different beaming patterns. In this work, we consider two transformations of fluxes and luminosities from the comoving to the observers' frames. The first case is similar to the conventional one-zone model, with the initial position located at the jet base. This blob then moves along the jet and expands along the jet edge. The second scenario involves a continuous jet. The injection at the jet base exhibits long-term stability. After a sufficiently long period, a stable, long-lasting jet will form, representing a steady component. For a moving blob, the index $n$ of $\delta$ is 4, in which one comes from the transformation of the frequency, one for the time, and two for the solid angle \citep{2013LNP...873.....G}. While for the continuous jet, $n=3$ since there is no time compression for a steady component \citep{1985ApJ...295..358L, 2009arXiv0909.2576M, 2020PhRvD.102h3028L}. With the injection relativistic proton energy distribution $\dot{Q}_{\rm inj}(\gamma)$, the differential luminosity spectrum of decayed $\gamma$ rays and neutrinos $\phi_{\gamma/\nu}(E_{\gamma/\nu}, M_{\rm BH})$, where $E_{\gamma/\nu}$ represents the energy of generated $\gamma$ rays and neutrinos, can be obtained with analytical approximations developed in \cite{2006PhRvD..74c4018K}.

\subsection{Diffuse neutrino emission}
The cumulative diffuse neutrino flux is given as
\begin{equation}
\begin{split}
E_{\nu}^2\Phi_{\nu} =&\frac{cE_{\nu}^2}{4\pi H_0}\int_{M_{\rm BH}}\int_{z}\int_{L}\frac{\varrho(z,L)}{\sqrt{\Omega_{\rm m}(1+z)^3+\Omega_{\rm{\Lambda}}}}\\
&\times \phi_{\nu}(E_{\nu}, M_{\rm BH})\rho(z, M_{\rm BH}) dLdzdM_{\rm BH},
\end{split}
\end{equation} 
where $\varrho(z,L)$ represents the luminosity function at a specific band, and $\rho(z, M_{\rm BH})$ represents the probability distribution of $M_{\rm BH}$ at different redshifts ($z$). In order to compile a comprehensive collection of jetted AGNs, we employ luminosity functions at different bands to represent the evolution of jetted AGNs with varying jet directions. For blazars, where the jet is directed towards the observer, we use the $\gamma$-ray luminosity function to represent their cosmological evolution \citep{2015ApJ...800L..27A}. For AGNs where the jet is not directed towards the observer, their evolution is described using the radio luminosity function. \cite{2001MNRAS.322..536W} suggested that a simple dual-population model can aptly describe the evolution of radio AGNs. In this model, the low-luminosity AGN (LLAGN) population is associated with radio galaxies having weak emission lines, and the high-luminosity AGN (HLAGN) population is associated with radio galaxies and quasars having strong emission lines. \cite{2020A&A...638A..46S} found that this dual-population model aligns well with the results of many radio surveys. In our calculation, we adopt {\it Model C} in \cite{2001MNRAS.322..536W}. In the $p-p$ jet model, $M_{\rm BH}$ has an impact on the upper limit of neutrino flux that the jet can produce, since the Eddington luminosity sets the upper limit of jet power. To incorporate the influence of $M_{\rm BH}$ on our model's results, we employ the black hole mass function \citep{2013ApJ...764...45K} to ascertain the probability distribution $\rho(z, M_{\rm BH})$ of $M_{\rm BH}$. It should be noted that the sample used to derive the black hole mass function in \cite{2013ApJ...764...45K} is comprised of broad-line AGNs. The evolution of $M_{\rm BH}$ in narrow-line AGNs may deviate from that of broad-line AGNs, potentially influencing our results.

Similarly, the corresponding diffuse $\gamma$-ray flux from $\pi^0$ decay can be evaluated as
\begin{equation}
\begin{split}
E_{\gamma}^2\Phi_{\gamma} =&\frac{cE_{\gamma}^2}{4\pi H_0}\int_{M_{\rm BH}}\int_{z}\int_{L}\frac{\varrho(z,L)}{\sqrt{\Omega_{\rm m}(1+z)^3+\Omega_{\rm{\Lambda}}}}\\
&\times \phi_{\gamma}(E_{\gamma}, M_{\rm BH})\rho(z, M_{\rm BH})e^{-\tau(z, E_{\gamma})} dLdzdM_{\rm BH},
\end{split}
\end{equation} 
where $\tau(z, E_{\gamma})$ is the optical depth for the extragalactic background light (EBL) absorption \citep{2021MNRAS.507.5144S}. It is necessary to check if $\pi^0$ decayed $\gamma$ rays would contribute significantly to the extragalatic $\gamma$-ray background as well.

\begin{table}
\setlength\tabcolsep{7pt}
\caption{Summary of model parameters. The superscripts of $l_0$ represent the different beaming patterns considered in the modeling.}
\centering
\begin{tabular}{ccc}
\hline\hline																			
	&	blazars	&	radio AGNs	\\
$\theta_{\rm open}$	&	$2^\circ$	&	$2^\circ$	\\
$\theta_{\rm obs}$	&	$\leq1/\Gamma$	&	$30^\circ$	\\
$\Gamma$	&	10	&	5	\\
$l_{\rm max}$	&	1 kpc	&	1 kpc	\\
$\gamma_{\rm min}$	&	1	&	1	\\
$\gamma_{\rm max}$	&	$1~\rm PeV/m_pc^2$	&	$1~\rm PeV/m_pc^2$	\\
$\alpha$	&	2.2	&	2	\\
$\xi$	&	0.05	&	0.77	\\
$\chi_{\rm p}$	&	0.5	&	0.5	\\
\hline
$l_0^{\rm n=4}$	&	$3.5\times10^{-3}~\rm pc$	&	$3.5\times10^{-3}~\rm pc$	\\
$l_0^{\rm n=3}$	&	$1.8\times10^{-4}~\rm pc$	&	$2.7\times10^{-3}~\rm pc$	\\
\hline
\label{parameters}
\end{tabular}
\end{table}

\section{Results}\label{result}
\subsection{Cumulative emissions from $p-p$ interactions}\label{3.1}
We employ the $p-p$ jet model described above to study the possible contribution to the DNB and the diffuse extragalactic $\gamma$-ray background (EGB). There are some free parameters in the model, where $\gamma_{\rm min}$, $\gamma_{\rm max}$, $\alpha$, $\xi$, $\chi_{\rm p}$ are used to describe the energy distribution of injected relativistic protons, and $\theta_{\rm open}$, $\theta_{\rm obs}$, $\Gamma$, $l_0$, $l_{\rm max}$ are used to describe the jet properties. Here, we set $\gamma_{\rm min} = 1$ since it has a marginal impact on the modeling result. In principle, both $\chi_{\rm p}$ and $\xi$ are free parameters, and they have similar functions, which are to adjust the power ratio of relativistic and cold protons. For simplicity, we fix $\chi_{\rm p}$ at 0.5 as discussed by \cite{2022A&A...659A.184L} and adjust $\xi$ to fit the background spectra. From Eq.~(\ref{analytical}), it can be seen that the influence of $l_{\rm max}$ on the result is minor. Hence, for the sake of simplicity, we set $l_{\rm max}=1~\rm kpc$. Guided by radio observation of pc-scale jets \citep{2013A&A...558A.144C}, we boldly set $\theta_{\rm open}=2^\circ$, even though $\theta_{\rm open}$ of each AGNs' jet is certainly different. For the blazar, given its jet is oriented towards the observer, we have $\delta \approx 2\Gamma$. Here we set $\Gamma=10$ as indicated by radio observation \citep{2018ApJ...866..137L}. For radio AGNs, $\Gamma$ and $\theta_{\rm obs}$ are considered as free parameters. For simplicity, we set $\Gamma=5$ and $\theta_{\rm obs}=30^\circ$ as typical values. When considering the Doppler boosting, two beaming patterns are considered in this work as mentioned below Eq.~(\ref{analytical}), namely a continuous jet, which can be regarded as a steady component, and a moving blob. For blazars, when considering the case of a moving blob, the adopted value of $\Gamma$ does not affect the observed spectrum flux, but it does influence the observed energy of the particles. With $\Gamma=10$, the energy of observed neutrinos $E_{\nu}^{\rm obs}$ is basically equal to that of the primary protons $E_{\rm rel, p}$, i.e., $E_{\nu}^{\rm obs}=0.05\times E_{\rm rel, p}\delta/(1+z)$, since $\delta \approx 2\Gamma$. Assuming that the maximum energy of primary protons is 1~PeV, PeV neutrinos are expected to be observed. The position of jet base $l_0$ has a significant impact on the result because it directly affects the $p-p$ interaction efficiency. From Eq.~(\ref{analytical}), it can be seen that the most important contribution of high-energy neutrinos is from the jet base, excluding the significant contribution from large-scale jets. This is consistent with the conclusion of \cite{2014PhRvD..89l3005B} on radio galaxies. In the following simulation of the DNB and EGB, both beaming patterns are considered separately. Under the premise of not significantly adjusting parameters and being able to obtain the same results, the only difference in the selection of free parameters under the two beaming patterns is $l_0$, to compensate for the difference in the index of $\delta$. 

\begin{figure}[htbp]
\includegraphics[width=0.5\textwidth]{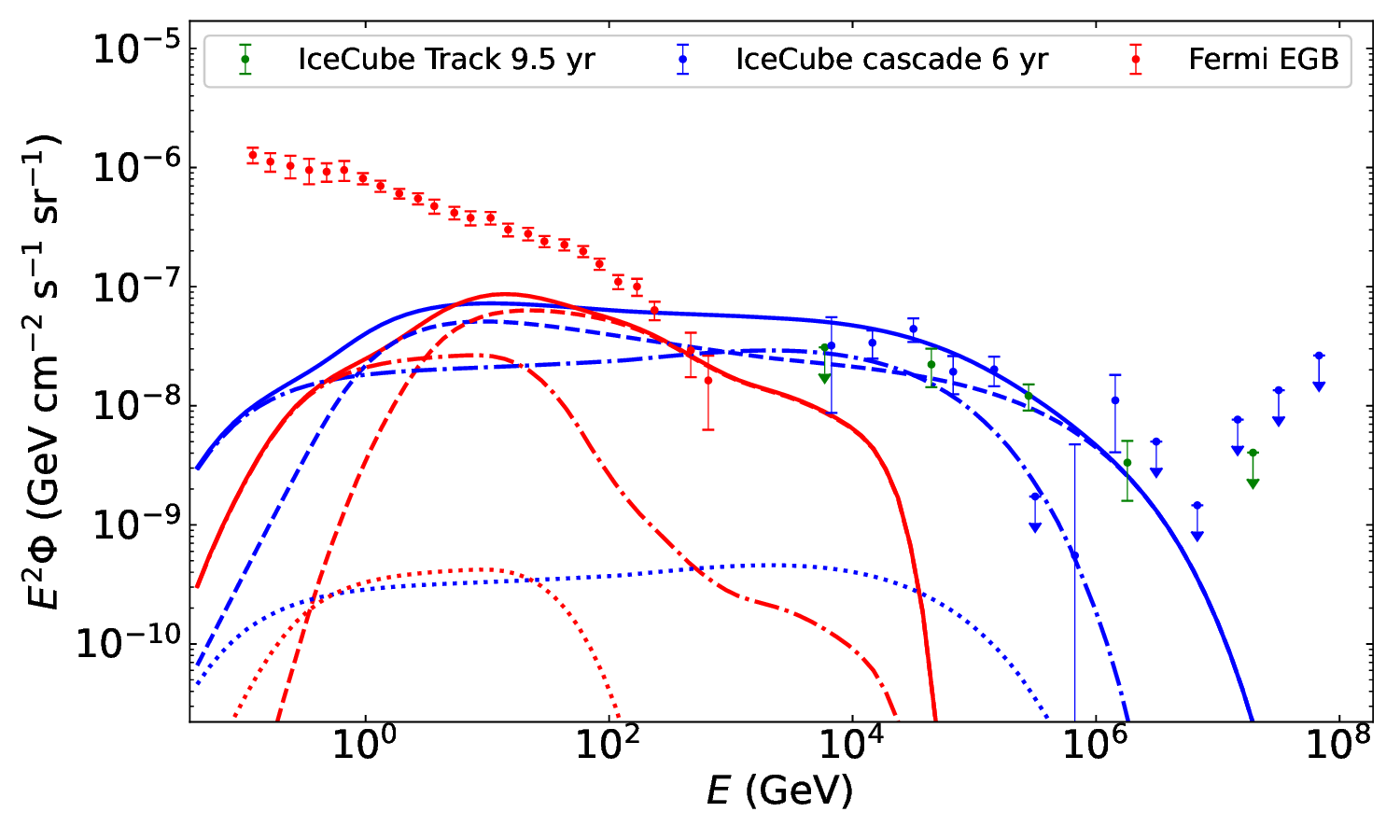}
\caption{Contribution of the $p-p$ interactions from AGN jets to the DNB and EGB. Red, blue, and green data points are the EGB spectrum measured by $Fermi$-LAT \citep{2015ApJ...799...86A}, cascade neutrino spectrum given by 6 year IceCube observation \citep{2020PhRvL.125l1104A}, and muon neutrino spectrum given by 9.5 year IceCube observations \citep{2022ApJ...928...50A}, respectively. Blue and red curves represent the $p-p$ model predicted neutrino and $\gamma$-ray spectra, respectively. Solid, dashed, dotted, and dot-dashed curves represent emissions from all jetted AGNs, blazars, HLAGNs and LLAGNs, respectively. 
\label{diffuse}}
\end{figure}

It is noteworthy that when considering the jet base as the most efficient neutrino production region, the transparency of this region to $\gamma$ rays is a significant issue that could impose varying degrees of constraints on the modeling results. Given that the jet base's proximity to the accretion disk, UV photons peak around 10 eV from the accretion disk, as well as the X-ray photons from 0.1~keV to 100~keV produced by the hot corona that envelops the accretion disk \citep[e.g.,][]{2018MNRAS.480.1819R}, could potentially absorb $\gamma$ rays of energy $E_{\gamma}\approx 2(m_{\rm e}c^2)^2/10~\rm eV=52~\rm GeV$ and $\gamma$ rays in energy range from $\sim$5.2 MeV to $\sim$5.2 GeV. If $\gamma$ rays are absorbed or not depends on the energy densities of these external photons in the comoving frame. The most of the EGB discovered by $Fermi$-LAT \citep{2015ApJ...799...86A} is contributed by resolved objects and the remaining component of EGB is the unresolved isotropic $\gamma$-ray background (IGRB). If the neutrino production region is opaque to $\gamma$ rays, then EGB will impose constraints on the EM emission from AGN jets. On the other hand, if the neutrino production region is transparent to $\gamma$ rays, then the IGRB will provide more direct constraints. This is because $\gamma$ rays escaping from the neutrino production region will interact with the EBL and the cosmic microwave background (CMB) during the propagation, leading to $\gamma-\gamma$ annihilation that produces electron-positron pairs. These secondary pairs, influenced by the intergalactic magnetic field, will have their paths deflected, thereby contributing to the unresolved IGRB via inverse Compton scattering \citep{1994ApJ...423L...5A, 1997ApJ...487L...9C, 2020PhRvD.101h3004W}. Conducting a detailed study on the opacity of each AGN jet base is extremely challenging, as it requires knowledge of each AGN's disk luminosity, the distance between the jet base and the SMBH, as well as the actual extent of the hot corona, which are all currently poorly understood. In this work, we simply assume that all jet bases are either $\gamma$-ray opaque or transparent, so that EGB and IGRB can respectively impose more reasonable constraints on the model predicted contribution to the DNB.

Let us first assume that the most significant neutrino production region near the jet base is opaque to $\gamma$ rays. With the parameter sets given in Table~\ref{parameters}, the model predicted contributions to DNB and EGB with the data points detected by IceCube and $Fermi$-LAT are shown in Fig.~\ref{diffuse}. It can be seen that neutrino background in tens of TeV range is mainly from LLAGNs, with blazars also making a significant contribution. On the other hand, $\sim \rm PeV$ neutrino background is predominantly contributed by blazars due to the stronger Doppler effect, resulting in higher observed neutrino energies. For the contribution of $\pi^0$ decayed $\gamma$ rays to the EGB, our results suggest an almost negligible contribution from radio AGNs, while blazars only contribute in the TeV range. This result is not inconsistent with the prevailing consensus that the GeV $\gamma$ rays from blazars is primarily a result of leptonic processes. Please note that the cascade emissions generated in-source for the $\gamma$-ray opaque neutrino production region are not taken into account here. The absorbed $\gamma$ rays will be re-distributed in the MeV-GeV band depending on the ratio of energy densities of magnetic field to soft photons in-source \citep{2019ApJ...881...46R}. Given that the flux of TeV-PeV photons is significantly lower than the EGB in the MeV-GeV band, the consideration of cascade emission does not impact our results.

\begin{figure}[htbp]
\includegraphics[width=0.48\textwidth]{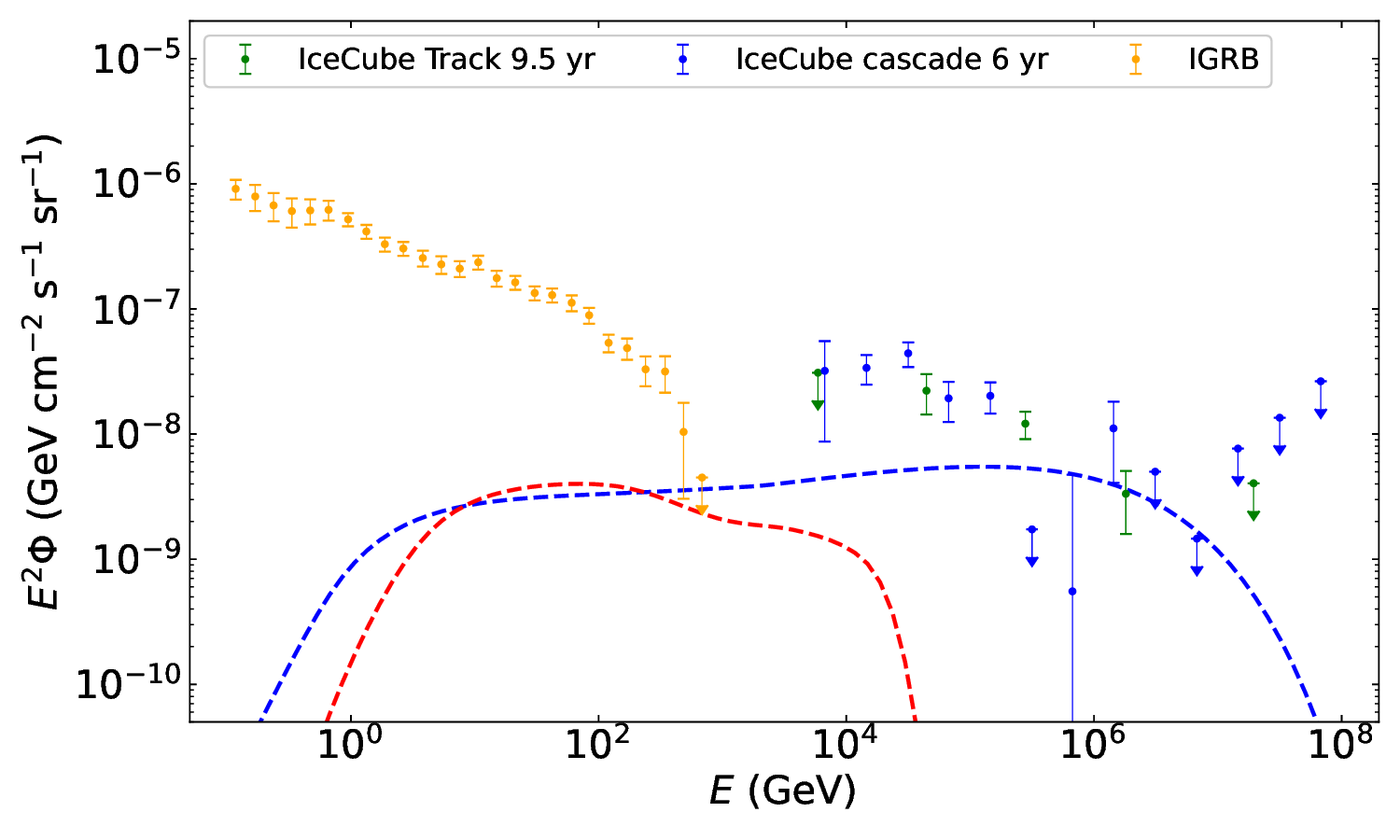}
\caption{Contribution of the $p-p$ interactions from blazar jets to the DNB and IGRB. Orange data points are the IGRB spectrum given by $Fermi$-LAT \citep{2015ApJ...799...86A}. Other data points and line styles have the same meaning as in Fig.~\ref{diffuse}.
\label{IGRB}}
\end{figure}

In the scenario where the neutrino production region is transparent to $\gamma$ rays, these high-energy photons escape the jet and propagate in the intergalactic space. During propagation, $\gamma$ rays interacts with EBL and CMB photons, leading to cascade emission that contribute to IGRB. Analytical studies have shown that the final remnant cascade spectrum exhibits a universal form, characterized by a flat $\gamma$-ray spectrum $E_\gamma^2dN/dE_\gamma \propto E_\gamma^{0\sim0.1}$ \citep[e.g.,][]{2016PhRvD..94b3007B}. It implies that IGRB provides a robust upper limit on the flux of absorbed $\gamma$ rays in propagation. The orange data points in Fig.~\ref{IGRB} show the IGRB spectrum measured by $Fermi$-LAT, which is approximately an order of magnitude lower in flux than the EGB at sub-TeV energies. Consequently, the $\pi^0$ decayed diffuse $\gamma$-ray emission depicted in Fig.~\ref{diffuse} substantially overshoot the sub-TeV IGRB, implying that the previously obtained results can no longer account for the DNB. By comparing the spectra of IGRB and DNB, it can be seen that only the PeV neutrino flux does not exceed the upper limit set by the highest energy data point of IGRB. Therefore, radio AGNs are unlikely to contribute significantly to the tens of TeV neutrino background, while PeV neutrinos may still be predominantly produced by blazars. In the modeling of blazars, we set $\xi=0.01$, $\alpha=2$, and $\gamma_{\rm max}=5~\rm PeV/m_{p}c^2$, and the remaining parameters are consistent with those presented in Table~\ref{parameters}. The modeling results are shown in Fig.~\ref{IGRB}. We suggest that due to the constraints of IGRB, blazars could only potentially account for the DNB around PeV band. Recently, \cite{2022ApJ...933..190F} suggested that from a model-independent perspective, if relativistic protons has a large minimum energy and a soft spectral index $\alpha\gtrsim2.5$, then $\gamma$-ray transparent sources may explain the DNB. However, we do not adopt such an approach here because an extremely high minimum proton energy is unphysical.

\subsection{Comparison with $p-\gamma$ interactions}
The main focus of this study is to investigate whether overlooked $p-p$ interactions in AGN jets could contribute to the DNB. So we did not delve into the detailed study of $p-\gamma$ interactions, which has been extensively examined in many previous studies (see Section~\ref{intro}). In the AGN environment, due to the presence of abundant soft photon fields, such as the jet inside, the accretion disk, the broad-line region (BLR), and the dusty torus (DT), $p-\gamma$ interactions are widely discussed as the main mechanism for high-energy neutrino production. However, in this work, we propose that $p-p$ interactions may be more likely to contribute significantly to the currently detected TeV--PeV DNB. To further explain this point, let us briefly review the $p-\gamma$ interactions efficiency in jets and compare it with the $p-p$ interactions efficiency.

As illustrated in Section~\ref{framework}, $p-p$ interactions are considered to occur throughout the jet, and the interactions efficiency would decrease due to the decreasing number density of cold protons $n_{\rm H}$ with increasing distance $l$. Similarly, the $p-\gamma$ interactions efficiency at different positions in the jet varies as well, and its evolutionary pattern is more complex compared to the $p-p$ interactions, because different photon fields dominate at different locations. For external photon fields, the accretion disk, BLR and DT are considered in our calculation; their energy densities in the comoving frame as functions of the jet position $l$ can be approximately written as \citep{2009ApJ...704...38S, 2012ApJ...754..114H}
\begin{equation}
\begin{split}
u_{\rm disk}= 0.28 \big(\frac{\Gamma^2L_{\rm disk}}{4\pi (l\Gamma)^2c}\big)\big(\frac{R_{\rm S}}{l\Gamma}\big)\\
u_{\rm BLR} = \frac{\eta_{\rm BLR} \Gamma^2L_{\rm disk}}{4\pi l_{\rm BLR}^2c[1+(l\Gamma/l_{\rm BLR})^3]}\\
u_{\rm DT}= \frac{\eta_{\rm DT} \Gamma^2L_{\rm disk}}{4\pi l_{\rm DT}^2c[1+(l\Gamma/l_{\rm DT})^4]},
\end{split}
\end{equation}
respectively, where $L_{\rm disk}$ is the disk luminosity, $\eta_{\rm BLR} = \eta_{\rm DT}$ = 0.1 are the fractions of $L_{\rm disk}$ reprocessed into BLR and DT radiation, respectively. The characteristic distances of BLR and DT in the AGN frame are estimated as $l_{\rm BLR}$ = 0.1 ($L_{\rm disk}/10^{46}$ ergs$^{-1})^{1/2}$ pc and $l_{\rm DT}$ = 2.5($L_{\rm disk}/10^{46}$ergs$^{-1})^{1/2}$ pc \citep[e.g.,][]{2008MNRAS.387.1669G}. To derive their number density distributions, the radiation spectrum of accretion disk is approximated as an isotropic blackbody with a peak at $2\times 10^{15}\Gamma$ Hz, and the radiation spectra of both BLR and DT is taken as an isotropic graybody with a peak at $2\times 10^{15}\Gamma$ Hz \citep{2008MNRAS.386..945T} and $3\times 10^{13}\Gamma$ Hz \citep{2007ApJ...660..117C} in the comoving frame, respectively. For the non-thermal photons in the jet, i.e., non-thermal photons from primary electrons, the estimation of their energy density is intricate. This complexity arises due to variations in the spectral index, peak frequency, and peak luminosity of each AGNs' spectrum. In addition, the observed luminosity is Doppler amplified, usually resulting in a weaker energy density compared to that of external photons in the jet comoving frame, $u_{\rm jet}(E)=L_{\rm jet}(E)/(4\pi R_l^2c\delta^n)$, where $u_{\rm jet}(E)$ represents the energy density of non-thermal photons and $L_{\rm jet}(E)$ represents the luminosity density of non-thermal emission.

\begin{figure}[htbp]
\includegraphics[width=0.5\textwidth]{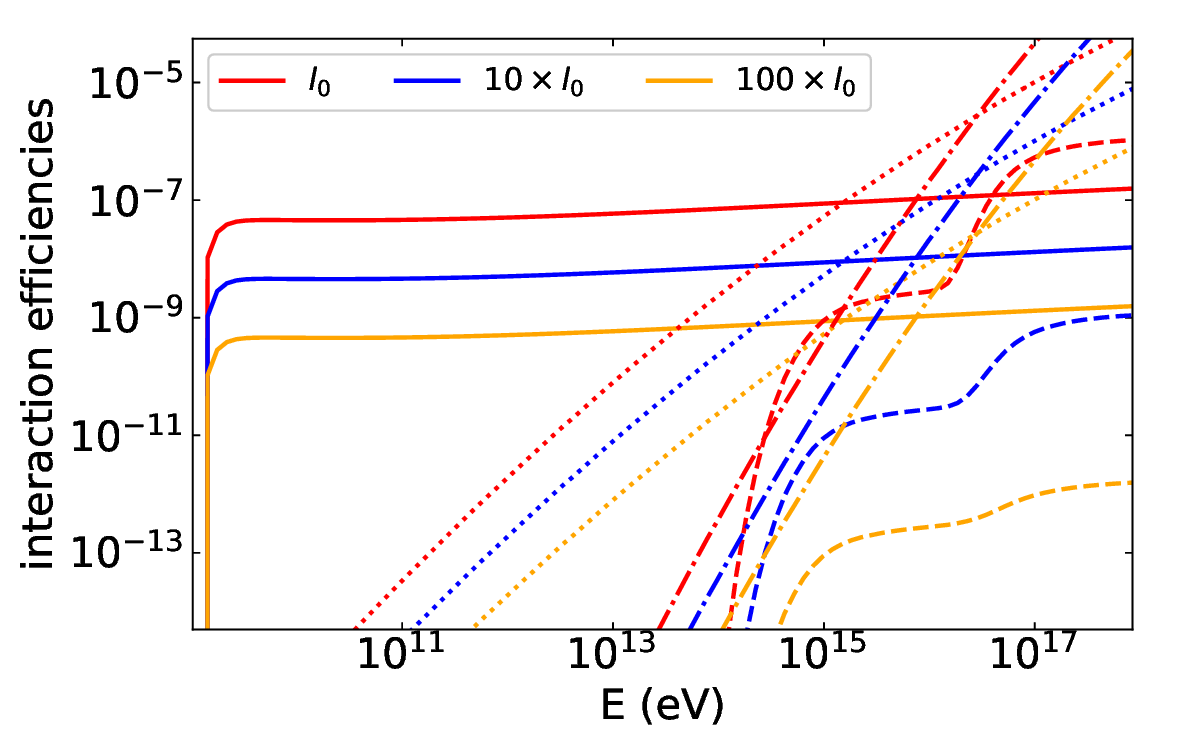}
\caption{The $p-p$ (solid curves) and $p-\gamma$ interaction efficiencies as functions of the proton energy in the comoving frame. Red, blue, and orange curves represent the distances between emitting region and the SMBH are $l_0$, $10\times l_0$, and $100\times l_0$, respectively. Dashed curves are $p-\gamma$ interaction efficiencies contributed by external photon fields including the accretion disk, BLR and DT. Dot-dashed and dotted curves represent $p-\gamma$ interaction efficiencies contributed by non-thermal photons of 3C 279 and Mrk 421, which are the typical representative of the powerful and weak jetted AGNs. Their spectra used in our calculation are historical spectra taken from the SSDC SED Builder Tool (\url{https://tools.ssdc.asi.it/}) of the Italian Space Agency \citep{2011arXiv1103.0749S}.
\label{rates}}
\end{figure}

For $p-\gamma$ interactions, the influx of external photons into the jet usually enhances the interaction efficiency, as the energy density of the external photons is amplified by a factor of $\Gamma^2$. So here we discuss the $p-\gamma$ interactions efficiency contributed by external photons first. For the $p-p$ jet model proposed in this work, it is expected that nearby AGNs would generate a higher neutrino flux on the premise that the geometric configurations of the AGN jets are comparably similar. According to \cite{2014ApJ...780...73A} and \cite{2001MNRAS.322..536W}, less powerful AGNs, including BL Lacs and LLAGNs, are more dominant in the nearby Universe. So here we check if $p-\gamma$ interactions would be more efficient for these nearby AGNs. For the less powerful AGNs, we set $L_{\rm disk}=10^{41}$~erg/s as a typical value. Other relevant parameters are taken as those outlined for blazars in Table~\ref{parameters}. The obtained efficiencies of $p-\gamma$ (dashed curves) and $p-p$ interactions (solid curves) are shown in Fig.~\ref{rates}. It can be seen that as $l$ increases the $p-\gamma$ interactions efficiency decays much more rapidly than that of the $p-p$ interactions. Most importantly, the $p-p$ interactions is significantly more efficient in producing neutrinos in the TeV-PeV range. The $p-\gamma$ interactions is only more efficient at the jet base in producing $>$ sub-EeV neutrinos. Of course, if one considers powerful AGNs with a stronger $L_{\rm disk}$, the $p-\gamma$ interactions efficiency would further increase. However, since the external photon fields' radiation spectra are blackbody or graybody spectra with certain characteristic frequencies, the increased efficiency curve of the $p-\gamma$ interactions rises almost vertically. Consequently, the energy of neutrinos most effectively produced by the $p-\gamma$ interactions remains larger than the PeV level. Some previous detailed and comprehensive studies on the $p-\gamma$ interactions in AGN jets also confirmed this perspective \citep[e.g.,][]{2014PhRvD..90b3007M, 2020A&A...642A..92R, 2021PhRvL.126s1101R}. In addition, \cite{2021ApJ...906...51X} suggested that if the jet base is sufficiently close to the accretion disk, then the external photons from the hot corona would provide abundant targets for $p-\gamma$ interactions, making the jet base an opaque region for $\gamma$ rays. Given that the X-ray spectrum of the hot corona follows a power-law distribution, the production rate of TeV-PeV neutrinos could also be efficient. A definitive conclusion requires knowledge of the specific extent of the hot corona and the true position of jet base, both of which are currently poorly understood. The value of $l_0$ adopted in this work is about two orders of magnitude larger than that used in \cite{2021ApJ...906...51X}, so here we tentatively consider the contribution of the hot corona as potentially insignificant for $p-\gamma$ interactions \citep[to understand the contribution of the $p-\gamma$ interactions dominated by the hot corona, one may see Figure 5 in][]{2021ApJ...906...51X}. 

\begin{figure}[htbp]
\subfigure{
\includegraphics[width=0.5\textwidth]{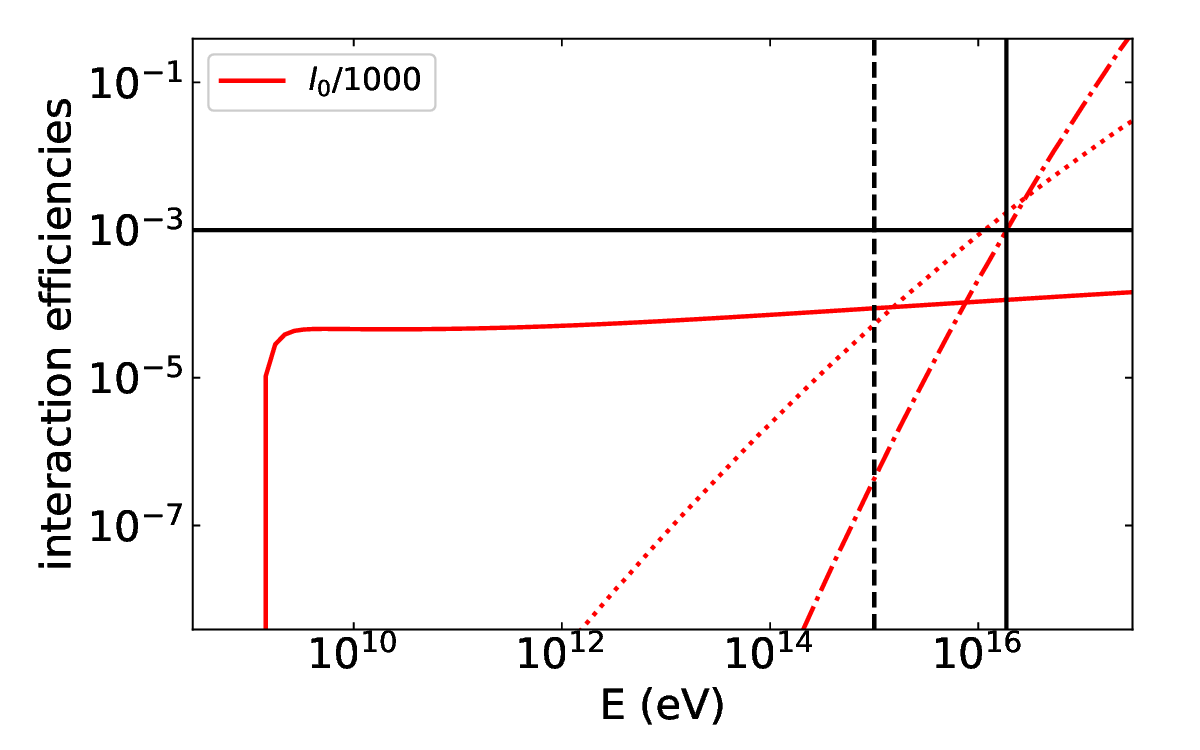}
}\hspace{-5mm}
\quad
\subfigure{
\includegraphics[width=0.5\textwidth]{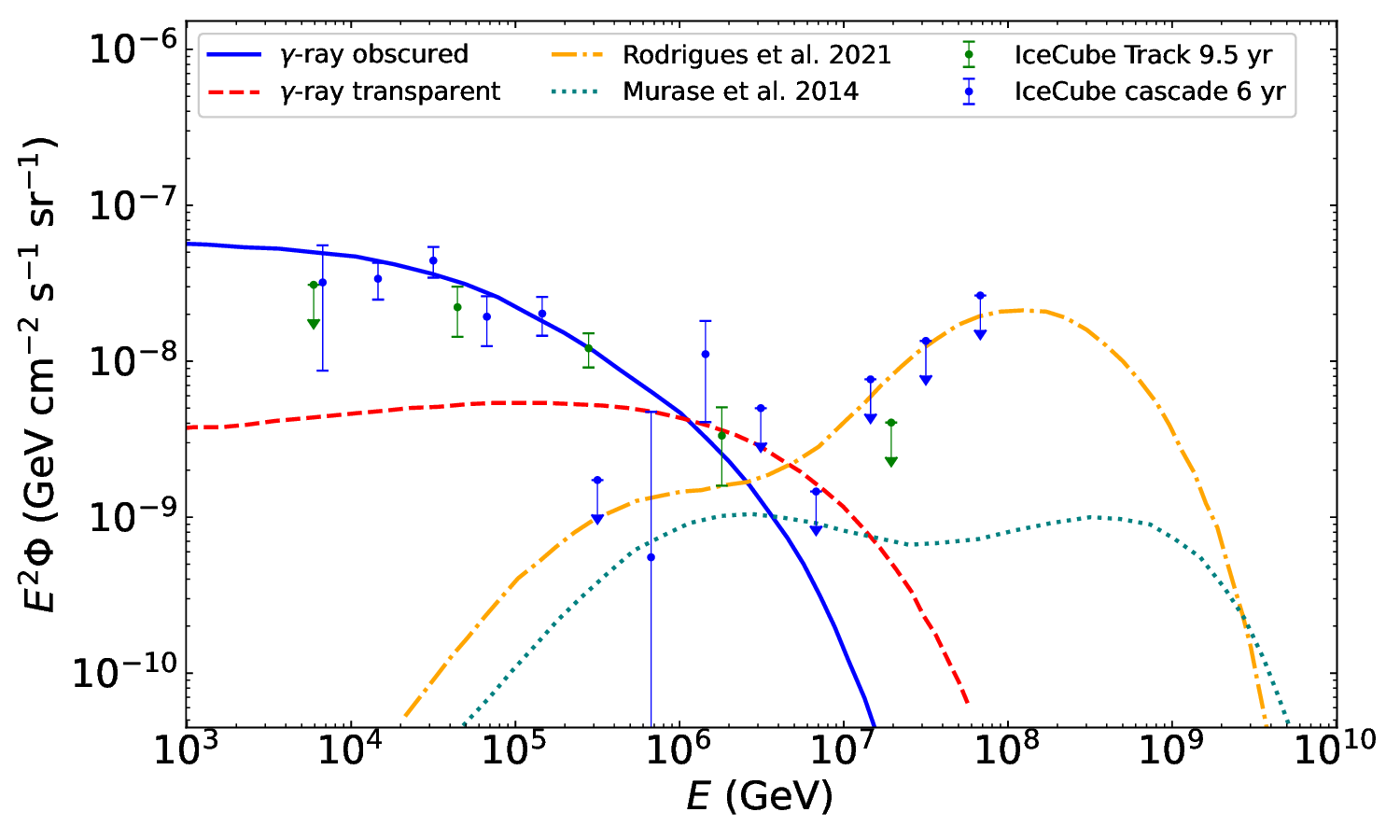}
}
\caption{Upper panel: The $p-p$ and $p-\gamma$ interaction efficiencies by reducing $l_0$ by a factor of one thousand. Solid and dashed black vertical lines correspond to the X-axes of 1 and 18.6 PeV, respectively. The vertical coordinate corresponding to the solid black horizontal line is $10^{-3}$. Other line styles have the same meaning as in Fig.~\ref{rates}.   Lower panel: The comparison between contributions of $p-p$ interactions and $p-\gamma$ interactions from jetted AGNs to the DNB. Blue solid and red dashed curves are the contributions of $p-p$ interactions in situations where the neutrino production region is either obscured or transparent to $\gamma$ rays, as shown in Fig.~\ref{diffuse} and Fig.~\ref{IGRB}. Orange dot-dashed and teal dotted curves represent the contributions of $p-\gamma$ interactions from \cite{2021PhRvL.126s1101R} and \cite{2014PhRvD..90b3007M}. Data points have the same meaning as in Fig.~\ref{diffuse}.
\label{comp}}
\end{figure}

Next, we discuss if the $p-\gamma$ interaction contributed by the internal non-thermal photons could be important. As mentioned before, due to the lack of a universal spectral shape for jetted AGNs, the estimation of the energy density distribution becomes complex. According to the synchrotron peak frequency $\nu_{\rm s}^{\rm p}$, jetted AGNs are divided into low synchrotron peaked (LSP; $\nu_{\rm s}^{\rm p}\lesssim 10^{14}~\rm Hz$), intermediate synchrotron peaked (ISP; $10^{14}~\rm Hz \lesssim \nu_{\rm s}^{\rm p}\lesssim 10^{15}~\rm Hz$), and high synchrotron peaked (HSP; $\nu_{\rm s}^{\rm p}\gtrsim 10^{15}~\rm Hz$) sources \citep{2010ApJ...716...30A}. For the sake of simplicity, we take 3C 279 and Mrk 421 as representatives of powerful LSP and weak HSP, respectively, to study their $p-\gamma$ interaction efficiencies. Following the relevant physical parameters listed in Table~\ref{parameters}, the derived $p-\gamma$ interaction efficiencies contributed by non-thermal photons for 3C 279 and Mrk 421 are shown as dot-dashed and dotted curves in Fig.~\ref{rates}. It can be seen that $p-\gamma$ interaction efficiencies contributed by non-thermal photons would be less efficient than those of $p-p$ interactions in TeV--PeV band. This is because the frequency of target photons required for generating TeV--PeV neutrinos through the $p-\gamma$ interactions is from $\sim 10^{18}~\rm Hz$ to $\sim 10^{21}~\rm Hz$ ($\nu \approx 1.5\times 10^{21}~\rm Hz(\frac{1~TeV}{E_{\rm \nu}^{\rm obs}})(\frac{\delta}{20})^2$), surpassing the synchrotron peak frequency. These target photons exhibit a soft spectrum, resulting in a low photon number density. In the energy range above PeV, $p-\gamma$ interactions become more efficient due to the higher number density of lower-frequency target photons. The different slopes of the $p-\gamma$ interaction efficiency curves for 3C 279 and Mrk 421 are caused by their distinct non-thermal spectral shapes. To further elucidate the argument that $p-\gamma$ interactions are less efficient than $p-p$ interactions in TeV--PeV range, we can reduce the size of the emission region to enhance efficiencies (on the premise that the non-thermal photon spectrum and $\delta$ remain unchanged), rendering the emission region opaque to $\gamma$ rays above the TeV band. This corresponds to the scenario where the neutrino production region is opaque to $\gamma$ rays as studied in Section~\ref{3.1}. Since the non-thermal photons in the jet serve as targets for both $p-\gamma$ interactions and internal $\gamma-\gamma$ absorption, the efficiency of $p-\gamma$ interactions $f_{\rm ph}$ is interconnected with the $\gamma-\gamma$ optical depth $\tau_{\gamma\gamma}$, i.e., $f_{\rm ph}(E_{\rm p,c})\simeq 10^{-3}\tau_{\gamma\gamma}(E_{\rm c})$, where $E_{\rm p,c}\simeq 5.7\times 10^5E_{\rm c}$ \citep{2019ApJ...871...81X}. As shown in Fig.~\ref{diffuse}, we have $E_{\rm c}\approx650/\delta~\rm GeV$ for the EGB, then $E_{\rm p,c}\approx 18.6~\rm PeV$. By reducing $l_0$ by a factor of one thousand, the corresponding $p-p$ interactions efficiency, and $p-\gamma$ interactions efficiencies for 3C 279 and Mrk 421 are shown in the upper panel of Fig.~\ref{comp}. It can be seen that $f_{\rm ph}(E_{\rm p,c}) \approx 10^{-3}$ (black vertical and horizontal lines), resulting in a $\gamma$-ray obscured emitting region since $\tau_{\gamma\gamma}(E_{\rm c}) \approx 1$. By comparing with the results in Fig.~\ref{rates}, it also can be found that the relative positions of these three interaction efficiency curves remain unchanged. In other words, $p-p$ interactions continue to be more efficient in generating TeV--PeV neutrinos. This is because both $p-p$ and $p-\gamma$ interaction efficiencies are both inversely proportional to $R_l$ \citep[i.e., $f_{\rm pp/ph}\propto R_l^{-1}$, see Eq.~\ref{analytical} for $p-p$ interactions and][for $p-\gamma$ interactions]{2012ApJ...755..147D, 2019ApJ...871...81X}. In addition to the analytical study based on the representative AGNs above, previous multimessenger studies have employed the blazar sequence paradigm \citep{1998MNRAS.299..433F} to derive the non-thermal photons spectra of jets. These investigations have also meticulously examined the potential contribution of $p-\gamma$ interactions to the DNB and suggested that the $p-\gamma$ interactions in jetted AGNs are more likely to produce PeV-EeV neutrinos \citep{2003ApJ...586...79A, 2014PhRvD..90b3007M, 2018ApJ...854...54R, 2020A&A...642A..92R, 2021PhRvL.126s1101R}. To provide a more direct comparison of the contributions of $p-\gamma$ and $p-p$ interactions to the DNB, we contrast the diffuse neutrino intensity resulting from $p-\gamma$ interactions contributed by both the non-thermal photons and external photons obtained from \cite{2014PhRvD..90b3007M} and \cite{2021PhRvL.126s1101R} with those arising from $p-p$ interactions in the lower panel of Fig.~\ref{comp}. It can be seen that $p-p$ interactions dominate the contribution to the TeV-PeV energy range, while above the PeV energy range, $p-\gamma$ interactions are more significant.

Overall, here we suggest that under certain assumptions, the $p-p$ interactions may significantly contribute to the TeV-PeV DNB. Meanwhile, the $p-\gamma$ interactions, when more efficient, could play an important role in contributing to the PeV-EeV DNB, which has yet to be detected with a definitive spectrum.

\begin{figure}[htbp]
\includegraphics[width=0.5\textwidth]{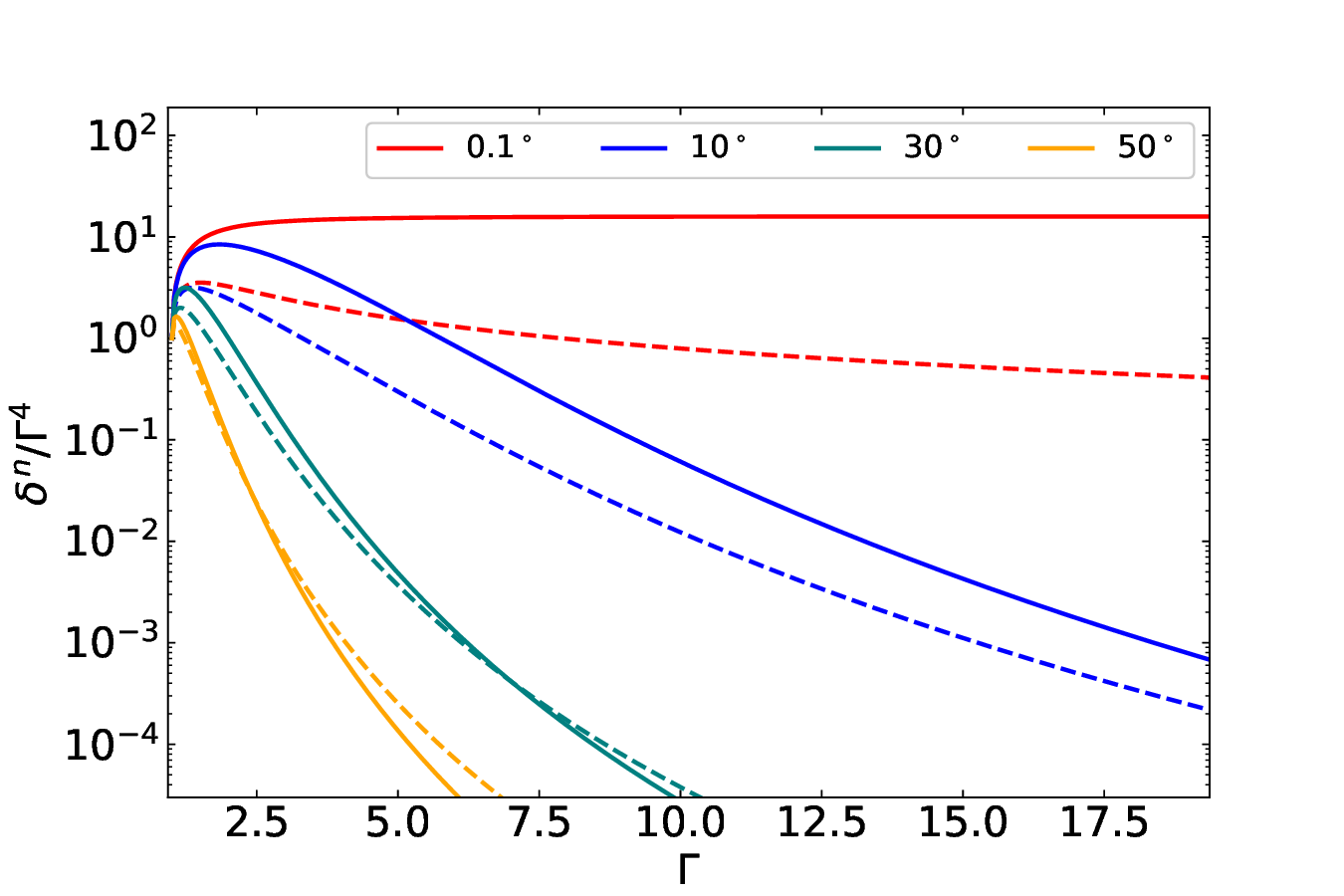}
\caption{The factor $\delta^n/\Gamma^4$ as a function of $\Gamma$ with different viewing angles. Solid and dashed curves mean $n=4$ and $n=3$ for different beaming patterns, respectively. The meaning of colors for all curves is explained in the inset legends.
\label{angle}}
\end{figure}

\section{Discussions}\label{sum}
Our calculations show that $p-p$ interactions in AGN jets have the potential to contribute towards DNB significantly. However, there are some caveats to be noted. As shown in Eq.~(\ref{analytical}), when calculating the observed neutrino flux from the $p-p$ interactions, a factor $\delta^n/\Gamma^4$ is incorporated. This factor reflects the constraints that the jet's total power does not surpass the Eddington luminosity and two beaming patterns are taken into consideration. 
In the case of considering the jet's emission is dominated by a moving blob that injected at the jet base, given that blazar's jet is directed towards the observer, then we have $(\delta/\Gamma)^4\approx 16$. As a result, the observed neutrino emission is virtually unaffected by the actual value of $\Gamma$, bolstering the reliability of the blazar's result. However, for radio AGNs with large viewing angles, the factor $(\delta/\Gamma)^4$ could be potentially small, leading to a substantial decrease the observed neutrino flux. In Fig.~\ref{angle}, we illustrate the variation of $\delta^n/\Gamma^4$ with $\Gamma$ under different viewing angles. It can be seen that as the viewing angle increases, the value of $(\delta/\Gamma)^4$ under larger viewing angles is typically less than unity, implying a substantial reduction in the observed neutrino flux. Consequently, if one still posits that radio AGNs could contribute to the DNB, it would suggest that the jet base of radio AGNs is proximal to the central SMBH, or that the jet's opening angle is exceedingly narrow, especially given that the value of $\xi$ employed in the current modeling is already close to the upper limit. In the case of considering that the entire jet emission can produce emission, i.e., it is in a steady component, the observed emission from the blazar would decrease because $n=3$. In the modeling of blazars, in order to achieve results equivalent to $n=4$, we move the jet base closer to the SMBH, which means the initial radius of the jet base is smaller. With the currently selected parameters, the transverse radius of the jet base is $1.9\times10^{13}~\rm cm$, which is equivalent to the $R_{\rm S}$ with $M_{\rm BH}=10^8~M_{\odot}$. For AGNs with larger $M_{\rm BH}$, in order to ensure that the transverse radius of the jet base is not less than the $R_{\rm S}$ \citep{2013ApJ...775..118N}, it is necessary to further adjust other parameters, such as reducing $\Gamma$ or increasing $\xi$. For radio AGNs, as can be seen from Fig.~\ref{angle}, the observed flux would not change significantly because when having a large $\theta_{\rm obs}$, the value of $\delta$ is close to unity. Besides the single moving blob and continuous jet, \cite{1997ApJ...484..108S} proposed another interesting scenario that considers many luminous blobs along the jet. In this scenario, it is assumed that the jet has $N$ blobs, each with the same radiative luminosity $L_{\rm blob}$. Even if the observed luminosity from a single blob is amplified by $\delta^4$, the total observed luminosity does not simply increase by $N\delta^4$ times the intrinsic luminosity of a blob, because the observer will see fewer blobs if $\theta_{\rm obs}<90^\circ$. Then the observed luminosity from blobs in the jet is given as $\frac{\delta^3}{\Gamma}NL_{\rm blob}$. In this scenario, more extreme parameters have to be introduced.

With the obtained neutrino luminosity, it is also possible to calculate the expected number of events from a single source. Considering that only two AGNs, namely TXS 0506+056 and NGC 1068, have detected neutrino multiplet so far, we maintain the assumption that the number of neutrino events from any single source does not exceed two. Convolving the IceCube effective area with the flux derived with our $p-p$ model, the model predicted number of neutrino event is given as $N_{\nu}=\int_{E_{\nu,{\rm min}}}^{E_{\nu,{\rm max}}}{\rm d}{E_{\nu}A_{\rm eff}\left(E_{\nu},\delta_{\rm decl}\right)\phi_\nu(E_{\nu}, M_{\rm BH})}\Delta T$, where $E_{\nu,{\rm min}}=100~\rm GeV$ and $E_{\nu,{\rm max}}=100~\rm PeV$ are the lower and upper bounds of the neutrino energy, respectively, $\Delta T=9.5\rm~year$ is the detection time interval, $A_{\rm eff}$ is the IceCube point-source effective area \citep{2019ICRC...36..851C}, which varies across four ranges of declination $\delta_{\rm decl}$, which are $-90^\circ \sim -30^\circ$, $-30^\circ \sim -5^\circ$, $-5^\circ \sim 30^\circ$, and $30^\circ \sim 90^\circ$, respectively. Please note that the multiplet check here is only for blazars, as previous discussions have shown that the parameter space for radio AGNs is already tight. Assuming that there are no significant differences in the geometric structure of each blazar jet, specifically the opening angle $\theta_{\rm open}$ and the location of the jet base $l_0$, it can be inferred that blazars in closer proximity would yield a higher neutrino flux as predicted by the $p-p$ jet model. Within the four ranges of $\delta_{\rm decl}$, the blazars with the smallest redshift ($z$) are PKS 0056-572 with $z=0.018$, 1RXS J022314.6-111741 with $z=0.042$, 4C +04.77 with $z=0.027$, and TXS 0149+710 with $z=0.022$, respectively. If setting $M_{\rm BH}=10^8~M_\odot$ as suggested in \cite{2013ApJ...764...45K}, for the nearby AGNs, the upper limits of $\xi$ for these four nearby blazars that cannot detect multiplet are 0.41, 0.38, 0.07, and 0.06, respectively. The value of $\xi$ adopted in our previous modeling of blazars is 0.05, which is lower than the upper limits derived above and further supports the reliability of explaining the DNB using the $p-p$ interactions in blazar jets. For nearby AGNs with larger masses of SMBH, the detection of multiplets can be avoided by fine-tuning the jet's opening angle or by positioning the base of the jet closer to the SMBH.

\section*{Acknowledgements}
R.X. thanks R.-Y. Liu for helpful discussions and comments. J.C.J. thanks S. Rakshit for helpful discussions and insights. This work is supported by the National Natural Science Foundation of China (NSFC) under the Grants No. 12203043 and 12203024. This paper is dedicated to Mr. Zhisheng Zhao, who passed away on 2023 February 5 in Laian, China.

\bibliography{ms.bib}{}

\begin{thebibliography}{}
\expandafter\ifx\csname natexlab\endcsname\relax\def\natexlab#1{#1}\fi
\providecommand{\url}[1]{\href{#1}{#1}}
\providecommand{\dodoi}[1]{doi:~\href{http://doi.org/#1}{\nolinkurl{#1}}}
\providecommand{\doeprint}[1]{\href{http://ascl.net/#1}{\nolinkurl{http://ascl.net/#1}}}
\providecommand{\doarXiv}[1]{\href{https://arxiv.org/abs/#1}{\nolinkurl{https://arxiv.org/abs/#1}}}

\bibitem[{{Aartsen} {et~al.}(2013){Aartsen}, {Abbasi}, {Abdou}, {Ackermann},
  {Adams}, {Aguilar}, {Ahlers}, {Altmann}, {Auffenberg}, {Bai}, {Baker},
  {Barwick}, {Baum}, {Bay}, {Beatty}, {Bechet}, {Becker Tjus}, {Becker},
  {Bell}, {Benabderrahmane}, {BenZvi}, {Berdermann}, {Berghaus}, {Berley},
  {Bernardini}, {Bernhard}, {Bertrand}, {Besson}, {Binder}, {Bindig}, {Bissok},
  {Blaufuss}, {Blumenthal}, {Boersma}, {Bohaichuk}, {Bohm}, {Bose},
  {B{\"o}ser}, {Botner}, {Brayeur}, {Bretz}, {Brown}, {Bruijn}, {Brunner},
  {Carson}, {Casey}, {Casier}, {Chirkin}, {Christov}, {Christy}, {Clark},
  {Clevermann}, {Coenders}, {Cohen}, {Cowen}, {Cruz Silva}, {Danninger},
  {Daughhetee}, {Davis}, {De Clercq}, {De Ridder}, {Desiati}, {de With},
  {DeYoung}, {D{\'\i}az-V{\'e}lez}, {Dunkman}, {Eagan}, {Eberhardt}, {Eisch},
  {Ellsworth}, {Euler}, {Evenson}, {Fadiran}, {Fazely}, {Fedynitch},
  {Feintzeig}, {Feusels}, {Filimonov}, {Finley}, {Fischer-Wasels}, {Flis},
  {Franckowiak}, {Franke}, {Frantzen}, {Fuchs}, {Gaisser}, {Gallagher},
  {Gerhardt}, {Gladstone}, {Gl{\"u}senkamp}, {Goldschmidt}, {Golup},
  {Gonzalez}, {Goodman}, {G{\'o}ra}, {Grant}, {Gro{\ss}}, {Gurtner}, {Ha}, {Haj
  Ismail}, {Hallen}, {Hallgren}, {Halzen}, {Hanson}, {Heereman}, {Heinen},
  {Helbing}, {Hellauer}, {Hickford}, {Hill}, {Hoffman}, {Hoffmann}, {Homeier},
  {Hoshina}, {Huelsnitz}, {Hulth}, {Hultqvist}, {Hussain}, {Ishihara},
  {Jacobi}, {Jacobsen}, {Jagielski}, {Japaridze}, {Jero}, {Jlelati},
  {Kaminsky}, {Kappes}, {Karg}, {Karle}, {Kelley}, {Kiryluk}, {Kislat},
  {Kl{\"a}s}, {Klein}, {K{\"o}hne}, {Kohnen}, {Kolanoski}, {K{\"o}pke},
  {Kopper}, {Kopper}, {Koskinen}, {Kowalski}, {Krasberg}, {Krings}, {Kroll},
  {Kunnen}, {Kurahashi}, {Kuwabara}, {Labare}, {Landsman}, {Larson},
  {Lesiak-Bzdak}, {Leuermann}, {Leute}, {L{\"u}nemann}, {Madsen}, {Maruyama},
  {Mase}, {Matis}, {McNally}, {Meagher}, {Merck}, {M{\'e}sz{\'a}ros}, {Meures},
  {Miarecki}, {Middell}, {Milke}, {Miller}, {Mohrmann}, {Montaruli}, {Morse},
  {Nahnhauer}, {Naumann}, {Niederhausen}, {Nowicki}, {Nygren}, {Obertacke},
  {Odrowski}, {Olivas}, {Olivo}, {O'Murchadha}, {Paul}, {Pepper}, {P{\'e}rez de
  los Heros}, {Pfendner}, {Pieloth}, {Pinat}, {Pirk}, {Posselt}, {Price},
  {Przybylski}, {R{\"a}del}, {Rameez}, {Rawlins}, {Redl}, {Reimann}, {Resconi},
  {Rhode}, {Ribordy}, {Richman}, {Riedel}, {Rodrigues}, {Rott}, {Ruhe},
  {Ruzybayev}, {Ryckbosch}, {Saba}, {Salameh}, {Sander}, {Santander}, {Sarkar},
  {Schatto}, {Scheel}, {Scheriau}, {Schmidt}, {Schmitz}, {Schoenen},
  {Sch{\"o}neberg}, {Sch{\"o}nwald}, {Schukraft}, {Schulte}, {Schulz},
  {Seckel}, {Sestayo}, {Seunarine}, {Sheremata}, {Smith}, {Soiron}, {Soldin},
  {Spiczak}, {Spiering}, {Stamatikos}, {Stanev}, {Stasik}, {Stezelberger},
  {Stokstad}, {St{\"o}{\ss}l}, {Strahler}, {Str{\"o}m}, {Sullivan}, {Taavola},
  {Taboada}, {Tamburro}, {Ter-Antonyan}, {Te{\v{s}}i{\'c}}, {Tilav}, {Toale},
  {Toscano}, {Usner}, {van der Drift}, {van Eijndhoven}, {Van Overloop}, {van
  Santen}, {Vehring}, {Voge}, {Vraeghe}, {Walck}, {Waldenmaier}, {Wallraff},
  {Wasserman}, {Weaver}, {Wellons}, {Wendt}, {Westerhoff}, {Whitehorn},
  {Wiebe}, {Wiebusch}, {Williams}, {Wissing}, {Wolf}, {Wood}, {Woschnagg},
  {Xu}, {Xu}, {Xu}, {Yanez}, {Yodh}, {Yoshida}, {Zarzhitsky}, {Ziemann},
  {Zierke}, {Zilles}, \& {Zoll}}]{Aartsen2013PhR103A}
{Aartsen}, M.~G., {Abbasi}, R., {Abdou}, Y., {et~al.} 2013, \prl, 111, 021103,
  \dodoi{10.1103/PhysRevLett.111.021103}

\bibitem[{{Aartsen} {et~al.}(2017{\natexlab{a}}){Aartsen}, {Abraham},
  {Ackermann}, {Adams}, {Aguilar}, {Ahlers}, {Ahrens}, {Altmann}, {Andeen},
  {Anderson}, {Ansseau}, {Anton}, {Archinger}, {Arg{\"u}elles}, {Auffenberg},
  {Axani}, {Bai}, {Barwick}, {Baum}, {Bay}, {Beatty}, {Becker Tjus}, {Becker},
  {BenZvi}, {Berley}, {Bernardini}, {Bernhard}, {Besson}, {Binder}, {Bindig},
  {Bissok}, {Blaufuss}, {Blot}, {Bohm}, {B{\"o}rner}, {Bos}, {Bose},
  {B{\"o}ser}, {Botner}, {Braun}, {Brayeur}, {Bretz}, {Bron}, {Burgman},
  {Carver}, {Casier}, {Cheung}, {Chirkin}, {Christov}, {Clark}, {Classen},
  {Coenders}, {Collin}, {Conrad}, {Cowen}, {Cross}, {Day}, {de Andr{\'e}}, {De
  Clercq}, {del Pino Rosendo}, {Dembinski}, {De Ridder}, {Desiati}, {de Vries},
  {de Wasseige}, {de With}, {DeYoung}, {D{\'\i}az-V{\'e}lez}, {di Lorenzo},
  {Dujmovic}, {Dumm}, {Dunkman}, {Eberhardt}, {Ehrhardt}, {Eichmann}, {Eller},
  {Euler}, {Evenson}, {Fahey}, {Fazely}, {Feintzeig}, {Felde}, {Filimonov},
  {Finley}, {Flis}, {F{\"o}sig}, {Franckowiak}, {Friedman}, {Fuchs}, {Gaisser},
  {Gallagher}, {Gerhardt}, {Ghorbani}, {Giang}, {Gladstone}, {Glauch},
  {Gl{\"u}senkamp}, {Goldschmidt}, {Golup}, {Gonzalez}, {Grant}, {Griffith},
  {Haack}, {Haj Ismail}, {Hallgren}, {Halzen}, {Hansen}, {Hansmann}, {Hanson},
  {Hebecker}, {Heereman}, {Helbing}, {Hellauer}, {Hickford}, {Hignight},
  {Hill}, {Hoffman}, {Hoffmann}, {Holzapfel}, {Hoshina}, {Huang}, {Huber},
  {Hultqvist}, {In}, {Ishihara}, {Jacobi}, {Japaridze}, {Jeong}, {Jero},
  {Jones}, {Jurkovic}, {Kappes}, {Karg}, {Karle}, {Katz}, {Kauer}, {Keivani},
  {Kelley}, {Kheirandish}, {Kim}, {Kintscher}, {Kiryluk}, {Kittler}, {Klein},
  {Kohnen}, {Koirala}, {Kolanoski}, {Konietz}, {K{\"o}pke}, {Kopper}, {Kopper},
  {Koskinen}, {Kowalski}, {Krings}, {Kroll}, {Kr{\"u}ckl}, {Kr{\"u}ger},
  {Kunnen}, {Kunwar}, {Kurahashi}, {Kuwabara}, {Labare}, {Lanfranchi},
  {Larson}, {Lauber}, {Lennarz}, {Lesiak-Bzdak}, {Leuermann}, {Lu},
  {L{\"u}nemann}, {Madsen}, {Maggi}, {Mahn}, {Mancina}, {Mandelartz},
  {Maruyama}, {Mase}, {Maunu}, {McNally}, {Meagher}, {Medici}, {Meier}, {Meli},
  {Menne}, {Merino}, {Meures}, {Miarecki}, {Mohrmann}, {Montaruli}, {Moulai},
  {Nahnhauer}, {Naumann}, {Neer}, {Niederhausen}, {Nowicki}, {Nygren},
  {Obertacke Pollmann}, {Olivas}, {O'Murchadha}, {Palczewski}, {Pandya},
  {Pankova}, {Peiffer}, {Penek}, {Pepper}, {P{\'e}rez de los Heros}, {Pieloth},
  {Pinat}, {Price}, {Przybylski}, {Quinnan}, {Raab}, {R{\"a}del}, {Rameez},
  {Rawlins}, {Reimann}, {Relethford}, {Relich}, {Resconi}, {Rhode}, {Richman},
  {Riedel}, {Robertson}, {Rongen}, {Rott}, {Ruhe}, {Ryckbosch}, {Rysewyk},
  {Sabbatini}, {Sanchez Herrera}, {Sandrock}, {Sandroos}, {Sarkar},
  {Satalecka}, {Schlunder}, {Schmidt}, {Schoenen}, {Sch{\"o}neberg},
  {Schumacher}, {Seckel}, {Seunarine}, {Soldin}, {Song}, {Spiczak}, {Spiering},
  {Stanev}, {Stasik}, {Stettner}, {Steuer}, {Stezelberger}, {Stokstad},
  {St{\"o}ssl}, {Str{\"o}m}, {Strotjohann}, {Sullivan}, {Sutherland},
  {Taavola}, {Taboada}, {Tatar}, {Tenholt}, {Ter-Antonyan}, {Terliuk},
  {Te{\v{s}}i{\'c}}, {Tilav}, {Toale}, {Tobin}, {Toscano}, {Tosi},
  {Tselengidou}, {Turcati}, {Unger}, {Usner}, {Vandenbroucke}, {van
  Eijndhoven}, {Vanheule}, {van Rossem}, {van Santen}, {Veenkamp}, {Vehring},
  {Voge}, {Vogel}, {Vraeghe}, {Walck}, {Wallace}, {Wallraff}, {Wandkowsky},
  {Weaver}, {Weiss}, {Wendt}, {Westerhoff}, {Whelan}, {Wickmann}, {Wiebe},
  {Wiebusch}, {Wille}, {Williams}, {Wills}, {Wolf}, {Wood}, {Woolsey},
  {Woschnagg}, {Xu}, {Xu}, {Xu}, {Yanez}, {Yodh}, {Yoshida}, {Zoll}, \&
  {IceCube Collaboration}}]{Aartsen2017ApJ.51A}
{Aartsen}, M.~G., {Abraham}, K., {Ackermann}, M., {et~al.} 2017{\natexlab{a}},
  \apj, 835, 151, \dodoi{10.3847/1538-4357/835/2/151}

\bibitem[{{Aartsen} {et~al.}(2017{\natexlab{b}}){Aartsen}, {Ackermann},
  {Adams}, {Aguilar}, {Ahlers}, {Ahrens}, {Samarai}, {Altmann}, {Andeen},
  {Anderson}, {Ansseau}, {Anton}, {Arg{\"u}elles}, {Auffenberg}, {Axani},
  {Bagherpour}, {Bai}, {Barron}, {Barwick}, {Baum}, {Bay}, {Beatty}, {Becker
  Tjus}, {Becker}, {BenZvi}, {Berley}, {Bernardini}, {Besson}, {Binder},
  {Bindig}, {Blaufuss}, {Blot}, {Bohm}, {B{\"o}rner}, {Bos}, {Bose},
  {B{\"o}ser}, {Botner}, {Bourbeau}, {Bradascio}, {Braun}, {Brayeur},
  {Brenzke}, {Bretz}, {Bron}, {Burgman}, {Carver}, {Casey}, {Casier}, {Cheung},
  {Chirkin}, {Christov}, {Clark}, {Classen}, {Coenders}, {Collin}, {Conrad},
  {Cowen}, {Cross}, {Day}, {de Andr{\'e}}, {De Clercq}, {DeLaunay},
  {Dembinski}, {De Ridder}, {Desiati}, {de Vries}, {de Wasseige}, {de With},
  {DeYoung}, {D{\'\i}az-V{\'e}lez}, {di Lorenzo}, {Dujmovic}, {Dumm},
  {Dunkman}, {Eberhardt}, {Ehrhardt}, {Eichmann}, {Eller}, {Evenson}, {Fahey},
  {Fazely}, {Felde}, {Filimonov}, {Finley}, {Flis}, {Franckowiak}, {Friedman},
  {Fuchs}, {Gaisser}, {Gallagher}, {Gerhardt}, {Ghorbani}, {Giang}, {Glauch},
  {Gl{\"u}senkamp}, {Goldschmidt}, {Gonzalez}, {Grant}, {Griffith}, {Haack},
  {Hallgren}, {Halzen}, {Hanson}, {Hebecker}, {Heereman}, {Helbing},
  {Hellauer}, {Hickford}, {Hignight}, {Hill}, {Hoffman}, {Hoffmann},
  {Hokanson-Fasig}, {Hoshina}, {Huang}, {Huber}, {Hultqvist}, {In}, {Ishihara},
  {Jacobi}, {Japaridze}, {Jeong}, {Jero}, {Jones}, {Kalacynski}, {Kang},
  {Kappes}, {Karg}, {Karle}, {Katz}, {Kauer}, {Keivani}, {Kelley},
  {Kheirandish}, {Kim}, {Kim}, {Kintscher}, {Kiryluk}, {Kittler}, {Klein},
  {Kohnen}, {Koirala}, {Kolanoski}, {K{\"o}pke}, {Kopper}, {Kopper},
  {Koschinsky}, {Koskinen}, {Kowalski}, {Krings}, {Kroll}, {Kr{\"u}ckl},
  {Kunnen}, {Kunwar}, {Kurahashi}, {Kuwabara}, {Kyriacou}, {Labare},
  {Lanfranchi}, {Larson}, {Lauber}, {Lennarz}, {Lesiak-Bzdak}, {Leuermann},
  {Liu}, {Lu}, {L{\"u}nemann}, {Luszczak}, {Madsen}, {Maggi}, {Mahn},
  {Mancina}, {Maruyama}, {Mase}, {Maunu}, {McNally}, {Meagher}, {Medici},
  {Meier}, {Menne}, {Merino}, {Meures}, {Miarecki}, {Micallef}, {Moment{\'e}},
  {Montaruli}, {Moore}, {Moulai}, {Nahnhauer}, {Nakarmi}, {Naumann}, {Neer},
  {Niederhausen}, {Nowicki}, {Nygren}, {Obertacke Pollmann}, {Olivas},
  {O'Murchadha}, {Palczewski}, {Pandya}, {Pankova}, {Peiffer}, {Pepper},
  {P{\'e}rez de los Heros}, {Pieloth}, {Pinat}, {Plum}, {Price}, {Przybylski},
  {Raab}, {R{\"a}del}, {Rameez}, {Rawlins}, {Reimann}, {Relethford}, {Relich},
  {Resconi}, {Rhode}, {Richman}, {Robertson}, {Rongen}, {Rott}, {Ruhe},
  {Ryckbosch}, {Rysewyk}, {S{\"a}lzer}, {Sanchez Herrera}, {Sandrock},
  {Sandroos}, {Sarkar}, {Sarkar}, {Satalecka}, {Schlunder}, {Schmidt},
  {Schneider}, {Schoenen}, {Sch{\"o}neberg}, {Schumacher}, {Seckel},
  {Seunarine}, {Soldin}, {Song}, {Spiczak}, {Spiering}, {Stachurska}, {Stanev},
  {Stasik}, {Stettner}, {Steuer}, {Stezelberger}, {Stokstad}, {St{\"o}{\ss}l},
  {Strotjohann}, {Sullivan}, {Sutherland}, {Taboada}, {Tatar}, {Tenholt},
  {Ter-Antonyan}, {Terliuk}, {Te{\v{s}}i{\'c}}, {Tilav}, {Toale}, {Tobin},
  {Toscano}, {Tosi}, {Tselengidou}, {Tung}, {Turcati}, {Turley}, {Ty}, {Unger},
  {Usner}, {Vandenbroucke}, {Van Driessche}, {van Eijndhoven}, {Vanheule}, {van
  Santen}, {Vehring}, {Vogel}, {Vraeghe}, {Walck}, {Wallace}, {Wallraff},
  {Wandler}, {Wandkowsky}, {Waza}, {Weaver}, {Weiss}, {Wendt}, {Westerhoff},
  {Whelan}, {Wickmann}, {Wiebe}, {Wiebusch}, {Wille}, {Williams}, {Wills},
  {Wolf}, {Wood}, {Wood}, {Woolsey}, {Woschnagg}, {Xu}, {Xu}, {Xu}, {Yanez},
  {Yodh}, {Yoshida}, {Yuan}, {Zoll}, \& {IceCube
  Collaboration}}]{Aartsen2017ApJ..67A}
{Aartsen}, M.~G., {Ackermann}, M., {Adams}, J., {et~al.} 2017{\natexlab{b}},
  \apj, 849, 67, \dodoi{10.3847/1538-4357/aa8dfb}

\bibitem[{{Aartsen} {et~al.}(2017{\natexlab{c}}){Aartsen}, {Abraham},
  {Ackermann}, {Adams}, {Aguilar}, {Ahlers}, {Ahrens}, {Altmann}, {Andeen},
  {Anderson}, {Ansseau}, {Anton}, {Archinger}, {Arguelles}, {Arlen},
  {Auffenberg}, {Axani}, {Bai}, {Barwick}, {Baum}, {Bay}, {Beatty}, {Becker
  Tjus}, {Becker}, {BenZvi}, {Berghaus}, {Berley}, {Bernardini}, {Bernhard},
  {Besson}, {Binder}, {Bindig}, {Bissok}, {Blaufuss}, {Blot}, {Boersma},
  {Bohm}, {B{\"o}rner}, {Bos}, {Bose}, {B{\"o}ser}, {Botner}, {Braun},
  {Brayeur}, {Bretz}, {Burgman}, {Casey}, {Casier}, {Cheung}, {Chirkin},
  {Christov}, {Clark}, {Classen}, {Coenders}, {Collin}, {Conrad}, {Cowen},
  {Cruz Silva}, {Daughhetee}, {Davis}, {Day}, {de Andr{\'e}}, {De Clercq}, {del
  Pino Rosendo}, {Dembinski}, {De Ridder}, {Desiati}, {de Vries}, {de
  Wasseige}, {de With}, {DeYoung}, {D{\'\i}az-V{\'e}lez}, {di Lorenzo},
  {Dujmovic}, {Dumm}, {Dunkman}, {Eberhardt}, {Ehrhardt}, {Eichmann}, {Euler},
  {Evenson}, {Fahey}, {Fazely}, {Feintzeig}, {Felde}, {Filimonov}, {Finley},
  {Flis}, {F{\"o}sig}, {Franckowiak}, {Fuchs}, {Gaisser}, {Gaior}, {Gallagher},
  {Gerhardt}, {Ghorbani}, {Giang}, {Gladstone}, {Glagla}, {Gl{\"u}senkamp},
  {Goldschmidt}, {Golup}, {Gonzalez}, {G{\'o}ra}, {Grant}, {Griffith}, {Haack},
  {Haj Ismail}, {Hallgren}, {Halzen}, {Hansen}, {Hansmann}, {Hansmann},
  {Hanson}, {Hebecker}, {Heereman}, {Helbing}, {Hellauer}, {Hickford},
  {Hignight}, {Hill}, {Hoffman}, {Hoffmann}, {Holzapfel}, {Homeier}, {Hoshina},
  {Huang}, {Huber}, {Huelsnitz}, {Hultqvist}, {In}, {Ishihara}, {Jacobi},
  {Japaridze}, {Jeong}, {Jero}, {Jones}, {Jurkovic}, {Kappes}, {Karg}, {Karle},
  {Katz}, {Kauer}, {Keivani}, {Kelley}, {Kemp}, {Kheirandish}, {Kim},
  {Kintscher}, {Kiryluk}, {Kittler}, {Klein}, {Kohnen}, {Koirala}, {Kolanoski},
  {Konietz}, {K{\"o}pke}, {Kopper}, {Kopper}, {Koskinen}, {Kowalski}, {Krings},
  {Kroll}, {Kr{\"u}ckl}, {Kr{\"u}ger}, {Kunnen}, {Kunwar}, {Kurahashi},
  {Kuwabara}, {Labare}, {Lanfranchi}, {Larson}, {Lennarz}, {Lesiak-Bzdak},
  {Leuermann}, {Leuner}, {Lu}, {L{\"u}nemann}, {Madsen}, {Maggi}, {Mahn},
  {Mancina}, {Mandelartz}, {Maruyama}, {Mase}, {Maunu}, {McNally}, {Meagher},
  {Medici}, {Meier}, {Meli}, {Menne}, {Merino}, {Meures}, {Miarecki},
  {Middell}, {Mohrmann}, {Montaruli}, {Moulai}, {Nahnhauer}, {Naumann}, {Neer},
  {Niederhausen}, {Nowicki}, {Nygren}, {Obertacke Pollmann}, {Olivas},
  {Omairat}, {O'Murchadha}, {Palczewski}, {Pandya}, {Pankova}, {Penek},
  {Pepper}, {P{\'e}rez de los Heros}, {Pfendner}, {Pieloth}, {Pinat},
  {Posselt}, {Price}, {Przybylski}, {Quinnan}, {Raab}, {R{\"a}del}, {Rameez},
  {Rawlins}, {Reimann}, {Relich}, {Resconi}, {Rhode}, {Richman}, {Riedel},
  {Robertson}, {Rongen}, {Rott}, {Ruhe}, {Ryckbosch}, {Rysewyk}, {Sabbatini},
  {Sanchez Herrera}, {Sandrock}, {Sandroos}, {Sarkar}, {Satalecka}, {Schimp},
  {Schlunder}, {Schmidt}, {Schoenen}, {Sch{\"o}neberg}, {Sch{\"o}nwald},
  {Schumacher}, {Seckel}, {Seunarine}, {Soldin}, {Song}, {Spiczak}, {Spiering},
  {Stahlberg}, {Stamatikos}, {Stanev}, {Stasik}, {Steuer}, {Stezelberger},
  {Stokstad}, {St{\"o}{\ss}l}, {Str{\"o}m}, {Strotjohann}, {Sullivan},
  {Sutherland}, {Taavola}, {Taboada}, {Tatar}, {Ter-Antonyan}, {Terliuk},
  {Te{\v{s}}i{\'c}}, {Tilav}, {Toale}, {Tobin}, {Toscano}, {Tosi},
  {Tselengidou}, {Turcati}, {Unger}, {Usner}, {Vallecorsa}, {Vandenbroucke},
  {van Eijndhoven}, {Vanheule}, {van Rossem}, {van Santen}, {Veenkamp},
  {Vehring}, {Voge}, {Vraeghe}, {Walck}, {Wallace}, {Wallraff}, {Wandkowsky},
  {Weaver}, {Wendt}, {Westerhoff}, {Whelan}, {Wickmann}, {Wiebe}, {Wiebusch},
  {Wille}, {Williams}, {Wills}, {Wissing}, {Wolf}, {Wood}, {Woolsey},
  {Woschnagg}, {Xu}, {Xu}, {Xu}, {Yanez}, {Yodh}, {Yoshida}, {Zoll}, \&
  {IceCube Collaboration}}]{2017ApJ...835...45A}
{Aartsen}, M.~G., {Abraham}, K., {Ackermann}, M., {et~al.} 2017{\natexlab{c}},
  \apj, 835, 45, \dodoi{10.3847/1538-4357/835/1/45}

\bibitem[{{Aartsen} {et~al.}(2020{\natexlab{a}}){Aartsen}, {Ackermann},
  {Adams}, {Aguilar}, {Ahlers}, {Ahrens}, {Alispach}, {Andeen}, {Anderson},
  {Ansseau}, {Anton}, {Arg{\"u}elles}, {Auffenberg}, {Axani}, {Backes},
  {Bagherpour}, {Bai}, {Balagopal}, {Barbano}, {Barwick}, {Bastian}, {Baum},
  {Baur}, {Bay}, {Beatty}, {Becker}, {Becker Tjus}, {BenZvi}, {Berley},
  {Bernardini}, {Besson}, {Binder}, {Bindig}, {Blaufuss}, {Blot}, {Bohm},
  {B{\"o}rner}, {B{\"o}ser}, {Botner}, {B{\"o}ttcher}, {Bourbeau}, {Bourbeau},
  {Bradascio}, {Braun}, {Bron}, {Brostean-Kaiser}, {Burgman}, {Buscher},
  {Busse}, {Carver}, {Chen}, {Cheung}, {Chirkin}, {Choi}, {Clark}, {Classen},
  {Coleman}, {Collin}, {Conrad}, {Coppin}, {Correa}, {Cowen}, {Cross}, {Dave},
  {De Clercq}, {DeLaunay}, {Dembinski}, {Deoskar}, {De Ridder}, {Desiati}, {de
  Vries}, {de Wasseige}, {de With}, {DeYoung}, {Diaz}, {D{\'\i}az-V{\'e}lez},
  {Dujmovic}, {Dunkman}, {Dvorak}, {Eberhardt}, {Ehrhardt}, {Eller}, {Engel},
  {Evenson}, {Fahey}, {Fazely}, {Felde}, {Filimonov}, {Finley}, {Fox},
  {Franckowiak}, {Friedman}, {Fritz}, {Gaisser}, {Gallagher}, {Ganster},
  {Garrappa}, {Gerhardt}, {Ghorbani}, {Glauch}, {Gl{\"u}senkamp},
  {Goldschmidt}, {Gonzalez}, {Grant}, {Griffith}, {Griswold}, {G{\"u}nder},
  {G{\"u}nd{\"u}z}, {Haack}, {Hallgren}, {Halliday}, {Halve}, {Halzen},
  {Hanson}, {Haungs}, {Hebecker}, {Heereman}, {Heix}, {Helbing}, {Hellauer},
  {Henningsen}, {Hickford}, {Hignight}, {Hill}, {Hoffman}, {Hoffmann},
  {Hoinka}, {Hokanson-Fasig}, {Hoshina}, {Huang}, {Huber}, {Huber},
  {Hultqvist}, {H{\"u}nnefeld}, {Hussain}, {In}, {Iovine}, {Ishihara},
  {Japaridze}, {Jeong}, {Jero}, {Jones}, {Jonske}, {Joppe}, {Kang}, {Kang},
  {Kappes}, {Kappesser}, {Karg}, {Karl}, {Karle}, {Katz}, {Kauer}, {Kelley},
  {Kheirandish}, {Kim}, {Kintscher}, {Kiryluk}, {Kittler}, {Klein}, {Koirala},
  {Kolanoski}, {K{\"o}pke}, {Kopper}, {Kopper}, {Koskinen}, {Kowalski},
  {Krings}, {Kr{\"u}ckl}, {Kulacz}, {Kurahashi}, {Kyriacou}, {Labare},
  {Lanfranchi}, {Larson}, {Lauber}, {Lazar}, {Leonard}, {Leszczy{\'n}ska},
  {Leuermann}, {Liu}, {Lohfink}, {Lozano Mariscal}, {Lu}, {Lucarelli},
  {L{\"u}nemann}, {Luszczak}, {Lyu}, {Ma}, {Madsen}, {Maggi}, {Mahn}, {Makino},
  {Mallik}, {Mallot}, {Mancina}, {Mari{\c{s}}}, {Maruyama}, {Mase}, {Matis},
  {Maunu}, {McNally}, {Meagher}, {Medici}, {Medina}, {Meier}, {Meighen-Berger},
  {Menne}, {Merino}, {Meures}, {Micallef}, {Mockler}, {Moment{\'e}},
  {Montaruli}, {Moore}, {Morse}, {Moulai}, {Muth}, {Nagai}, {Naumann}, {Neer},
  {Niederhausen}, {Nisa}, {Nowicki}, {Nygren}, {Obertacke Pollmann}, {Oehler},
  {Olivas}, {O'Murchadha}, {O'Sullivan}, {Palczewski}, {Pandya}, {Pankova},
  {Park}, {Peiffer}, {P{\'e}rez de los Heros}, {Philippen}, {Pieloth}, {Pinat},
  {Pizzuto}, {Plum}, {Porcelli}, {Price}, {Przybylski}, {Raab}, {Raissi},
  {Rameez}, {Rauch}, {Rawlins}, {Rea}, {Reimann}, {Relethford}, {Renschler},
  {Renzi}, {Resconi}, {Rhode}, {Richman}, {Robertson}, {Rongen}, {Rott},
  {Ruhe}, {Ryckbosch}, {Rysewyk}, {Safa}, {Sanchez Herrera}, {Sandrock},
  {Sandroos}, {Santander}, {Sarkar}, {Sarkar}, {Satalecka}, {Schaufel},
  {Schieler}, {Schlunder}, {Schmidt}, {Schneider}, {Schneider}, {Schr{\"o}der},
  {Schumacher}, {Sclafani}, {Seckel}, {Seunarine}, {Shefali}, {Silva},
  {Snihur}, {Soedingrekso}, {Soldin}, {Song}, {Spiczak}, {Spiering},
  {Stachurska}, {Stamatikos}, {Stanev}, {Stein}, {Steinm{\"u}ller}, {Stettner},
  {Steuer}, {Stezelberger}, {Stokstad}, {St{\"o}{\ss}l}, {Strotjohann},
  {St{\"u}rwald}, {Stuttard}, {Sullivan}, {Taboada}, {Tenholt}, {Ter-Antonyan},
  {Terliuk}, {Tilav}, {Tollefson}, {Tomankova}, {T{\"o}nnis}, {Toscano},
  {Tosi}, {Trettin}, {Tselengidou}, {Tung}, {Turcati}, {Turcotte}, {Turley},
  {Ty}, {Unger}, {Unland Elorrieta}, {Usner}, {Vandenbroucke}, {Van Driessche},
  {van Eijk}, {van Eijndhoven}, {Vanheule}, {van Santen}, {Vraeghe}, {Walck},
  {Wallace}, {Wallraff}, {Wandkowsky}, {Watson}, {Weaver}, {Weindl}, {Weiss},
  {Weldert}, {Wendt}, {Werthebach}, {Whelan}, {Whitehorn}, {Wiebe}, {Wiebusch},
  {Wille}, {Williams}, {Wills}, {Wolf}, {Wood}, {Wood}, {Woschnagg}, {Wrede},
  {Xu}, {Xu}, {Xu}, {Yanez}, {Yodh}, {Yoshida}, {Yuan}, \&
  {Z{\"o}cklein}}]{2020PhRvL.124e1103A}
{Aartsen}, M.~G., {Ackermann}, M., {Adams}, J., {et~al.} 2020{\natexlab{a}},
  \prl, 124, 051103, \dodoi{10.1103/PhysRevLett.124.051103}

\bibitem[{{Aartsen} {et~al.}(2020{\natexlab{b}}){Aartsen}, {Ackermann},
  {Adams}, {Aguilar}, {Ahlers}, {Ahrens}, {Alispach}, {Andeen}, {Anderson},
  {Ansseau}, {Anton}, {Arg{\"u}elles}, {Auffenberg}, {Axani}, {Backes},
  {Bagherpour}, {Bai}, {Balagopal V.}, {Barbano}, {Barwick}, {Bastian}, {Baum},
  {Baur}, {Bay}, {Beatty}, {Becker}, {Becker Tjus}, {BenZvi}, {Berley},
  {Bernardini}, {Besson}, {Binder}, {Bindig}, {Blaufuss}, {Blot}, {Bohm},
  {B{\"o}ser}, {Botner}, {B{\"o}ttcher}, {Bourbeau}, {Bourbeau}, {Bradascio},
  {Braun}, {Bron}, {Brostean-Kaiser}, {Burgman}, {Buscher}, {Busse}, {Carver},
  {Chen}, {Cheung}, {Chirkin}, {Choi}, {Clark}, {Classen}, {Coleman}, {Collin},
  {Conrad}, {Coppin}, {Correa}, {Cowen}, {Cross}, {Dave}, {De Clercq},
  {DeLaunay}, {Dembinski}, {Deoskar}, {De Ridder}, {Desiati}, {de Vries}, {de
  Wasseige}, {de With}, {DeYoung}, {Diaz}, {D{\'\i}az-V{\'e}lez}, {Dujmovic},
  {Dunkman}, {Dvorak}, {Eberhardt}, {Ehrhardt}, {Eller}, {Engel}, {Evenson},
  {Fahey}, {Fazely}, {Felde}, {Filimonov}, {Finley}, {Fox}, {Franckowiak},
  {Friedman}, {Fritz}, {Gaisser}, {Gallagher}, {Ganster}, {Garrappa},
  {Gerhardt}, {Ghorbani}, {Glauch}, {Gl{\"u}senkamp}, {Goldschmidt},
  {Gonzalez}, {Grant}, {Gr{\'e}goire}, {Griffith}, {Griswold}, {G{\"u}nder},
  {G{\"u}nd{\"u}z}, {Haack}, {Hallgren}, {Halliday}, {Halve}, {Halzen},
  {Hanson}, {Haungs}, {Hebecker}, {Heereman}, {Heix}, {Helbing}, {Hellauer},
  {Henningsen}, {Hickford}, {Hignight}, {Hill}, {Hoffman}, {Hoffmann},
  {Hoinka}, {Hokanson-Fasig}, {Hoshina}, {Huang}, {Huber}, {Huber},
  {Hultqvist}, {H{\"u}nnefeld}, {Hussain}, {In}, {Iovine}, {Ishihara},
  {Jansson}, {Japaridze}, {Jeong}, {Jero}, {Jones}, {Jonske}, {Joppe}, {Kang},
  {Kang}, {Kappes}, {Kappesser}, {Karg}, {Karl}, {Karle}, {Katz}, {Kauer},
  {Kelley}, {Kheirandish}, {Kim}, {Kintscher}, {Kiryluk}, {Kittler}, {Klein},
  {Koirala}, {Kolanoski}, {K{\"o}pke}, {Kopper}, {Kopper}, {Koskinen},
  {Kowalski}, {Krings}, {Kr{\"u}ckl}, {Kulacz}, {Kurahashi}, {Kyriacou},
  {Lanfranchi}, {Larson}, {Lauber}, {Lazar}, {Leonard}, {Lesiak-Bzdak},
  {Leszczy{\'n}ska}, {Leuermann}, {Liu}, {Lohfink}, {Lozano Mariscal}, {Lu},
  {Lucarelli}, {L{\"u}nemann}, {Luszczak}, {Lyu}, {Ma}, {Madsen}, {Maggi},
  {Mahn}, {Makino}, {Mallik}, {Mallot}, {Mancina}, {Mari{\c{s}}}, {Maruyama},
  {Mase}, {Maunu}, {McNally}, {Meagher}, {Medici}, {Medina}, {Meier},
  {Meighen-Berger}, {Merino}, {Meures}, {Micallef}, {Mockler}, {Moment{\'e}},
  {Montaruli}, {Moore}, {Morse}, {Moulai}, {Muth}, {Nagai}, {Naumann}, {Neer},
  {Niederhausen}, {Nisa}, {Nowicki}, {Nygren}, {Obertacke Pollmann}, {Oehler},
  {Olivas}, {O'Murchadha}, {O'Sullivan}, {Palczewski}, {Pandya}, {Pankova},
  {Park}, {Peiffer}, {P{\'e}rez de los Heros}, {Philippen}, {Pieloth},
  {Pieper}, {Pinat}, {Pizzuto}, {Plum}, {Porcelli}, {Price}, {Przybylski},
  {Raab}, {Raissi}, {Rameez}, {Rauch}, {Rawlins}, {Rea}, {Rehman}, {Reimann},
  {Relethford}, {Renschler}, {Renzi}, {Resconi}, {Rhode}, {Richman},
  {Robertson}, {Rongen}, {Rott}, {Ruhe}, {Ryckbosch}, {Rysewyk}, {Safa},
  {Sanchez Herrera}, {Sandrock}, {Sandroos}, {Santander}, {Sarkar}, {Sarkar},
  {Satalecka}, {Schaufel}, {Schieler}, {Schlunder}, {Schmidt}, {Schneider},
  {Schneider}, {Schr{\"o}der}, {Schumacher}, {Sclafani}, {Seckel}, {Seunarine},
  {Shefali}, {Silva}, {Snihur}, {Soedingrekso}, {Soldin}, {Song}, {Spiczak},
  {Spiering}, {Stachurska}, {Stamatikos}, {Stanev}, {Stein}, {Stettner},
  {Steuer}, {Stezelberger}, {Stokstad}, {St{\"o}{\ss}l}, {Strotjohann},
  {St{\"u}rwald}, {Stuttard}, {Sullivan}, {Taboada}, {Tenholt}, {Ter-Antonyan},
  {Terliuk}, {Tilav}, {Tollefson}, {Tomankova}, {T{\"o}nnis}, {Toscano},
  {Tosi}, {Trettin}, {Tselengidou}, {Tung}, {Turcati}, {Turcotte}, {Turley},
  {Ty}, {Unger}, {Unland Elorrieta}, {Usner}, {Vandenbroucke}, {Van Driessche},
  {van Eijk}, {van Eijndhoven}, {van Santen}, {Verpoest}, {Vraeghe}, {Walck},
  {Wallace}, {Wallraff}, {Wandkowsky}, {Watson}, {Weaver}, {Weindl}, {Weiss},
  {Weldert}, {Wendt}, {Werthebach}, {Whelan}, {Whitehorn}, {Wiebe}, {Wiebusch},
  {Wille}, {Williams}, {Wills}, {Wolf}, {Wood}, {Wood}, {Woschnagg}, {Wrede},
  {Xu}, {Xu}, {Xu}, {Yanez}, {Yodh}, {Yoshida}, {Yuan}, {Z{\"o}cklein}, \&
  {IceCube Collaboration}}]{2020PhRvL.125l1104A}
---. 2020{\natexlab{b}}, \prl, 125, 121104,
  \dodoi{10.1103/PhysRevLett.125.121104}

\bibitem[{{Abbasi} {et~al.}(2022){Abbasi}, {Ackermann}, {Adams}, {Aguilar},
  {Ahlers}, {Ahrens}, {Alameddine}, {Alispach}, {Alves}, {Amin}, {Andeen},
  {Anderson}, {Anton}, {Arg{\"u}elles}, {Ashida}, {Axani}, {Bai}, {Balagopal
  V.}, {Barbano}, {Barwick}, {Bastian}, {Basu}, {Baur}, {Bay}, {Beatty},
  {Becker}, {Tjus}, {Bellenghi}, {BenZvi}, {Berley}, {Bernardini}, {Besson},
  {Binder}, {Bindig}, {Blaufuss}, {Blot}, {Boddenberg}, {Bontempo}, {Borowka},
  {B{\"o}ser}, {Botner}, {B{\"o}ttcher}, {Bourbeau}, {Bradascio}, {Braun},
  {Brinson}, {Bron}, {Brostean-Kaiser}, {Browne}, {Burgman}, {Burley}, {Busse},
  {Campana}, {Carnie-Bronca}, {Chen}, {Chen}, {Chirkin}, {Choi}, {Clark},
  {Clark}, {Classen}, {Coleman}, {Collin}, {Conrad}, {Coppin}, {Correa},
  {Cowen}, {Cross}, {Dappen}, {Dave}, {De Clercq}, {DeLaunay}, {L{\'o}pez},
  {Dembinski}, {Deoskar}, {Desai}, {Desiati}, {de Vries}, {de Wasseige}, {de
  With}, {DeYoung}, {Diaz}, {D{\'\i}az-V{\'e}lez}, {Dittmer}, {Dujmovic},
  {Dunkman}, {DuVernois}, {Dvorak}, {Ehrhardt}, {Eller}, {Engel}, {Erpenbeck},
  {Evans}, {Evenson}, {Fan}, {Fazely}, {Feigl}, {Fiedlschuster}, {Fienberg},
  {Filimonov}, {Finley}, {Fischer}, {Fox}, {Franckowiak}, {Friedman}, {Fritz},
  {F{\"u}rst}, {Gaisser}, {Gallagher}, {Ganster}, {Garcia}, {Garrappa},
  {Gerhardt}, {Ghadimi}, {Glaser}, {Glauch}, {Gl{\"u}senkamp}, {Gonzalez},
  {Goswami}, {Grant}, {Gr{\'e}goire}, {Griswold}, {G{\"u}nther}, {Gutjahr},
  {Haack}, {Hallgren}, {Halliday}, {Halve}, {Halzen}, {Minh}, {Hanson},
  {Hardin}, {Harnisch}, {Haungs}, {Hebecker}, {Helbing}, {Henningsen},
  {Hettinger}, {Hickford}, {Hignight}, {Hill}, {Hill}, {Hoffman}, {Hoffmann},
  {Hokanson-Fasig}, {Hoshina}, {Huang}, {Huber}, {Huber}, {Hultqvist},
  {H{\"u}nnefeld}, {Hussain}, {Hymon}, {In}, {Iovine}, {Ishihara}, {Jansson},
  {Japaridze}, {Jeong}, {Jin}, {Jones}, {Kang}, {Kang}, {Kang}, {Kappes},
  {Kappesser}, {Kardum}, {Karg}, {Karl}, {Karle}, {Katz}, {Kauer},
  {Kellermann}, {Kelley}, {Kheirandish}, {Kin}, {Kintscher}, {Kiryluk},
  {Klein}, {Koirala}, {Kolanoski}, {Kontrimas}, {K{\"o}pke}, {Kopper},
  {Kopper}, {Koskinen}, {Koundal}, {Kovacevich}, {Kowalski}, {Kozynets}, {Kun},
  {Kurahashi}, {Lad}, {Gualda}, {Lanfranchi}, {Larson}, {Lauber}, {Lazar},
  {Lee}, {Leonard}, {Leszczy{\'n}ska}, {Li}, {Lincetto}, {Liu}, {Liubarska},
  {Lohfink}, {Mariscal}, {Lu}, {Lucarelli}, {Ludwig}, {Luszczak}, {Lyu}, {Ma},
  {Madsen}, {Mahn}, {Makino}, {Mancina}, {Mari{\c{s}}}, {Martinez-Soler},
  {Maruyama}, {Mase}, {McElroy}, {McNally}, {Mead}, {Meagher}, {Mechbal},
  {Medina}, {Meier}, {Meighen-Berger}, {Micallef}, {Mockler}, {Montaruli},
  {Moore}, {Morse}, {Moulai}, {Naab}, {Nagai}, {Naumann}, {Necker}, {Nguy?n},
  {Niederhausen}, {Nisa}, {Nowicki}, {Pollmann}, {Oehler}, {Oeyen}, {Olivas},
  {O'Sullivan}, {Pandya}, {Pankova}, {Park}, {Parker}, {Paudel}, {Paul}, {de
  los Heros}, {Peters}, {Peterson}, {Philippen}, {Pieper}, {Pittermann},
  {Pizzuto}, {Plum}, {Popovych}, {Porcelli}, {Rodriguez}, {Price}, {Pries},
  {Przybylski}, {Raab}, {Raissi}, {Rameez}, {Rawlins}, {Rea}, {Rehman},
  {Reichherzer}, {Reimann}, {Renzi}, {Resconi}, {Reusch}, {Rhode}, {Richman},
  {Riedel}, {Roberts}, {Robertson}, {Roellinghoff}, {Rongen}, {Rott}, {Ruhe},
  {Ryckbosch}, {Cantu}, {Safa}, {Saffer}, {Herrera}, {Sandrock}, {Sandroos},
  {Santander}, {Sarkar}, {Sarkar}, {Satalecka}, {Schaufel}, {Schieler},
  {Schindler}, {Schmidt}, {Schneider}, {Schneider}, {Schr{\"o}der},
  {Schumacher}, {Schwefer}, {Sclafani}, {Seckel}, {Seunarine}, {Sharma},
  {Shefali}, {Silva}, {Skrzypek}, {Smithers}, {Snihur}, {Soedingrekso},
  {Soldin}, {Spannfellner}, {Spiczak}, {Spiering}, {Stachurska}, {Stamatikos},
  {Stanev}, {Stein}, {Stettner}, {Steuer}, {Stezelberger}, {St{\"u}rwald},
  {Stuttard}, {Sullivan}, {Taboada}, {Ter-Antonyan}, {Tilav}, {Tischbein},
  {Tollefson}, {T{\"o}nnis}, {Toscano}, {Tosi}, {Trettin}, {Tselengidou},
  {Tung}, {Turcati}, {Turcotte}, {Turley}, {Twagirayezu}, {Ty}, {Elorrieta},
  {Valtonen-Mattila}, {Vandenbroucke}, {van Eijndhoven}, {Vannerom}, {van
  Santen}, {Verpoest}, {Walck}, {Watson}, {Weaver}, {Weigel}, {Weindl},
  {Weiss}, {Weldert}, {Wendt}, {Werthebach}, {Weyrauch}, {Whitehorn},
  {Wiebusch}, {Williams}, {Wolf}, {Woschnagg}, {Wrede}, {Wulff}, {Xu}, {Yanez},
  {Yoshida}, {Yu}, {Yuan}, {Zhang}, {Zhelnin}, \& {IceCube
  Collaboration}}]{2022ApJ...928...50A}
{Abbasi}, R., {Ackermann}, M., {Adams}, J., {et~al.} 2022, \apj, 928, 50,
  \dodoi{10.3847/1538-4357/ac4d29}

\bibitem[{{Abdo} {et~al.}(2010){Abdo}, {Ackermann}, {Agudo}, {Ajello}, {Aller},
  {Aller}, {Angelakis}, {Arkharov}, {Axelsson}, {Bach}, {Baldini}, {Ballet},
  {Barbiellini}, {Bastieri}, {Baughman}, {Bechtol}, {Bellazzini}, {Benitez},
  {Berdyugin}, {Berenji}, {Blandford}, {Bloom}, {Boettcher}, {Bonamente},
  {Borgland}, {Bregeon}, {Brez}, {Brigida}, {Bruel}, {Burnett}, {Burrows},
  {Buson}, {Caliandro}, {Calzoletti}, {Cameron}, {Capalbi}, {Caraveo},
  {Carosati}, {Casandjian}, {Cavazzuti}, {Cecchi}, {{\c{C}}elik}, {Charles},
  {Chaty}, {Chekhtman}, {Chen}, {Chiang}, {Chincarini}, {Ciprini}, {Claus},
  {Cohen-Tanugi}, {Colafrancesco}, {Cominsky}, {Conrad}, {Costamante},
  {Cutini}, {D'ammando}, {Deitrick}, {D'Elia}, {Dermer}, {de Angelis}, {de
  Palma}, {Digel}, {Donnarumma}, {Silva}, {Drell}, {Dubois}, {Dultzin},
  {Dumora}, {Falcone}, {Farnier}, {Favuzzi}, {Fegan}, {Focke}, {Forn{\'e}},
  {Fortin}, {Frailis}, {Fuhrmann}, {Fukazawa}, {Funk}, {Fusco}, {G{\'o}mez},
  {Gargano}, {Gasparrini}, {Gehrels}, {Germani}, {Giebels}, {Giglietto},
  {Giommi}, {Giordano}, {Giuliani}, {Glanzman}, {Godfrey}, {Grenier},
  {Gronwall}, {Grove}, {Guillemot}, {Guiriec}, {Gurwell}, {Hadasch},
  {Hanabata}, {Harding}, {Hayashida}, {Hays}, {Healey}, {Heidt}, {Hiriart},
  {Horan}, {Hoversten}, {Hughes}, {Itoh}, {Jackson}, {J{\'o}hannesson},
  {Johnson}, {Johnson}, {Jorstad}, {Kadler}, {Kamae}, {Katagiri}, {Kataoka},
  {Kawai}, {Kennea}, {Kerr}, {Kimeridze}, {Kn{\"o}dlseder}, {Kocian},
  {Kopatskaya}, {Koptelova}, {Konstantinova}, {Kovalev}, {Kovalev},
  {Kurtanidze}, {Kuss}, {Lande}, {Larionov}, {Latronico}, {Leto}, {Lindfors},
  {Longo}, {Loparco}, {Lott}, {Lovellette}, {Lubrano}, {Madejski}, {Makeev},
  {Marchegiani}, {Marscher}, {Marshall}, {Max-Moerbeck}, {Mazziotta},
  {McConville}, {McEnery}, {Meurer}, {Michelson}, {Mitthumsiri}, {Mizuno},
  {Moiseev}, {Monte}, {Monzani}, {Morselli}, {Moskalenko}, {Murgia},
  {Nestoras}, {Nilsson}, {Nizhelsky}, {Nolan}, {Norris}, {Nuss}, {Ohsugi},
  {Ojha}, {Omodei}, {Orlando}, {Ormes}, {Osborne}, {Ozaki}, {Pacciani},
  {Padovani}, {Pagani}, {Page}, {Paneque}, {Panetta}, {Parent}, {Pasanen},
  {Pavlidou}, {Pelassa}, {Pepe}, {Perri}, {Pesce-Rollins}, {Piranomonte},
  {Piron}, {Pittori}, {Porter}, {Puccetti}, {Rahoui}, {Rain{\`o}}, {Raiteri},
  {Rando}, {Razzano}, {Reimer}, {Reimer}, {Reposeur}, {Richards}, {Ritz},
  {Rochester}, {Rodriguez}, {Romani}, {Ros}, {Roth}, {Roustazadeh}, {Ryde},
  {Sadrozinski}, {Sadun}, {Sanchez}, {Sander}, {Saz Parkinson}, {Scargle},
  {Sellerholm}, {Sgr{\`o}}, {Shaw}, {Sigua}, {Siskind}, {Smith}, {Smith},
  {Spandre}, {Spinelli}, {Starck}, {Stevenson}, {Stratta}, {Strickman},
  {Suson}, {Tajima}, {Takahashi}, {Takahashi}, {Takalo}, {Tanaka}, {Thayer},
  {Thayer}, {Thompson}, {Tibaldo}, {Torres}, {Tosti}, {Tramacere}, {Uchiyama},
  {Usher}, {Vasileiou}, {Verrecchia}, {Vilchez}, {Villata}, {Vitale}, {Waite},
  {Wang}, {Winer}, {Wood}, {Ylinen}, {Zensus}, {Zhekanis}, \&
  {Ziegler}}]{2010ApJ...716...30A}
{Abdo}, A.~A., {Ackermann}, M., {Agudo}, I., {et~al.} 2010, \apj, 716, 30,
  \dodoi{10.1088/0004-637X/716/1/30}

\bibitem[{{Ackermann} {et~al.}(2015){Ackermann}, {Ajello}, {Albert}, {Atwood},
  {Baldini}, {Ballet}, {Barbiellini}, {Bastieri}, {Bechtol}, {Bellazzini},
  {Bissaldi}, {Blandford}, {Bloom}, {Bottacini}, {Brandt}, {Bregeon}, {Bruel},
  {Buehler}, {Buson}, {Caliandro}, {Cameron}, {Caragiulo}, {Caraveo},
  {Cavazzuti}, {Cecchi}, {Charles}, {Chekhtman}, {Chiang}, {Chiaro}, {Ciprini},
  {Claus}, {Cohen-Tanugi}, {Conrad}, {Cuoco}, {Cutini}, {D'Ammando}, {de
  Angelis}, {de Palma}, {Dermer}, {Digel}, {Silva}, {Drell}, {Favuzzi},
  {Ferrara}, {Focke}, {Franckowiak}, {Fukazawa}, {Funk}, {Fusco}, {Gargano},
  {Gasparrini}, {Germani}, {Giglietto}, {Giommi}, {Giordano}, {Giroletti},
  {Godfrey}, {Gomez-Vargas}, {Grenier}, {Guiriec}, {Gustafsson}, {Hadasch},
  {Hayashi}, {Hays}, {Hewitt}, {Ippoliti}, {Jogler}, {J{\'o}hannesson},
  {Johnson}, {Johnson}, {Kamae}, {Kataoka}, {Kn{\"o}dlseder}, {Kuss},
  {Larsson}, {Latronico}, {Li}, {Li}, {Longo}, {Loparco}, {Lott}, {Lovellette},
  {Lubrano}, {Madejski}, {Manfreda}, {Massaro}, {Mayer}, {Mazziotta},
  {McEnery}, {Michelson}, {Mitthumsiri}, {Mizuno}, {Moiseev}, {Monzani},
  {Morselli}, {Moskalenko}, {Murgia}, {Nemmen}, {Nuss}, {Ohsugi}, {Omodei},
  {Orlando}, {Ormes}, {Paneque}, {Panetta}, {Perkins}, {Pesce-Rollins},
  {Piron}, {Pivato}, {Porter}, {Rain{\`o}}, {Rando}, {Razzano}, {Razzaque},
  {Reimer}, {Reimer}, {Reposeur}, {Ritz}, {Romani}, {S{\'a}nchez-Conde},
  {Schaal}, {Schulz}, {Sgr{\`o}}, {Siskind}, {Spandre}, {Spinelli}, {Strong},
  {Suson}, {Takahashi}, {Thayer}, {Thayer}, {Tibaldo}, {Tinivella}, {Torres},
  {Tosti}, {Troja}, {Uchiyama}, {Vianello}, {Werner}, {Winer}, {Wood}, {Wood},
  {Zaharijas}, \& {Zimmer}}]{2015ApJ...799...86A}
{Ackermann}, M., {Ajello}, M., {Albert}, A., {et~al.} 2015, \apj, 799, 86,
  \dodoi{10.1088/0004-637X/799/1/86}

\bibitem[{{Aharonian} {et~al.}(1994){Aharonian}, {Coppi}, \&
  {Voelk}}]{1994ApJ...423L...5A}
{Aharonian}, F.~A., {Coppi}, P.~S., \& {Voelk}, H.~J. 1994, \apjl, 423, L5,
  \dodoi{10.1086/187222}

\bibitem[{{Ajello} {et~al.}(2014){Ajello}, {Romani}, {Gasparrini}, {Shaw},
  {Bolmer}, {Cotter}, {Finke}, {Greiner}, {Healey}, {King}, {Max-Moerbeck},
  {Michelson}, {Potter}, {Rau}, {Readhead}, {Richards}, \&
  {Schady}}]{2014ApJ...780...73A}
{Ajello}, M., {Romani}, R.~W., {Gasparrini}, D., {et~al.} 2014, \apj, 780, 73,
  \dodoi{10.1088/0004-637X/780/1/73}

\bibitem[{{Ajello} {et~al.}(2015){Ajello}, {Gasparrini}, {S{\'a}nchez-Conde},
  {Zaharijas}, {Gustafsson}, {Cohen-Tanugi}, {Dermer}, {Inoue}, {Hartmann},
  {Ackermann}, {Bechtol}, {Franckowiak}, {Reimer}, {Romani}, \&
  {Strong}}]{2015ApJ...800L..27A}
{Ajello}, M., {Gasparrini}, D., {S{\'a}nchez-Conde}, M., {et~al.} 2015, \apjl,
  800, L27, \dodoi{10.1088/2041-8205/800/2/L27}

\bibitem[{{Albert} {et~al.}(2017){Albert}, {Andr{\'e}}, {Anghinolfi}, {Anton},
  {Ardid}, {Aubert}, {Avgitas}, {Baret}, {Barrios-Mart{\'\i}}, {Basa},
  {Belhorma}, {Bertin}, {Biagi}, {Bormuth}, {Bourret}, {Bouwhuis},
  {Br{\^a}nza{\c{s}}}, {Bruijn}, {Brunner}, {Busto}, {Capone}, {Caramete},
  {Carr}, {Celli}, {Cherkaoui El Moursli}, {Chiarusi}, {Circella}, {Coelho},
  {Coleiro}, {Coniglione}, {Costantini}, {Coyle}, {Creusot}, {D{\'\i}az},
  {Deschamps}, {de Bonis}, {Distefano}, {di Palma}, {Domi}, {Donzaud},
  {Dornic}, {Drouhin}, {Eberl}, {El Bojaddaini}, {El Khayati}, {Els{\"a}sser},
  {Enzenh{\"o}fer}, {Ettahiri}, {Fassi}, {Felis}, {Fusco}, {Galat{\`a}}, {Gay},
  {Giordano}, {Glotin}, {Gr{\'e}goire}, {Gracia Ruiz}, {Graf}, {Hallmann}, {van
  Haren}, {Heijboer}, {Hello}, {Hern{\'a}ndez-Rey}, {H{\"o}{\ss}l},
  {Hofest{\"a}dt}, {Hugon}, {Illuminati}, {James}, {de Jong}, {Jongen},
  {Kadler}, {Kalekin}, {Katz}, {Kie{\ss}ling}, {Kouchner}, {Kreter},
  {Kreykenbohm}, {Kulikovskiy}, {Lachaud}, {Lahmann}, {Lef{\`e}vre}, {Leonora},
  {Lotze}, {Loucatos}, {Marcelin}, {Margiotta}, {Marinelli},
  {Mart{\'\i}nez-Mora}, {Mele}, {Melis}, {Michael}, {Migliozzi}, {Moussa},
  {Navas}, {Nezri}, {Organokov}, {P{\v{a}}v{\v{a}}la{\c{s}}}, {Pellegrino},
  {Perrina}, {Piattelli}, {Popa}, {Pradier}, {Quinn}, {Racca}, {Riccobene},
  {S{\'a}nchez-Losa}, {Salda{\~n}a}, {Salvadori}, {Samtleben}, {Sanguineti},
  {Sapienza}, {Sch{\"u}ssler}, {Sieger}, {Spurio}, {Stolarczyk}, {Taiuti},
  {Tayalati}, {Trovato}, {Turpin}, {T{\"o}nnis}, {Vallage}, {van Elewyck},
  {Versari}, {Vivolo}, {Vizzoca}, {Wilms}, {Zornoza}, {Z{\'u}{\~n}iga}, \&
  {ANTARES Collaboration}}]{2017PhRvD..96h2001A}
{Albert}, A., {Andr{\'e}}, M., {Anghinolfi}, M., {et~al.} 2017, \prd, 96,
  082001, \dodoi{10.1103/PhysRevD.96.082001}

\bibitem[{{Albert} {et~al.}(2020){Albert}, {Andr{\'e}}, {Anghinolfi}, {Anton},
  {Ardid}, {Aubert}, {Aublin}, {Baret}, {Basa}, {Belhorma}, {Bertin}, {Biagi},
  {Bissinger}, {Boumaaza}, {Bourret}, {Bouta}, {Bouwhuis}, {Br{\^a}nza{\c{s}}},
  {Bruijn}, {Brunner}, {Busto}, {Capone}, {Caramete}, {Carr}, {Celli},
  {Chabab}, {Chau}, {El Moursli}, {Chiarusi}, {Circella}, {Coleiro}, {Colomer},
  {Coniglione}, {Costantini}, {Coyle}, {Creusot}, {D{\'\i}az}, {De Wasseige},
  {Deschamps}, {Distefano}, {di Palma}, {Domi}, {Donzaud}, {Dornic}, {Drouhin},
  {Eberl}, {Bojaddaini}, {Khayati}, {Els{\"a}sser}, {Enzenh{\"o}fer},
  {Ettahiri}, {Fassi}, {Fermani}, {Ferrara}, {Filippini}, {Fusco}, {Gay},
  {Glotin}, {Gozzini}, {Ruiz}, {Graf}, {Guidi}, {Hallmann}, {van Haren},
  {Heijboer}, {Hello}, {Hern{\'a}ndez-Rey}, {H{\"o}{\ss}l}, {Hofest{\"a}dt},
  {Illuminati}, {James}, {De Jong}, {De Jong}, {Jongen}, {Kadler}, {Kalekin},
  {Katz}, {Khan-Chowdhury}, {Kouchner}, {Kreter}, {Kreykenbohm}, {Kulikovskiy},
  {Lahmann}, {Breton}, {Lef{\`e}vre}, {Leonora}, {Levi}, {Lincetto},
  {Lopez-Coto}, {Loucatos}, {Maggi}, {Manczak}, {Marcelin}, {Margiotta},
  {Marinelli}, {Mart{\'\i}nez-Mora}, {Mele}, {Melis}, {Migliozzi}, {Moser},
  {Moussa}, {Muller}, {Nauta}, {Navas}, {Nezri}, {Nielsen},
  {Nu{\~n}ez-Casti{\~n}eyra}, {O'Fearraigh}, {Organokov},
  {P{\u{a}}v{\u{a}}la{\c{s}}}, {Pellegrino}, {Perrin-Terrin}, {Piattelli},
  {Poir{\`e}}, {Popa}, {Pradier}, {Quinn}, {Randazzo}, {Riccobene},
  {S{\'a}nchez-Losa}, {Salah-Eddine}, {Samtleben}, {Sanguineti}, {Sapienza},
  {Sch{\"u}ssler}, {Spurio}, {Stolarczyk}, {Strandberg}, {Taiuti}, {Tayalati},
  {Thakore}, {Tingay}, {Trovato}, {Vallage}, {van Elewyck}, {Versari}, {Viola},
  {Vivolo}, {Wilms}, {Zaborov}, {Zegarelli}, {Zornoza}, {Z{\'u}{\~n}iga},
  {ANTARES Collaboration}, {Aartsen}, {Ackermann}, {Adams}, {Aguilar},
  {Ahlers}, {Ahrens}, {Alispach}, {Andeen}, {Anderson}, {Ansseau}, {Anton},
  {Arg{\"u}elles}, {Auffenberg}, {Axani}, {Backes}, {Bagherpour}, {Bai},
  {Balagopal}, {Barbano}, {Barwick}, {Bastian}, {Baum}, {Baur}, {Bay},
  {Beatty}, {Becker}, {Tjus}, {Benzvi}, {Berley}, {Bernardini}, {Besson},
  {Binder}, {Bindig}, {Blaufuss}, {Blot}, {Bohm}, {B{\"o}ser}, {Botner},
  {B{\"o}ttcher}, {Bourbeau}, {Bourbeau}, {Bradascio}, {Braun}, {Bron},
  {Brostean-Kaiser}, {Burgman}, {Buscher}, {Busse}, {Carver}, {Chen}, {Cheung},
  {Chirkin}, {Choi}, {Clark}, {Classen}, {Coleman}, {Collin}, {Conrad},
  {Coppin}, {Correa}, {Cowen}, {Cross}, {Dave}, {Clercq}, {Delaunay},
  {Dembinski}, {Deoskar}, {De Ridder}, {Desiati}, {De Vries}, {De Wasseige},
  {De With}, {Deyoung}, {Diaz}, {D{\'\i}az-V{\'e}lez}, {Dujmovic}, {Dunkman},
  {Dvorak}, {Eberhardt}, {Ehrhardt}, {Eller}, {Engel}, {Evenson}, {Fahey},
  {Fazely}, {Felde}, {Filimonov}, {Finley}, {Fox}, {Franckowiak}, {Friedman},
  {Fritz}, {Gaisser}, {Gallagher}, {Ganster}, {Garrappa}, {Gerhardt},
  {Ghorbani}, {Glauch}, {Gl{\"u}senkamp}, {Goldschmidt}, {Gonzalez}, {Grant},
  {Gr{\'e}goire}, {Griffith}, {Griswold}, {G{\"u}nder}, {G{\"u}nd{\"u}z},
  {Haack}, {Hallgren}, {Halliday}, {Halve}, {Halzen}, {Hanson}, {Haungs},
  {Hebecker}, {Heereman}, {Heix}, {Helbing}, {Hellauer}, {Henningsen},
  {Hickford}, {Hignight}, {Hill}, {Hoffman}, {Hoffmann}, {Hoinka},
  {Hokanson-Fasig}, {Hoshina}, {Huang}, {Huber}, {Huber}, {Hultqvist},
  {H{\"u}nnefeld}, {Hussain}, {in}, {Iovine}, {Ishihara}, {Jansson},
  {Japaridze}, {Jeong}, {Jero}, {Jones}, {Jonske}, {Joppe}, {Kang}, {Kang},
  {Kappes}, {Kappesser}, {Karg}, {Karl}, {Karle}, {Katz}, {Kauer}, {Kelley},
  {Kheirandish}, {Kim}, {Kintscher}, {Kiryluk}, {Kittler}, {Klein}, {Koirala},
  {Kolanoski}, {K{\"o}pke}, {Kopper}, {Kopper}, {Koskinen}, {Kowalski},
  {Krings}, {Kr{\"u}ckl}, {Kulacz}, {Kurahashi}, {Kyriacou}, {Lanfranchi},
  {Larson}, {Lauber}, {Lazar}, {Leonard}, {Leszczy{\'n}ska}, {Leuermann},
  {Liu}, {Lohfink}, {Lozano Mariscal}, {Lu}, {Lucarelli}, {L{\"u}nemann},
  {Luszczak}, {Lyu}, {Ma}, {Madsen}, {Maggi}, {Mahn}, {Makino}, {Mallik},
  {Mallot}, {Mancina}, {Mari{\c{s}}}, {Maruyama}, {Mase}, {Maunu}, {McNally},
  {Meagher}, {Medici}, {Medina}, {Meier}, {Meighen-Berger}, {Merino}, {Meures},
  {Micallef}, {Mockler}, {Moment{\'e}}, {Montaruli}, {Moore}, {Morse},
  {Moulai}, {Muth}, {Nagai}, {Naumann}, {Neer}, {Niederhausen}, {Nisa},
  {Nowicki}, {Nygren}, {Obertacke Pollmann}, {Oehler}, {Olivas}, {O'Murchadha},
  {O'Sullivan}, {Palczewski}, {Pandya}, {Pankova}, {Park}, {Peiffer}, {de Los
  Heros}, {Philippen}, {Pieloth}, {Pieper}, {Pinat}, {Pizzuto}, {Plum},
  {Porcelli}, {Price}, {Przybylski}, {Raab}, {Raissi}, {Rameez}, {Rauch},
  {Rawlins}, {Rea}, {Reimann}, {Relethford}, {Renschler}, {Renzi}, {Resconi},
  {Rhode}, {Richman}, {Robertson}, {Rongen}, {Rott}, {Ruhe}, {Ryckbosch},
  {Rysewyk}, {Safa}, {Sanchez Herrera}, {Sandrock}, {Sandroos}, {Santander},
  {Sarkar}, {Sarkar}, {Satalecka}, {Schaufel}, {Schieler}, {Schlunder},
  {Schmidt}, {Schneider}, {Schneider}, {Schr{\"o}der}, {Schumacher},
  {Sclafani}, {Seckel}, {Seunarine}, {Shefali}, {Silva}, {Snihur},
  {Soedingrekso}, {Soldin}, {Song}, {Spiczak}, {Spiering}, {Stachurska},
  {Stamatikos}, {Stanev}, {Stein}, {Stettner}, {Steuer}, {Stezelberger},
  {Stokstad}, {St{\"o}{\ss}l}, {Strotjohann}, {St{\"u}rwald}, {Stuttard},
  {Sullivan}, {Taboada}, {Tenholt}, {Ter-Antonyan}, {Terliuk}, {Tilav},
  {Tollefson}, {Tomankova}, {T{\"o}nnis}, {Toscano}, {Tosi}, {Trettin},
  {Tselengidou}, {Tung}, {Turcati}, {Turcotte}, {Turley}, {Ty}, {Unger},
  {Unland Elorrieta}, {Usner}, {Vandenbroucke}, {van Driessche}, {van Eijk},
  {van Eijndhoven}, {van Santen}, {Verpoest}, {Vraeghe}, {Walck}, {Wallace},
  {Wallraff}, {Wandkowsky}, {Watson}, {Weaver}, {Weindl}, {Weiss}, {Weldert},
  {Wendt}, {Werthebach}, {Whelan}, {Whitehorn}, {Wiebe}, {Wiebusch}, {Wille},
  {Williams}, {Wills}, {Wolf}, {Wood}, {Wood}, {Woschnagg}, {Wrede}, {Xu},
  {Xu}, {Xu}, {Yanez}, {Yodh}, {Yoshida}, {Yuan}, {Z{\"o}cklein}, \& {Icecube
  Collaboration}}]{2020ApJ...892...92A}
---. 2020, \apj, 892, 92, \dodoi{10.3847/1538-4357/ab7afb}

\bibitem[{{Atoyan} \& {Dermer}(2003)}]{2003ApJ...586...79A}
{Atoyan}, A.~M., \& {Dermer}, C.~D. 2003, \apj, 586, 79, \dodoi{10.1086/346261}

\bibitem[{{Becker}(2008)}]{2008PhR...458..173B}
{Becker}, J.~K. 2008, \physrep, 458, 173, \dodoi{10.1016/j.physrep.2007.10.006}

\bibitem[{{Becker Tjus} {et~al.}(2014){Becker Tjus}, {Eichmann}, {Halzen},
  {Kheirandish}, \& {Saba}}]{2014PhRvD..89l3005B}
{Becker Tjus}, J., {Eichmann}, B., {Halzen}, F., {Kheirandish}, A., \& {Saba},
  S.~M. 2014, \prd, 89, 123005, \dodoi{10.1103/PhysRevD.89.123005}

\bibitem[{{Bennett} {et~al.}(2014){Bennett}, {Larson}, {Weiland}, \&
  {Hinshaw}}]{2014ApJ...794..135B}
{Bennett}, C.~L., {Larson}, D., {Weiland}, J.~L., \& {Hinshaw}, G. 2014, \apj,
  794, 135, \dodoi{10.1088/0004-637X/794/2/135}

\bibitem[{{Berezinsky} \& {Kalashev}(2016)}]{2016PhRvD..94b3007B}
{Berezinsky}, V., \& {Kalashev}, O. 2016, \prd, 94, 023007,
  \dodoi{10.1103/PhysRevD.94.023007}

\bibitem[{{Biermann} \& {Strittmatter}(1987)}]{1987ApJ...322..643B}
{Biermann}, P.~L., \& {Strittmatter}, P.~A. 1987, \apj, 322, 643,
  \dodoi{10.1086/165759}

\bibitem[{{Carver}(2019)}]{2019ICRC...36..851C}
{Carver}, T. 2019, in International Cosmic Ray Conference, Vol.~36, 36th
  International Cosmic Ray Conference (ICRC2019), 851,
  \dodoi{10.22323/1.358.0851}

\bibitem[{{Celotti} \& {Fabian}(1993)}]{1993MNRAS.264..228C}
{Celotti}, A., \& {Fabian}, A.~C. 1993, \mnras, 264, 228,
  \dodoi{10.1093/mnras/264.1.228}

\bibitem[{{Clausen-Brown} {et~al.}(2013){Clausen-Brown}, {Savolainen},
  {Pushkarev}, {Kovalev}, \& {Zensus}}]{2013A&A...558A.144C}
{Clausen-Brown}, E., {Savolainen}, T., {Pushkarev}, A.~B., {Kovalev}, Y.~Y., \&
  {Zensus}, J.~A. 2013, \aap, 558, A144, \dodoi{10.1051/0004-6361/201322203}

\bibitem[{{Cleary} {et~al.}(2007){Cleary}, {Lawrence}, {Marshall}, {Hao}, \&
  {Meier}}]{2007ApJ...660..117C}
{Cleary}, K., {Lawrence}, C.~R., {Marshall}, J.~A., {Hao}, L., \& {Meier}, D.
  2007, \apj, 660, 117, \dodoi{10.1086/511969}

\bibitem[{{Coppi} \& {Aharonian}(1997)}]{1997ApJ...487L...9C}
{Coppi}, P.~S., \& {Aharonian}, F.~A. 1997, \apjl, 487, L9,
  \dodoi{10.1086/310883}

\bibitem[{{Dermer} {et~al.}(2014){Dermer}, {Murase}, \&
  {Inoue}}]{2014JHEAp...3...29D}
{Dermer}, C.~D., {Murase}, K., \& {Inoue}, Y. 2014, Journal of High Energy
  Astrophysics, 3, 29, \dodoi{10.1016/j.jheap.2014.09.001}

\bibitem[{{Dermer} {et~al.}(2012){Dermer}, {Murase}, \&
  {Takami}}]{2012ApJ...755..147D}
{Dermer}, C.~D., {Murase}, K., \& {Takami}, H. 2012, \apj, 755, 147,
  \dodoi{10.1088/0004-637X/755/2/147}

\bibitem[{{Dermer} {et~al.}(2015){Dermer}, {Yan}, {Zhang}, {Finke}, \&
  {Lott}}]{2015ApJ...809..174D}
{Dermer}, C.~D., {Yan}, D., {Zhang}, L., {Finke}, J.~D., \& {Lott}, B. 2015,
  \apj, 809, 174, \dodoi{10.1088/0004-637X/809/2/174}

\bibitem[{{Di Mauro} {et~al.}(2014){Di Mauro}, {Calore}, {Donato}, {Ajello}, \&
  {Latronico}}]{pp2014ApJ_1D}
{Di Mauro}, M., {Calore}, F., {Donato}, F., {Ajello}, M., \& {Latronico}, L.
  2014, \apj, 780, 161, \dodoi{10.1088/0004-637X/780/2/161}

\bibitem[{{Eichmann} {et~al.}(2018){Eichmann}, {Rachen}, {Merten}, {van Vliet},
  \& {Becker Tjus}}]{eich2018JCAP36E}
{Eichmann}, B., {Rachen}, J.~P., {Merten}, L., {van Vliet}, A., \& {Becker
  Tjus}, J. 2018, \jcap, 2018, 036, \dodoi{10.1088/1475-7516/2018/02/036}

\bibitem[{{Fang} {et~al.}(2022){Fang}, {Gallagher}, \&
  {Halzen}}]{2022ApJ...933..190F}
{Fang}, K., {Gallagher}, J.~S., \& {Halzen}, F. 2022, \apj, 933, 190,
  \dodoi{10.3847/1538-4357/ac7649}

\bibitem[{{Fichet de Clairfontaine} {et~al.}(2023){Fichet de Clairfontaine},
  {Buson}, {Pfeiffer}, {Marchesi}, {Azzollini}, {Baghmanyan}, {Tramacere},
  {Barbano}, \& {Oswald}}]{2023ApJ...958L...2F}
{Fichet de Clairfontaine}, G., {Buson}, S., {Pfeiffer}, L., {et~al.} 2023,
  \apjl, 958, L2, \dodoi{10.3847/2041-8213/ad0644}

\bibitem[{{Fossati} {et~al.}(1998){Fossati}, {Maraschi}, {Celotti}, {Comastri},
  \& {Ghisellini}}]{1998MNRAS.299..433F}
{Fossati}, G., {Maraschi}, L., {Celotti}, A., {Comastri}, A., \& {Ghisellini},
  G. 1998, \mnras, 299, 433, \dodoi{10.1046/j.1365-8711.1998.01828.x}

\bibitem[{{Garrappa} {et~al.}(2019){Garrappa}, {Buson}, {Franckowiak},
  {Fermi-LAT Collaboration}, {Shappee}, {Beacom}, {Dong}, {Holoien},
  {Kochanek}, {Prieto}, {Stanek}, {Thompson}, {ASAS-SN Collaboration},
  {Aartsen}, {Ackermann}, {Adams}, {Aguilar}, {Ahlers}, {Ahrens}, {Alispach},
  {Andeen}, {Anderson}, {Ansseau}, {Anton}, {Arg{\"u}elles}, {Auffenberg},
  {Axani}, {Backes}, {Bagherpour}, {Bai}, {Barbano}, {Barwick}, {Baum}, {Bay},
  {Beatty}, {Becker}, {Becker Tjus}, {BenZvi}, {Berley}, {Bernardini},
  {Besson}, {Binder}, {Bindig}, {Blaufuss}, {Blot}, {Bohm}, {B{\"o}rner},
  {B{\"o}ser}, {Botner}, {Bourbeau}, {Bourbeau}, {Bradascio}, {Braun}, {Bretz},
  {Bron}, {Brostean-Kaiser}, {Burgman}, {Busse}, {Carver}, {Chen}, {Cheung},
  {Chirkin}, {Clark}, {Classen}, {Collin}, {Conrad}, {Coppin}, {Correa},
  {Cowen}, {Cross}, {Dave}, {de Andr{\'e}}, {De Clercq}, {DeLaunay},
  {Dembinski}, {Deoskar}, {De Ridder}, {Desiati}, {de Vries}, {de Wasseige},
  {de With}, {DeYoung}, {Diaz}, {D{\'\i}az-V{\'e}lez}, {Dujmovic}, {Dunkman},
  {Dvorak}, {Eberhardt}, {Ehrhardt}, {Eller}, {Evenson}, {Fahey}, {Fazely},
  {Felde}, {Filimonov}, {Finley}, {Franckowiak}, {Friedman}, {Fritz},
  {Gaisser}, {Gallagher}, {Ganster}, {Garrappa}, {Gerhardt}, {Ghorbani},
  {Glauch}, {Gl{\"u}senkamp}, {Goldschmidt}, {Gonzalez}, {Grant}, {Griffith},
  {G{\"u}nder}, {G{\"u}nd{\"u}z}, {Haack}, {Hallgren}, {Halve}, {Halzen},
  {Hanson}, {Hebecker}, {Heereman}, {Helbing}, {Hellauer}, {Henningsen},
  {Hickford}, {Hignight}, {Hill}, {Hoffman}, {Hoffmann}, {Hoinka},
  {Hokanson-Fasig}, {Hoshina}, {Huang}, {Huber}, {Hultqvist}, {H{\"u}nnefeld},
  {Hussain}, {In}, {Iovine}, {Ishihara}, {Jacobi}, {Japaridze}, {Jeong},
  {Jero}, {Jones}, {Kang}, {Kappes}, {Kappesser}, {Karg}, {Karl}, {Karle},
  {Katz}, {Kauer}, {Keivani}, {Kelley}, {Kheirandish}, {Kim}, {Kintscher},
  {Kiryluk}, {Kittler}, {Klein}, {Koirala}, {Kolanoski}, {K{\"o}pke}, {Kopper},
  {Kopper}, {Koskinen}, {Kowalski}, {Krings}, {Kr{\"u}ckl}, {Kulacz}, {Kunwar},
  {Kurahashi}, {Kyriacou}, {Labare}, {Lanfranchi}, {Larson}, {Lauber}, {Lazar},
  {Leonard}, {Leuermann}, {Liu}, {Lohfink}, {Lozano Mariscal}, {Lu},
  {Lucarelli}, {L{\"u}nemann}, {Luszczak}, {Madsen}, {Maggi}, {Mahn}, {Makino},
  {Mallot}, {Mancina}, {Mari{\c{s}}}, {Maruyama}, {Mase}, {Maunu}, {Meagher},
  {Medici}, {Medina}, {Meier}, {Meighen-Berger}, {Menne}, {Merino}, {Meures},
  {Miarecki}, {Micallef}, {Moment{\'e}}, {Montaruli}, {Moore}, {Moulai},
  {Nagai}, {Nahnhauer}, {Nakarmi}, {Naumann}, {Neer}, {Niederhausen},
  {Nowicki}, {Nygren}, {Obertacke Pollmann}, {Olivas}, {O'Murchadha},
  {O'Sullivan}, {Palczewski}, {Pandya}, {Pankova}, {Park}, {Peiffer},
  {P{\'e}rez de los Heros}, {Pieloth}, {Pinat}, {Pizzuto}, {Plum}, {Price},
  {Przybylski}, {Raab}, {Raissi}, {Rameez}, {Rauch}, {Rawlins}, {Rea},
  {Reimann}, {Relethford}, {Renzi}, {Resconi}, {Rhode}, {Richman}, {Robertson},
  {Rongen}, {Rott}, {Ruhe}, {Ryckbosch}, {Rysewyk}, {Safa}, {Sanchez Herrera},
  {Sandrock}, {Sandroos}, {Santander}, {Sarkar}, {Sarkar}, {Satalecka},
  {Schaufel}, {Schlunder}, {Schmidt}, {Schneider}, {Schneider}, {Schumacher},
  {Sclafani}, {Seckel}, {Seunarine}, {Silva}, {Snihur}, {Soedingrekso},
  {Soldin}, {Song}, {Spiczak}, {Spiering}, {Stachurska}, {Stamatikos},
  {Stanev}, {Stasik}, {Stein}, {Stettner}, {Steuer}, {Stezelberger},
  {Stokstad}, {St{\"o}{\ss}l}, {Strotjohann}, {Stuttard}, {Sullivan},
  {Sutherland}, {Taboada}, {Tenholt}, {Ter-Antonyan}, {Terliuk}, {Tilav},
  {Tomankova}, {T{\"o}nnis}, {Toscano}, {Tosi}, {Tselengidou}, {Tung},
  {Turcati}, {Turcotte}, {Turley}, {Ty}, {Unger}, {Unland Elorrieta}, {Usner},
  {Vandenbroucke}, {Van Driessche}, {van Eijk}, {van Eijndhoven}, {Vanheule},
  {van Santen}, {Vraeghe}, {Walck}, {Wallace}, {Wallraff}, {Wandkowsky},
  {Watson}, {Weaver}, {Weiss}, {Weldert}, {Wendt}, {Werthebach}, {Westerhoff},
  {Whelan}, {Whitehorn}, {Wiebe}, {Wiebusch}, {Wille}, {Williams}, {Wills},
  {Wolf}, {Wood}, {Wood}, {Woschnagg}, {Wrede}, {Xu}, {Xu}, {Xu}, {Yanez},
  {Yodh}, {Yoshida}, {Yuan}, \& {IceCube Collaboration}}]{2019ApJ...880..103G}
{Garrappa}, S., {Buson}, S., {Franckowiak}, A., {et~al.} 2019, \apj, 880, 103,
  \dodoi{10.3847/1538-4357/ab2ada}

\bibitem[{{Ghisellini}(2013)}]{2013LNP...873.....G}
{Ghisellini}, G. 2013, {Radiative Processes in High Energy Astrophysics}, Vol.
  873, \dodoi{10.1007/978-3-319-00612-3}

\bibitem[{{Ghisellini} \& {Tavecchio}(2008)}]{2008MNRAS.387.1669G}
{Ghisellini}, G., \& {Tavecchio}, F. 2008, \mnras, 387, 1669,
  \dodoi{10.1111/j.1365-2966.2008.13360.x}

\bibitem[{{Giommi} {et~al.}(2020){Giommi}, {Padovani}, {Oikonomou}, {Glauch},
  {Paiano}, \& {Resconi}}]{2020A&A...640L...4G}
{Giommi}, P., {Padovani}, P., {Oikonomou}, F., {et~al.} 2020, \aap, 640, L4,
  \dodoi{10.1051/0004-6361/202038423}

\bibitem[{{Grandi} {et~al.}(2016){Grandi}, {Capetti}, \&
  {Baldi}}]{2016MNRAS.457....2G}
{Grandi}, P., {Capetti}, A., \& {Baldi}, R.~D. 2016, \mnras, 457, 2,
  \dodoi{10.1093/mnras/stv2846}

\bibitem[{{Gu{\'e}pin} {et~al.}(2022){Gu{\'e}pin}, {Kotera}, \&
  {Oikonomou}}]{2022NatRP...4..697G}
{Gu{\'e}pin}, C., {Kotera}, K., \& {Oikonomou}, F. 2022, Nature Reviews
  Physics, 4, 697, \dodoi{10.1038/s42254-022-00504-9}

\bibitem[{{Hayasaki}(2021)}]{2021NatAs...5..436H}
{Hayasaki}, K. 2021, Nature Astronomy, 5, 436,
  \dodoi{10.1038/s41550-021-01309-z}

\bibitem[{{Hayashida} {et~al.}(2012){Hayashida}, {Madejski}, {Nalewajko},
  {Sikora}, {Wehrle}, {Ogle}, {Collmar}, {Larsson}, {Fukazawa}, {Itoh},
  {Chiang}, {Stawarz}, {Blandford}, {Richards}, {Max-Moerbeck}, {Readhead},
  {Buehler}, {Cavazzuti}, {Ciprini}, {Gehrels}, {Reimer}, {Szostek}, {Tanaka},
  {Tosti}, {Uchiyama}, {Kawabata}, {Kino}, {Sakimoto}, {Sasada}, {Sato},
  {Uemura}, {Yamanaka}, {Greiner}, {Kruehler}, {Rossi}, {Macquart}, {Bock},
  {Villata}, {Raiteri}, {Agudo}, {Aller}, {Aller}, {Arkharov}, {Bach},
  {Ben{\'\i}tez}, {Berdyugin}, {Blinov}, {Blumenthal}, {B{\"o}ttcher}, {Buemi},
  {Carosati}, {Chen}, {Di Paola}, {Dolci}, {Efimova}, {Forn{\'e}}, {G{\'o}mez},
  {Gurwell}, {Heidt}, {Hiriart}, {Jordan}, {Jorstad}, {Joshi}, {Kimeridze},
  {Konstantinova}, {Kopatskaya}, {Koptelova}, {Kurtanidze},
  {L{\"a}hteenm{\"a}ki}, {Lamerato}, {Larionov}, {Larionova}, {Larionova},
  {Leto}, {Lindfors}, {Marscher}, {McHardy}, {Molina}, {Morozova},
  {Nikolashvili}, {Nilsson}, {Reinthal}, {Roustazadeh}, {Sakamoto}, {Sigua},
  {Sillanp{\"a}{\"a}}, {Takalo}, {Tammi}, {Taylor}, {Tornikoski}, {Trigilio},
  {Troitsky}, \& {Umana}}]{2012ApJ...754..114H}
{Hayashida}, M., {Madejski}, G.~M., {Nalewajko}, K., {et~al.} 2012, \apj, 754,
  114, \dodoi{10.1088/0004-637X/754/2/114}

\bibitem[{{Hooper} {et~al.}(2016){Hooper}, {Linden}, \& {Lopez}}]{2016JCAP019H}
{Hooper}, D., {Linden}, T., \& {Lopez}, A. 2016, \jcap, 2016, 019,
  \dodoi{10.1088/1475-7516/2016/08/019}

\bibitem[{{IceCube Collaboration}(2013)}]{2013Sci...342E...1I}
{IceCube Collaboration}. 2013, Science, 342, 1242856,
  \dodoi{10.1126/science.1242856}

\bibitem[{{IceCube Collaboration} {et~al.}(2018{\natexlab{a}}){IceCube
  Collaboration}, {Aartsen}, {Ackermann}, {Adams}, {Aguilar}, {Ahlers},
  {Ahrens}, {Al Samarai}, {Altmann}, {Andeen}, {Anderson}, {Ansseau}, {Anton},
  {Arg{\"u}elles}, {Auffenberg}, {Axani}, {Bagherpour}, {Bai}, {Barron},
  {Barwick}, {Baum}, {Bay}, {Beatty}, {Becker Tjus}, {Becker}, {BenZvi},
  {Berley}, {Bernardini}, {Besson}, {Binder}, {Bindig}, {Blaufuss}, {Blot},
  {Bohm}, {B{\"o}rner}, {Bos}, {B{\"o}ser}, {Botner}, {Bourbeau}, {Bourbeau},
  {Bradascio}, {Braun}, {Brenzke}, {Bretz}, {Bron}, {Brostean-Kaiser},
  {Burgman}, {Busse}, {Carver}, {Cheung}, {Chirkin}, {Christov}, {Clark},
  {Classen}, {Coenders}, {Collin}, {Conrad}, {Coppin}, {Correa}, {Cowen},
  {Cross}, {Dave}, {Day}, {de Andr{\'e}}, {De Clercq}, {DeLaunay}, {Dembinski},
  {De Ridder}, {Desiati}, {de Vries}, {de Wasseige}, {de With}, {DeYoung},
  {D{\'\i}az-V{\'e}lez}, {di Lorenzo}, {Dujmovic}, {Dumm}, {Dunkman}, {Dvorak},
  {Eberhardt}, {Ehrhardt}, {Eichmann}, {Eller}, {Evenson}, {Fahey}, {Fazely},
  {Felde}, {Filimonov}, {Finley}, {Flis}, {Franckowiak}, {Friedman}, {Fritz},
  {Gaisser}, {Gallagher}, {Gerhardt}, {Ghorbani}, {Glauch}, {Gl{\"u}senkamp},
  {Goldschmidt}, {Gonzalez}, {Grant}, {Griffith}, {Haack}, {Hallgren},
  {Halzen}, {Hanson}, {Hebecker}, {Heereman}, {Helbing}, {Hellauer},
  {Hickford}, {Hignight}, {Hill}, {Hoffman}, {Hoffmann}, {Hoinka},
  {Hokanson-Fasig}, {Hoshina}, {Huang}, {Huber}, {Hultqvist}, {H{\"u}nnefeld},
  {Hussain}, {In}, {Iovine}, {Ishihara}, {Jacobi}, {Japaridze}, {Jeong},
  {Jero}, {Jones}, {Kalaczynski}, {Kang}, {Kappes}, {Kappesser}, {Karg},
  {Karle}, {Katz}, {Kauer}, {Keivani}, {Kelley}, {Kheirandish}, {Kim}, {Kim},
  {Kintscher}, {Kiryluk}, {Kittler}, {Klein}, {Koirala}, {Kolanoski},
  {K{\"o}pke}, {Kopper}, {Kopper}, {Koschinsky}, {Koskinen}, {Kowalski},
  {Krings}, {Kroll}, {Kr{\"u}ckl}, {Kunwar}, {Kurahashi}, {Kuwabara},
  {Kyriacou}, {Labare}, {Lanfranchi}, {Larson}, {Lauber}, {Leonard},
  {Lesiak-Bzdak}, {Leuermann}, {Liu}, {Lozano Mariscal}, {Lu}, {L{\"u}nemann},
  {Luszczak}, {Madsen}, {Maggi}, {Mahn}, {Mancina}, {Maruyama}, {Mase},
  {Maunu}, {Meagher}, {Medici}, {Meier}, {Menne}, {Merino}, {Meures},
  {Miarecki}, {Micallef}, {Moment{\'e}}, {Montaruli}, {Moore}, {Morse},
  {Moulai}, {Nahnhauer}, {Nakarmi}, {Naumann}, {Neer}, {Niederhausen},
  {Nowicki}, {Nygren}, {Obertacke Pollmann}, {Olivas}, {O'Murchadha},
  {O'Sullivan}, {Palczewski}, {Pandya}, {Pankova}, {Peiffer}, {Pepper},
  {P{\'e}rez de los Heros}, {Pieloth}, {Pinat}, {Plum}, {Price}, {Przybylski},
  {Raab}, {R{\"a}del}, {Rameez}, {Rauch}, {Rawlins}, {Rea}, {Reimann},
  {Relethford}, {Relich}, {Resconi}, {Rhode}, {Richman}, {Robertson}, {Rongen},
  {Rott}, {Ruhe}, {Ryckbosch}, {Rysewyk}, {Safa}, {S{\"a}lzer}, {Sanchez
  Herrera}, {Sandrock}, {Sandroos}, {Santander}, {Sarkar}, {Sarkar},
  {Satalecka}, {Schlunder}, {Schmidt}, {Schneider}, {Schoenen},
  {Sch{\"o}neberg}, {Schumacher}, {Sclafani}, {Seckel}, {Seunarine},
  {Soedingrekso}, {Soldin}, {Song}, {Spiczak}, {Spiering}, {Stachurska},
  {Stamatikos}, {Stanev}, {Stasik}, {Stein}, {Stettner}, {Steuer},
  {Stezelberger}, {Stokstad}, {St{\"o}{\ss}l}, {Strotjohann}, {Stuttard},
  {Sullivan}, {Sutherland}, {Taboada}, {Tatar}, {Tenholt}, {Ter-Antonyan},
  {Terliuk}, {Tilav}, {Toale}, {Tobin}, {Toennis}, {Toscano}, {Tosi},
  {Tselengidou}, {Tung}, {Turcati}, {Turley}, {Ty}, {Unger}, {Usner},
  {Vandenbroucke}, {Van Driessche}, {van Eijk}, {van Eijndhoven}, {Vanheule},
  {van Santen}, {Vogel}, {Vraeghe}, {Walck}, {Wallace}, {Wallraff}, {Wandler},
  {Wandkowsky}, {Waza}, {Weaver}, {Weiss}, {Wendt}, {Werthebach}, {Westerhoff},
  {Whelan}, {Whitehorn}, {Wiebe}, {Wiebusch}, {Wille}, {Williams}, {Wills},
  {Wolf}, {Wood}, {Wood}, {Woschnagg}, {Xu}, {Xu}, {Xu}, {Yanez}, {Yodh},
  {Yoshida}, {Yuan}, {Fermi-LAT Collaboration}, {Abdollahi}, {Ajello},
  {Angioni}, {Baldini}, {Ballet}, {Barbiellini}, {Bastieri}, {Bechtol},
  {Bellazzini}, {Berenji}, {Bissaldi}, {Blandford}, {Bonino}, {Bottacini},
  {Bregeon}, {Bruel}, {Buehler}, {Burnett}, {Burns}, {Buson}, {Cameron},
  {Caputo}, {Caraveo}, {Cavazzuti}, {Charles}, {Chen}, {Cheung}, {Chiang},
  {Chiaro}, {Ciprini}, {Cohen-Tanugi}, {Conrad}, {Costantin}, {Cutini},
  {D'Ammando}, {de Palma}, {Digel}, {Di Lalla}, {Di Mauro}, {Di Venere},
  {Dom{\'\i}nguez}, {Favuzzi}, {Franckowiak}, {Fukazawa}, {Funk}, {Fusco},
  {Gargano}, {Gasparrini}, {Giglietto}, {Giomi}, {Giommi}, {Giordano},
  {Giroletti}, {Glanzman}, {Green}, {Grenier}, {Grondin}, {Guiriec}, {Harding},
  {Hayashida}, {Hays}, {Hewitt}, {Horan}, {J{\'o}hannesson}, {Kadler},
  {Kensei}, {Kocevski}, {Krauss}, {Kreter}, {Kuss}, {La Mura}, {Larsson},
  {Latronico}, {Lemoine-Goumard}, {Li}, {Longo}, {Loparco}, {Lovellette},
  {Lubrano}, {Magill}, {Maldera}, {Malyshev}, {Manfreda}, {Mazziotta},
  {McEnery}, {Meyer}, {Michelson}, {Mizuno}, {Monzani}, {Morselli},
  {Moskalenko}, {Negro}, {Nuss}, {Ojha}, {Omodei}, {Orienti}, {Orlando},
  {Palatiello}, {Paliya}, {Perkins}, {Persic}, {Pesce-Rollins}, {Piron},
  {Porter}, {Principe}, {Rain{\`o}}, {Rando}, {Rani}, {Razzano}, {Razzaque},
  {Reimer}, {Reimer}, {Renault-Tinacci}, {Ritz}, {Rochester}, {Saz Parkinson},
  {Sgr{\`o}}, {Siskind}, {Spandre}, {Spinelli}, {Suson}, {Tajima}, {Takahashi},
  {Tanaka}, {Thayer}, {Thompson}, {Tibaldo}, {Torres}, {Torresi}, {Tosti},
  {Troja}, {Valverde}, {Vianello}, {Vogel}, {Wood}, {Wood}, {Zaharijas}, {MAGIC
  Collaboration}, {Ahnen}, {Ansoldi}, {Antonelli}, {Arcaro}, {Baack},
  {Babi{\'c}}, {Banerjee}, {Bangale}, {Barres de Almeida}, {Barrio}, {Becerra
  Gonz{\'a}lez}, {Bednarek}, {Bernardini}, {Berti}, {Bhattacharyya}, {Biland},
  {Blanch}, {Bonnoli}, {Carosi}, {Carosi}, {Ceribella}, {Chatterjee}, {Colak},
  {Colin}, {Colombo}, {Contreras}, {Cortina}, {Covino}, {Cumani}, {Da Vela},
  {Dazzi}, {De Angelis}, {De Lotto}, {Delfino}, {Delgado}, {Di Pierro},
  {Dom{\'\i}nguez}, {Dominis Prester}, {Dorner}, {Doro}, {Einecke},
  {Elsaesser}, {Fallah Ramazani}, {Fern{\'a}ndez-Barral}, {Fidalgo}, {Foffano},
  {Pfrang}, {Fonseca}, {Font}, {Franceschini}, {Fruck}, {Galindo}, {Gallozzi},
  {Garc{\'\i}a L{\'o}pez}, {Garczarczyk}, {Gaug}, {Giammaria}, {Godinovi{\'c}},
  {Gora}, {Guberman}, {Hadasch}, {Hahn}, {Hassan}, {Hayashida}, {Herrera},
  {Hose}, {Hrupec}, {Inoue}, {Ishio}, {Konno}, {Kubo}, {Kushida}, {Lelas},
  {Lindfors}, {Lombardi}, {Longo}, {L{\'o}pez}, {Maggio}, {Majumdar},
  {Makariev}, {Maneva}, {Manganaro}, {Mannheim}, {Maraschi}, {Mariotti},
  {Mart{\'\i}nez}, {Masuda}, {Mazin}, {Minev}, {M}, {Mirzoyan}, {Moralejo},
  {Moreno}, {Moretti}, {Nagayoshi}, {Neustroev}, {Niedzwiecki}, {Nievas
  Rosillo}, {Nigro}, {Nilsson}, {Ninci}, {Nishijima}, {Noda}, {Nogu{\'e}s},
  {Paiano}, {Palacio}, {Paneque}, {Paoletti}, {Paredes}, {Pedaletti},
  {Peresano}, {Persic}, {Prada Moroni}, {Prandini}, {Puljak}, {Rodriguez
  Garcia}, {Reichardt}, {Rhode}, {Rib{\'o}}, {Rico}, {Righi}, {Rugliancich},
  {Saito}, {Satalecka}, {Schweizer}, {Sitarek}, {{\v{S}}nidaric
  {\textasciiacute}}, {Sobczynska}, {Stamerra}, {Strzys}, {Suri{\'c}},
  {Takahashi}, {Tavecchio}, {Temnikov}, {Terzi{\'c}}, {Teshima},
  {Torres-Alb{\`a}}, {Treves}, {Tsujimoto}, {Vanzo}, {Vazquez Acosta}, {Vovk},
  {Ward}, {Will}, {S}, {Zaric {\textasciiacute}}, {AGILE Team}, {Lucarelli},
  {Tavani}, {Piano}, {Donnarumma}, {Pittori}, {Verrecchia}, {Barbiellini},
  {Bulgarelli}, {Caraveo}, {Cattaneo}, {Colafrancesco}, {Costa}, {Di Cocco},
  {Ferrari}, {Gianotti}, {Giuliani}, {Lipari}, {Mereghetti}, {Morselli},
  {Pacciani}, {Paoletti}, {Parmiggiani}, {Pellizzoni}, {Picozza}, {Pilia},
  {Rappoldi}, {Trois}, {Vercellone}, {Vittorini}, {ASAS-SN Team}, {Stanek},
  {Franckowiak}, {Kochanek}, {Beacom}, {Thompson}, {Holoien}, {Dong}, {Prieto},
  {Shappee}, {Holmbo}, {HAWC Collaboration}, {Abeysekara}, {Albert}, {Alfaro},
  {Alvarez}, {Arceo}, {Arteaga-Vel{\'a}zquez}, {Avila Rojas}, {Ayala Solares},
  {Becerril}, {Belmont-Moreno}, {Bernal}, {Caballero-Mora}, {Capistr{\'a}n},
  {Carrami{\~n}ana}, {Casanova}, {Castillo}, {Cotti}, {Cotzomi}, {Couti{\~n}o
  de Le{\'o}n}, {De Le{\'o}n}, {De la Fuente}, {Diaz Hernandez}, {Dichiara},
  {Dingus}, {DuVernois}, {D{\'\i}az-V{\'e}lez}, {Ellsworth}, {Engel},
  {Fiorino}, {Fleischhack}, {Fraija}, {Garc{\'\i}a-Gonz{\'a}lez}, {Garfias},
  {Gonz{\'a}lez Mu{\~n}oz}, {Gonz{\'a}lez}, {Goodman}, {Hampel-Arias},
  {Harding}, {Hernandez}, {Hona}, {Hueyotl-Zahuantitla}, {Hui},
  {H{\"u}ntemeyer}, {Iriarte}, {Jardin-Blicq}, {Joshi}, {Kaufmann}, {Kunde},
  {Lara}, {Lauer}, {Lee}, {Lennarz}, {Le{\'o}n Vargas}, {Linnemann},
  {Longinotti}, {Luis-Raya}, {Luna-Garc{\'\i}a}, {Malone}, {Marinelli},
  {Martinez}, {Martinez-Castellanos}, {Mart{\'\i}nez-Castro},
  {Mart{\'\i}nez-Huerta}, {Matthews}, {Miranda-Romagnoli}, {Moreno},
  {Mostaf{\'a}}, {Nayerhoda}, {Nellen}, {Newbold}, {Nisa}, {Noriega-Papaqui},
  {Pelayo}, {Pretz}, {P{\'e}rez-P{\'e}rez}, {Ren}, {Rho}, {Rivi{\`e}re},
  {Rosa-Gonz{\'a}lez}, {Rosenberg}, {Ruiz-Velasco}, {Ruiz-Velasco}, {Salesa
  Greus}, {Sandoval}, {Schneider}, {Schoorlemmer}, {Sinnis}, {Smith},
  {Springer}, {Surajbali}, {Tibolla}, {Tollefson}, {Torres}, {Villase{\~n}or},
  {Weisgarber}, {Werner}, {Yapici}, {Gaurang}, {Zepeda}, {Zhou}, {{\'A}lvarez},
  {H.~E.~S.~S. Collaboration}, {Abdalla}, {Ang{\"u}ner}, {Armand}, {Backes},
  {Becherini}, {Berge}, {B{\"o}ttcher}, {Boisson}, {Bolmont}, {Bonnefoy},
  {Bordas}, {Brun}, {B{\"u}chele}, {Bulik}, {Caroff}, {Carosi}, {Casanova},
  {Cerruti}, {Chakraborty}, {Chandra}, {Chen}, {Colafrancesco}, {Davids},
  {Deil}, {Devin}, {Djannati-Ata{\"\i}}, {Egberts}, {Emery}, {Eschbach},
  {Fiasson}, {Fontaine}, {Funk}, {F{\"u}{\ss}ling}, {Gallant}, {Gat{\'e}},
  {Giavitto}, {Glawion}, {Glicenstein}, {Gottschall}, {Grondin}, {Haupt},
  {Henri}, {Hinton}, {Hoischen}, {Holch}, {Huber}, {Jamrozy}, {Jankowsky},
  {Jankowsky}, {Jouvin}, {Jung-Richardt}, {Kerszberg}, {Kh{\'e}lifi}, {King},
  {Klepser}, {Kluz {\textasciiacute}niak}, {Komin}, {Kraus}, {Lefaucheur},
  {Lemi{\`e}re}, {Lemoine-Goumard}, {Lenain}, {Leser}, {Lohse},
  {L{\'o}pez-Coto}, {Lorentz}, {Lypova}, {Marandon}, {Guillem
  Mart{\'\i}-Devesa}, {Maurin}, {Mitchell}, {Moderski}, {Mohamed}, {Mohrmann},
  {Moulin}, {Murach}, {de Naurois}, {Niederwanger}, {Niemiec}, {Oakes},
  {O'Brien}, {Ohm}, {Ostrowski}, {Oya}, {Panter}, {Parsons}, {Perennes},
  {Piel}, {Pita}, {Poireau}, {Priyana Noel}, {Prokoph}, {P{\"u}hlhofer},
  {Quirrenbach}, {Raab}, {Rauth}, {Renaud}, {Rieger}, {Rinchiuso}, {Romoli},
  {Rowell}, {Rudak}, {Sasaki}, {Sanchez}, {Schlickeiser}, {Sch{\"u}ssler},
  {Schulz}, {Schwanke}, {Seglar-Arroyo}, {Shafi}, {Simoni}, {Sol}, {Stegmann},
  {Steppa}, {Tavernier}, {Taylor}, {Tiziani}, {Trichard}, {Tsirou}, {van
  Eldik}, {van Rensburg}, {van Soelen}, {Veh}, {Vincent}, {Voisin}, {Wagner},
  {Wagner}, {Wierzcholska}, {Zanin}, {Zdziarski}, {Zech}, {Ziegler}, {Zorn},
  {{\.Z}ywucka}, {INTEGRAL Team}, {Savchenko}, {Ferrigno}, {Bazzano}, {Diehl},
  {Kuulkers}, {Laurent}, {Mereghetti}, {Natalucci}, {Panessa}, {Rodi},
  {Ubertini}, {Kanata}, Teams, {Morokuma}, {Ohta}, {Tanaka}, {Mori},
  {Yamanaka}, {Kawabata}, {Utsumi}, {Nakaoka}, {Kawabata}, {Nagashima},
  {Yoshida}, {Matsuoka}, {Itoh}, {Kapteyn Team}, {Keel}, {Liverpool Telescope
  Team}, {Copperwheat}, {Steele}, {Swift/NuSTAR Team}, {Cenko}, {Cowen},
  {DeLaunay}, {Evans}, {Fox}, {Keivani}, {Kennea}, {Marshall}, {Osborne},
  {Santander}, {Tohuvavohu}, {Turley}, {VERITAS Collaboration}, {Abeysekara},
  {Archer}, {Benbow}, {Bird}, {Brill}, {Brose}, {Buchovecky}, {Buckley},
  {Bugaev}, {Christiansen}, {Connolly}, {Cui}, {Daniel}, {Errando}, {Falcone},
  {Feng}, {Finley}, {Fortson}, {Furniss}, {Gueta}, {H{\"u}tten}, {Hervet},
  {Hughes}, {Humensky}, {Johnson}, {Kaaret}, {Kar}, {Kelley-Hoskins},
  {Kertzman}, {Kieda}, {Krause}, {Krennrich}, {Kumar}, {Lang}, {Lin}, {Maier},
  {McArthur}, {Moriarty}, {Mukherjee}, {Nieto}, {O'Brien}, {Ong}, {Otte},
  {Park}, {Petrashyk}, {Pohl}, {Popkow}, {Pueschel}, {Quinn}, {Ragan},
  {Reynolds}, {Richards}, {Roache}, {Rulten}, {Sadeh}, {Santander}, {Scott},
  {Sembroski}, {Shahinyan}, {Sushch}, {Tr{\'e}panier}, {Tyler}, {Vassiliev},
  {Wakely}, {Weinstein}, {Wells}, {Wilcox}, {Wilhelm}, {Williams}, {Zitzer},
  {VLA/B Team}, {Tetarenko}, {Kimball}, {Miller-Jones}, \&
  {Sivakoff}}]{2018Sci...361.1378I}
{IceCube Collaboration}, {Aartsen}, M.~G., {Ackermann}, M., {et~al.}
  2018{\natexlab{a}}, Science, 361, eaat1378, \dodoi{10.1126/science.aat1378}

\bibitem[{{IceCube Collaboration} {et~al.}(2018{\natexlab{b}}){IceCube
  Collaboration}, {Aartsen}, {Ackermann}, {Adams}, {Aguilar}, {Ahlers},
  {Ahrens}, {Samarai}, {Altmann}, {Andeen}, {Anderson}, {Ansseau}, {Anton},
  {Arg{\"u}elles}, {Arsioli}, {Auffenberg}, {Axani}, {Bagherpour}, {Bai},
  {Barron}, {Barwick}, {Baum}, {Bay}, {Beatty}, {Becker Tjus}, {Becker},
  {BenZvi}, {Berley}, {Bernardini}, {Besson}, {Binder}, {Bindig}, {Blaufuss},
  {Blot}, {Bohm}, {B{\"o}rner}, {Bos}, {B{\"o}ser}, {Botner}, {Bourbeau},
  {Bourbeau}, {Bradascio}, {Braun}, {Brenzke}, {Bretz}, {Bron},
  {Brostean-Kaiser}, {Burgman}, {Busse}, {Carver}, {Cheung}, {Chirkin},
  {Christov}, {Clark}, {Classen}, {Coenders}, {Collin}, {Conrad}, {Coppin},
  {Correa}, {Cowen}, {Cross}, {Dave}, {Day}, {de Andr{\'e}}, {De Clercq},
  {DeLaunay}, {Dembinski}, {DeRidder}, {Desiati}, {de Vries}, {de Wasseige},
  {de With}, {DeYoung}, {D{\'\i}az-V{\'e}lez}, {di Lorenzo}, {Dujmovic},
  {Dumm}, {Dunkman}, {Dvorak}, {Eberhardt}, {Ehrhardt}, {Eichmann}, {Eller},
  {Evenson}, {Fahey}, {Fazely}, {Felde}, {Filimonov}, {Finley}, {Flis},
  {Franckowiak}, {Friedman}, {Fritz}, {Gaisser}, {Gallagher}, {Gerhardt},
  {Ghorbani}, {Giommi}, {Glauch}, {Gl{\"u}senkamp}, {Goldschmidt}, {Gonzalez},
  {Grant}, {Griffith}, {Haack}, {Hallgren}, {Halzen}, {Hanson}, {Hebecker},
  {Heereman}, {Helbing}, {Hellauer}, {Hickford}, {Hignight}, {Hill}, {Hoffman},
  {Hoffmann}, {Hoinka}, {Hokanson-Fasig}, {Hoshina}, {Huang}, {Huber},
  {Hultqvist}, {H{\"u}nnefeld}, {Hussain}, {In}, {Iovine}, {Ishihara},
  {Jacobi}, {Japaridze}, {Jeong}, {Jero}, {Jones}, {Kalaczynski}, {Kang},
  {Kappes}, {Kappesser}, {Karg}, {Karle}, {Katz}, {Kauer}, {Keivani}, {Kelley},
  {Kheirandish}, {Kim}, {Kim}, {Kintscher}, {Kiryluk}, {Kittler}, {Klein},
  {Koirala}, {Kolanoski}, {K{\"o}pke}, {Kopper}, {Kopper}, {Koschinsky},
  {Koskinen}, {Kowalski}, {Krammer}, {Krings}, {Kroll}, {Kr{\"u}ckl}, {Kunwar},
  {Kurahashi}, {Kuwabara}, {Kyriacou}, {Labare}, {Lanfranchi}, {Larson},
  {Lauber}, {Leonard}, {Lesiak-Bzdak}, {Leuermann}, {Liu}, {Lozano Mariscal},
  {Lu}, {L{\"u}nemann}, {Luszczak}, {Madsen}, {Maggi}, {Mahn}, {Mancina},
  {Maruyama}, {Mase}, {Maunu}, {Meagher}, {Medici}, {Meier}, {Menne}, {Merino},
  {Meures}, {Miarecki}, {Micallef}, {Moment{\'e}}, {Montaruli}, {Moore},
  {Morse}, {Moulai}, {Nahnhauer}, {Nakarmi}, {Naumann}, {Neer}, {Niederhausen},
  {Nowicki}, {Nygren}, {Obertacke Pollmann}, {Olivas}, {O'Murchadha},
  {O'Sullivan}, {Padovani}, {Palczewski}, {Pandya}, {Pankova}, {Peiffer},
  {Pepper}, {P{\'e}rez de los Heros}, {Pieloth}, {Pinat}, {Plum}, {Price},
  {Przybylski}, {Raab}, {R{\"a}del}, {Rameez}, {Rawlins}, {Rea}, {Reimann},
  {Relethford}, {Relich}, {Resconi}, {Rhode}, {Richman}, {Robertson}, {Rongen},
  {Rott}, {Ruhe}, {Ryckbosch}, {Rysewyk}, {Safa}, {Sahakyan}, {S{\"a}lzer},
  {Sanchez Herrera}, {Sandrock}, {Sandroos}, {Santander}, {Sarkar}, {Sarkar},
  {Satalecka}, {Schlunder}, {Schmidt}, {Schneider}, {Schoenen},
  {Sch{\"o}neberg}, {Schumacher}, {Sclafani}, {Seckel}, {Seunarine},
  {Soedingrekso}, {Soldin}, {Song}, {Spiczak}, {Spiering}, {Stachurska},
  {Stamatikos}, {Stanev}, {Stasik}, {Stettner}, {Steuer}, {Stezelberger},
  {Stokstad}, {St{\"o}{\ss}l}, {Strotjohann}, {Stuttard}, {Sullivan},
  {Sutherland}, {Taboada}, {Tatar}, {Tenholt}, {Ter-Antonyan}, {Terliuk},
  {Tilav}, {Toale}, {Tobin}, {Toennis}, {Toscano}, {Tosi}, {Tselengidou},
  {Tung}, {Turcati}, {Turley}, {Ty}, {Unger}, {Usner}, {Vandenbroucke}, {Van
  Driessche}, {van Eijk}, {van Eijndhoven}, {Vanheule}, {van Santen}, {Vogel},
  {Vraeghe}, {Walck}, {Wallace}, {Wallraff}, {Wandler}, {Wandkowsky}, {Waza},
  {Weaver}, {Weiss}, {Wendt}, {Werthebach}, {Westerhoff}, {Whelan},
  {Whitehorn}, {Wiebe}, {Wiebusch}, {Wille}, {Williams}, {Wills}, {Wolf},
  {Wood}, {Wood}, {Woschnagg}, {Xu}, {Xu}, {Xu}, {Yanez}, {Yodh}, {Yoshida}, \&
  {Yuan}}]{2018Sci...361..147I}
---. 2018{\natexlab{b}}, Science, 361, 147, \dodoi{10.1126/science.aat2890}

\bibitem[{{IceCube Collaboration} {et~al.}(2022){IceCube Collaboration},
  {Abbasi}, {Ackermann}, {Adams}, {Aguilar}, {Ahlers}, {Ahrens}, {Alameddine},
  {Alispach}, {Alves}, {Amin}, {Andeen}, {Anderson}, {Anton}, {Arg{\"u}elles},
  {Ashida}, {Axani}, {Bai}, {Balagopal}, {Barbano}, {Barwick}, {Bastian},
  {Basu}, {Baur}, {Bay}, {Beatty}, {Becker}, {Becker Tjus}, {Bellenghi},
  {Benzvi}, {Berley}, {Bernardini}, {Besson}, {Binder}, {Bindig}, {Blaufuss},
  {Blot}, {Boddenberg}, {Bontempo}, {Borowka}, {B{\"o}ser}, {Botner},
  {B{\"o}ttcher}, {Bourbeau}, {Bradascio}, {Braun}, {Brinson}, {Bron},
  {Brostean-Kaiser}, {Browne}, {Burgman}, {Burley}, {Busse}, {Campana},
  {Carnie-Bronca}, {Chen}, {Chen}, {Chirkin}, {Choi}, {Clark}, {Clark},
  {Classen}, {Coleman}, {Collin}, {Conrad}, {Coppin}, {Correa}, {Cowen},
  {Cross}, {Dappen}, {Dave}, {de Clercq}, {Delaunay}, {Delgado L{\'o}pez},
  {Dembinski}, {Deoskar}, {Desai}, {Desiati}, {de Vries}, {de Wasseige}, {de
  With}, {Deyoung}, {Diaz}, {D{\'\i}az-V{\'e}lez}, {Dittmer}, {Dujmovic},
  {Dunkman}, {Duvernois}, {Dvorak}, {Ehrhardt}, {Eller}, {Engel}, {Erpenbeck},
  {Evans}, {Evenson}, {Fan}, {Fazely}, {Fedynitch}, {Feigl}, {Fiedlschuster},
  {Fienberg}, {Filimonov}, {Finley}, {Fischer}, {Fox}, {Franckowiak},
  {Friedman}, {Fritz}, {F{\"u}rst}, {Gaisser}, {Gallagher}, {Ganster},
  {Garcia}, {Garrappa}, {Gerhardt}, {Ghadimi}, {Glaser}, {Glauch},
  {Gl{\"u}senkamp}, {Goldschmidt}, {Gonzalez}, {Goswami}, {Grant},
  {Gr{\'e}goire}, {Griswold}, {G{\"u}nther}, {Gutjahr}, {Haack}, {Hallgren},
  {Halliday}, {Halve}, {Halzen}, {Hanson}, {Hardin}, {Harnisch}, {Haungs},
  {Hebecker}, {Helbing}, {Henningsen}, {Hettinger}, {Hickford}, {Hignight},
  {Hill}, {Hill}, {Hoffman}, {Hoffmann}, {Hokanson-Fasig}, {Hoshina}, {Huang},
  {Huber}, {Huber}, {Hultqvist}, {H{\"u}nnefeld}, {Hussain}, {Hymon}, {in},
  {Iovine}, {Ishihara}, {Jansson}, {Japaridze}, {Jeong}, {Jin}, {Jones},
  {Kang}, {Kang}, {Kang}, {Kappes}, {Kappesser}, {Kardum}, {Karg}, {Karl},
  {Karle}, {Katz}, {Kauer}, {Kellermann}, {Kelley}, {Kheirandish}, {Kin},
  {Kintscher}, {Kiryluk}, {Klein}, {Koirala}, {Kolanoski}, {Kontrimas},
  {K{\"o}pke}, {Kopper}, {Kopper}, {Koskinen}, {Koundal}, {Kovacevich},
  {Kowalski}, {Kozynets}, {Kun}, {Kurahashi}, {Lad}, {Lagunas Gualda},
  {Lanfranchi}, {Larson}, {Lauber}, {Lazar}, {Lee}, {Leonard},
  {Leszczy{\'n}ska}, {Li}, {Lincetto}, {Liu}, {Liubarska}, {Lohfink}, {Lozano
  Mariscal}, {Lu}, {Lucarelli}, {Ludwig}, {Luszczak}, {Lyu}, {Ma}, {Madsen},
  {Mahn}, {Makino}, {Mancina}, {Mari{\c{s}}}, {Martinez-Soler}, {Maruyama},
  {Mase}, {McElroy}, {McNally}, {Mead}, {Meagher}, {Mechbal}, {Medina},
  {Meier}, {Meighen-Berger}, {Micallef}, {Mockler}, {Montaruli}, {Moore},
  {Morse}, {Moulai}, {Naab}, {Nagai}, {Nahnhauer}, {Naumann}, {Necker},
  {Nguyen}, {Niederhausen}, {Nisa}, {Nowicki}, {Nygren}, {Obertack},
  {Pollmann}, {Oehler}, {Oeyen}, {Olivas}, {O'Sullivan}, {Pandya}, {Pankova},
  {Park}, {Parker}, {Paudel}, {Paul}, {P{\'e}rez de Los Heros}, {Peters},
  {Peterson}, {Philippen}, {Pieper}, {Pittermann}, {Pizzuto}, {Plum},
  {Popovych}, {Porcelli}, {Prado Rodriguez}, {Price}, {Pries}, {Przybylski},
  {Rack-Helleis}, {Raissi}, {Rameez}, {Rawlins}, {Rea}, {Rehman},
  {Reichherzer}, {Reimann}, {Renzi}, {Resconi}, {Reusch}, {Rhode}, {Richman},
  {Riedel}, {Roberts}, {Robertson}, {Roellinghoff}, {Rongen}, {Rott}, {Ruhe},
  {Ryckbosch}, {Rysewyk Cantu}, {Safa}, {Saffer}, {Sanchez Herrera},
  {Sandrock}, {Sandroos}, {Santander}, {Sarkar}, {Sarkar}, {Satalecka},
  {Schaufel}, {Schieler}, {Schindler}, {Schmidt}, {Schneider}, {Schneider},
  {Schr{\"o}der}, {Schumacher}, {Schwefer}, {Sclafani}, {Seckel}, {Seunarine},
  {Sharma}, {Shefali}, {Silva}, {Skrzypek}, {Smithers}, {Snihur},
  {Soedingrekso}, {Soldin}, {Spannfellner}, {Spiczak}, {Spiering},
  {Stachurska}, {Stamatikos}, {Stanev}, {Stein}, {Stettner}, {Steuer},
  {Stezelberger}, {Stokstad}, {St{\"u}rwald}, {Stuttard}, {Sullivan},
  {Taboada}, {Ter-Antonyan}, {Tilav}, {Tischbein}, {Tollefson}, {T{\"o}nnis},
  {Toscano}, {Tosi}, {Trettin}, {Tselengidou}, {Tung}, {Turcati}, {Turcotte},
  {Turley}, {Twagirayezu}, {Ty}, {Unland Elorrieta}, {Valtonen-Mattila},
  {Vandenbroucke}, {van Eijndhoven}, {Vannerom}, {van Santen}, {Verpoest},
  {Walck}, {Watson}, {Weaver}, {Weigel}, {Weindl}, {Weiss}, {Weldert}, {Wendt},
  {Werthebach}, {Weyrauch}, {Whitehorn}, {Wiebusch}, {Williams}, {Wolf},
  {Woschnagg}, {Wrede}, {Wulff}, {Xu}, {Yanez}, {Yoshida}, {Yu}, {Yuan},
  {Zhangan}, \& {Zhelnin}}]{2022Sci...378..538I}
{IceCube Collaboration}, {Abbasi}, R., {Ackermann}, M., {et~al.} 2022, Science,
  378, 538, \dodoi{10.1126/science.abg3395}

\bibitem[{{Inoue} {et~al.}(2019){Inoue}, {Khangulyan}, {Inoue}, \&
  {Doi}}]{2019ApJ...880...40I}
{Inoue}, Y., {Khangulyan}, D., {Inoue}, S., \& {Doi}, A. 2019, \apj, 880, 40,
  \dodoi{10.3847/1538-4357/ab2715}

\bibitem[{{Jiang} {et~al.}(2024){Jiang}, {Liao}, {Wang}, {Xue}, {Jiang}, \&
  {Wang}}]{2024ApJ...965L...2J}
{Jiang}, X., {Liao}, N.-H., {Wang}, Y.-B., {et~al.} 2024, \apjl, 965, L2,
  \dodoi{10.3847/2041-8213/ad36b9}

\bibitem[{{Kadler} {et~al.}(2016){Kadler}, {Krau{\ss}}, {Mannheim}, {Ojha},
  {M{\"u}ller}, {Schulz}, {Anton}, {Baumgartner}, {Beuchert}, {Buson},
  {Carpenter}, {Eberl}, {Edwards}, {Eisenacher Glawion}, {Els{\"a}sser},
  {Gehrels}, {Gr{\"a}fe}, {Gulyaev}, {Hase}, {Horiuchi}, {James}, {Kappes},
  {Kappes}, {Katz}, {Kreikenbohm}, {Kreter}, {Kreykenbohm}, {Langejahn},
  {Leiter}, {Litzinger}, {Longo}, {Lovell}, {McEnery}, {Natusch}, {Phillips},
  {Pl{\"o}tz}, {Quick}, {Ros}, {Stecker}, {Steinbring}, {Stevens}, {Thompson},
  {Tr{\"u}stedt}, {Tzioumis}, {Weston}, {Wilms}, \&
  {Zensus}}]{2016NatPh..12..807K}
{Kadler}, M., {Krau{\ss}}, F., {Mannheim}, K., {et~al.} 2016, Nature Physics,
  12, 807, \dodoi{10.1038/nphys3715}

\bibitem[{{Kelly} \& {Shen}(2013)}]{2013ApJ...764...45K}
{Kelly}, B.~C., \& {Shen}, Y. 2013, \apj, 764, 45,
  \dodoi{10.1088/0004-637X/764/1/45}

\bibitem[{{Kelner} {et~al.}(2006){Kelner}, {Aharonian}, \&
  {Bugayov}}]{2006PhRvD..74c4018K}
{Kelner}, S.~R., {Aharonian}, F.~A., \& {Bugayov}, V.~V. 2006, \prd, 74,
  034018, \dodoi{10.1103/PhysRevD.74.034018}

\bibitem[{{Kurahashi} {et~al.}(2022){Kurahashi}, {Murase}, \&
  {Santander}}]{2022ARNPS..72..365K}
{Kurahashi}, N., {Murase}, K., \& {Santander}, M. 2022, Annual Review of
  Nuclear and Particle Science, 72, 365,
  \dodoi{10.1146/annurev-nucl-011122-061547}

\bibitem[{{Li} {et~al.}(2022){Li}, {Xue}, {Long}, {Wang}, {Nagataki}, {Yan}, \&
  {Wang}}]{2022A&A...659A.184L}
{Li}, W.-J., {Xue}, R., {Long}, G.-B., {et~al.} 2022, \aap, 659, A184,
  \dodoi{10.1051/0004-6361/202142051}

\bibitem[{{Liao} {et~al.}(2022){Liao}, {Sheng}, {Jiang}, {Chang}, {Wang}, {Xu},
  {Shu}, {Fan}, \& {Wang}}]{2022ApJ...932L..25L}
{Liao}, N.-H., {Sheng}, Z.-F., {Jiang}, N., {et~al.} 2022, \apjl, 932, L25,
  \dodoi{10.3847/2041-8213/ac756f}

\bibitem[{{Lind} \& {Blandford}(1985)}]{1985ApJ...295..358L}
{Lind}, K.~R., \& {Blandford}, R.~D. 1985, \apj, 295, 358,
  \dodoi{10.1086/163380}

\bibitem[{{Liodakis} {et~al.}(2018){Liodakis}, {Hovatta}, {Huppenkothen},
  {Kiehlmann}, {Max-Moerbeck}, \& {Readhead}}]{2018ApJ...866..137L}
{Liodakis}, I., {Hovatta}, T., {Huppenkothen}, D., {et~al.} 2018, \apj, 866,
  137, \dodoi{10.3847/1538-4357/aae2b7}

\bibitem[{{Liu} {et~al.}(2020){Liu}, {Xi}, \& {Wang}}]{2020PhRvD.102h3028L}
{Liu}, R.-Y., {Xi}, S.-Q., \& {Wang}, X.-Y. 2020, \prd, 102, 083028,
  \dodoi{10.1103/PhysRevD.102.083028}

\bibitem[{{Luashvili} {et~al.}(2023){Luashvili}, {Boisson}, {Zech},
  {Arrieta-Lobo}, \& {Kynoch}}]{2023MNRAS.523..404L}
{Luashvili}, A., {Boisson}, C., {Zech}, A., {Arrieta-Lobo}, M., \& {Kynoch}, D.
  2023, \mnras, 523, 404, \dodoi{10.1093/mnras/stad1393}

\bibitem[{{Marscher}(2009)}]{2009arXiv0909.2576M}
{Marscher}, A.~P. 2009, arXiv e-prints, arXiv:0909.2576,
  \dodoi{10.48550/arXiv.0909.2576}

\bibitem[{{Moharana} \& {Razzaque}(2015)}]{moharana2015JCAP}
{Moharana}, R., \& {Razzaque}, S. 2015, \jcap, 2015, 014,
  \dodoi{10.1088/1475-7516/2015/08/014}

\bibitem[{{Moharana} \& {Razzaque}(2016)}]{moharana2016JCAP}
---. 2016, \jcap, 2016, 021, \dodoi{10.1088/1475-7516/2016/12/021}

\bibitem[{{Murase} {et~al.}(2013){Murase}, {Ahlers}, \&
  {Lacki}}]{2013PhRvD..88l1301M}
{Murase}, K., {Ahlers}, M., \& {Lacki}, B.~C. 2013, \prd, 88, 121301,
  \dodoi{10.1103/PhysRevD.88.121301}

\bibitem[{{Murase} {et~al.}(2016){Murase}, {Guetta}, \&
  {Ahlers}}]{2016PhRvL.116g1101M}
{Murase}, K., {Guetta}, D., \& {Ahlers}, M. 2016, \prl, 116, 071101,
  \dodoi{10.1103/PhysRevLett.116.071101}

\bibitem[{{Murase} {et~al.}(2014){Murase}, {Inoue}, \&
  {Dermer}}]{2014PhRvD..90b3007M}
{Murase}, K., {Inoue}, Y., \& {Dermer}, C.~D. 2014, \prd, 90, 023007,
  \dodoi{10.1103/PhysRevD.90.023007}

\bibitem[{{Murase} {et~al.}(2020){Murase}, {Kimura}, \&
  {M{\'e}sz{\'a}ros}}]{2020PhRvL.125a1101M}
{Murase}, K., {Kimura}, S.~S., \& {M{\'e}sz{\'a}ros}, P. 2020, \prl, 125,
  011101, \dodoi{10.1103/PhysRevLett.125.011101}

\bibitem[{{Murase} \& {Stecker}(2022)}]{2022arXiv220203381M}
{Murase}, K., \& {Stecker}, F.~W. 2022, arXiv e-prints, arXiv:2202.03381,
  \dodoi{10.48550/arXiv.2202.03381}

\bibitem[{{Nakamura} \& {Asada}(2013)}]{2013ApJ...775..118N}
{Nakamura}, M., \& {Asada}, K. 2013, \apj, 775, 118,
  \dodoi{10.1088/0004-637X/775/2/118}

\bibitem[{{Oikonomou} {et~al.}(2021){Oikonomou}, {Petropoulou}, {Murase},
  {Tohuvavohu}, {Vasilopoulos}, {Buson}, \& {Santander}}]{2021JCAP...10..082O}
{Oikonomou}, F., {Petropoulou}, M., {Murase}, K., {et~al.} 2021, \jcap, 2021,
  082, \dodoi{10.1088/1475-7516/2021/10/082}

\bibitem[{{Paliya} {et~al.}(2020){Paliya}, {B{\"o}ttcher}, {Olmo-Garc{\'\i}a},
  {Dom{\'\i}nguez}, {Gil de Paz}, {Franckowiak}, {Garrappa}, \&
  {Stein}}]{2020ApJ...902...29P}
{Paliya}, V.~S., {B{\"o}ttcher}, M., {Olmo-Garc{\'\i}a}, A., {et~al.} 2020,
  \apj, 902, 29, \dodoi{10.3847/1538-4357/abb46e}

\bibitem[{{Panessa} {et~al.}(2019){Panessa}, {Baldi}, {Laor}, {Padovani},
  {Behar}, \& {McHardy}}]{2019NatAs...3..387P}
{Panessa}, F., {Baldi}, R.~D., {Laor}, A., {et~al.} 2019, Nature Astronomy, 3,
  387, \dodoi{10.1038/s41550-019-0765-4}

\bibitem[{{Petropoulou} \& {Dermer}(2016)}]{2016ApJ...825L..11P}
{Petropoulou}, M., \& {Dermer}, C.~D. 2016, \apjl, 825, L11,
  \dodoi{10.3847/2041-8205/825/1/L11}

\bibitem[{{Petropoulou} {et~al.}(2020){Petropoulou}, {Oikonomou},
  {Mastichiadis}, {Murase}, {Padovani}, {Vasilopoulos}, \&
  {Giommi}}]{2020ApJ...899..113P}
{Petropoulou}, M., {Oikonomou}, F., {Mastichiadis}, A., {et~al.} 2020, \apj,
  899, 113, \dodoi{10.3847/1538-4357/aba8a0}

\bibitem[{{Prince} {et~al.}(2024){Prince}, {Das}, {Gupta}, {Majumdar}, \&
  {Czerny}}]{2024MNRAS.527.8746P}
{Prince}, R., {Das}, S., {Gupta}, N., {Majumdar}, P., \& {Czerny}, B. 2024,
  \mnras, 527, 8746, \dodoi{10.1093/mnras/stad3804}

\bibitem[{{Reimer} {et~al.}(2019){Reimer}, {B{\"o}ttcher}, \&
  {Buson}}]{2019ApJ...881...46R}
{Reimer}, A., {B{\"o}ttcher}, M., \& {Buson}, S. 2019, \apj, 881, 46,
  \dodoi{10.3847/1538-4357/ab2bff}

\bibitem[{{Reusch} {et~al.}(2022){Reusch}, {Stein}, {Kowalski}, {van Velzen},
  {Franckowiak}, {Lunardini}, {Murase}, {Winter}, {Miller-Jones}, {Kasliwal},
  {Gilfanov}, {Garrappa}, {Paliya}, {Ahumada}, {Anand}, {Barbarino}, {Bellm},
  {Brinnel}, {Buson}, {Cenko}, {Coughlin}, {De}, {Dekany}, {Frederick},
  {Gal-Yam}, {Gezari}, {Giroletti}, {Graham}, {Karambelkar}, {Kimura}, {Kong},
  {Kool}, {Laher}, {Medvedev}, {Necker}, {Nordin}, {Perley}, {Rigault},
  {Rusholme}, {Schulze}, {Schweyer}, {Singer}, {Sollerman}, {Strotjohann},
  {Sunyaev}, {van Santen}, {Walters}, {Zhang}, \&
  {Zimmerman}}]{2022PhRvL.128v1101R}
{Reusch}, S., {Stein}, R., {Kowalski}, M., {et~al.} 2022, \prl, 128, 221101,
  \dodoi{10.1103/PhysRevLett.128.221101}

\bibitem[{{Ricci} {et~al.}(2018){Ricci}, {Ho}, {Fabian}, {Trakhtenbrot},
  {Koss}, {Ueda}, {Lohfink}, {Shimizu}, {Bauer}, {Mushotzky}, {Schawinski},
  {Paltani}, {Lamperti}, {Treister}, \& {Oh}}]{2018MNRAS.480.1819R}
{Ricci}, C., {Ho}, L.~C., {Fabian}, A.~C., {et~al.} 2018, \mnras, 480, 1819,
  \dodoi{10.1093/mnras/sty1879}

\bibitem[{{Righi} {et~al.}(2020){Righi}, {Palladino}, {Tavecchio}, \&
  {Vissani}}]{2020A&A...642A..92R}
{Righi}, C., {Palladino}, A., {Tavecchio}, F., \& {Vissani}, F. 2020, \aap,
  642, A92, \dodoi{10.1051/0004-6361/202038301}

\bibitem[{{Rodrigues} {et~al.}(2018){Rodrigues}, {Fedynitch}, {Gao},
  {Boncioli}, \& {Winter}}]{2018ApJ...854...54R}
{Rodrigues}, X., {Fedynitch}, A., {Gao}, S., {Boncioli}, D., \& {Winter}, W.
  2018, \apj, 854, 54, \dodoi{10.3847/1538-4357/aaa7ee}

\bibitem[{{Rodrigues} {et~al.}(2021{\natexlab{a}}){Rodrigues}, {Garrappa},
  {Gao}, {Paliya}, {Franckowiak}, \& {Winter}}]{2021ApJ...912...54R}
{Rodrigues}, X., {Garrappa}, S., {Gao}, S., {et~al.} 2021{\natexlab{a}}, \apj,
  912, 54, \dodoi{10.3847/1538-4357/abe87b}

\bibitem[{{Rodrigues} {et~al.}(2021{\natexlab{b}}){Rodrigues}, {Heinze},
  {Palladino}, {van Vliet}, \& {Winter}}]{2021PhRvL.126s1101R}
{Rodrigues}, X., {Heinze}, J., {Palladino}, A., {van Vliet}, A., \& {Winter},
  W. 2021{\natexlab{b}}, \prl, 126, 191101,
  \dodoi{10.1103/PhysRevLett.126.191101}

\bibitem[{{Sahakyan} {et~al.}(2023){Sahakyan}, {Giommi}, {Padovani},
  {Petropoulou}, {B{\'e}gu{\'e}}, {Boccardi}, \&
  {Gasparyan}}]{2023MNRAS.519.1396S}
{Sahakyan}, N., {Giommi}, P., {Padovani}, P., {et~al.} 2023, \mnras, 519, 1396,
  \dodoi{10.1093/mnras/stac3607}

\bibitem[{{Sahu} \& {Miranda}(2015)}]{2015EPJC...75..273S}
{Sahu}, S., \& {Miranda}, L.~S. 2015, European Physical Journal C, 75, 273,
  \dodoi{10.1140/epjc/s10052-015-3519-1}

\bibitem[{{Saldana-Lopez} {et~al.}(2021){Saldana-Lopez}, {Dom{\'\i}nguez},
  {P{\'e}rez-Gonz{\'a}lez}, {Finke}, {Ajello}, {Primack}, {Paliya}, \&
  {Desai}}]{2021MNRAS.507.5144S}
{Saldana-Lopez}, A., {Dom{\'\i}nguez}, A., {P{\'e}rez-Gonz{\'a}lez}, P.~G.,
  {et~al.} 2021, \mnras, 507, 5144, \dodoi{10.1093/mnras/stab2393}

\bibitem[{{Sikora} {et~al.}(1997){Sikora}, {Madejski}, {Moderski}, \&
  {Poutanen}}]{1997ApJ...484..108S}
{Sikora}, M., {Madejski}, G., {Moderski}, R., \& {Poutanen}, J. 1997, \apj,
  484, 108, \dodoi{10.1086/304305}

\bibitem[{{Sikora} {et~al.}(2009){Sikora}, {Stawarz}, {Moderski}, {Nalewajko},
  \& {Madejski}}]{2009ApJ...704...38S}
{Sikora}, M., {Stawarz}, {\L}., {Moderski}, R., {Nalewajko}, K., \& {Madejski},
  G.~M. 2009, \apj, 704, 38, \dodoi{10.1088/0004-637X/704/1/38}

\bibitem[{{Stein} {et~al.}(2021){Stein}, {van Velzen}, {Kowalski},
  {Franckowiak}, {Gezari}, {Miller-Jones}, {Frederick}, {Sfaradi},
  {Bietenholz}, {Horesh}, {Fender}, {Garrappa}, {Ahumada}, {Andreoni},
  {Belicki}, {Bellm}, {B{\"o}ttcher}, {Brinnel}, {Burruss}, {Cenko},
  {Coughlin}, {Cunningham}, {Drake}, {Farrar}, {Feeney}, {Foley}, {Gal-Yam},
  {Golkhou}, {Goobar}, {Graham}, {Hammerstein}, {Helou}, {Hung}, {Kasliwal},
  {Kilpatrick}, {Kong}, {Kupfer}, {Laher}, {Mahabal}, {Masci}, {Necker},
  {Nordin}, {Perley}, {Rigault}, {Reusch}, {Rodriguez}, {Rojas-Bravo},
  {Rusholme}, {Shupe}, {Singer}, {Sollerman}, {Soumagnac}, {Stern}, {Taggart},
  {van Santen}, {Ward}, {Woudt}, \& {Yao}}]{2021NatAs...5..510S}
{Stein}, R., {van Velzen}, S., {Kowalski}, M., {et~al.} 2021, Nature Astronomy,
  5, 510, \dodoi{10.1038/s41550-020-01295-8}

\bibitem[{{Stratta} {et~al.}(2011){Stratta}, {Capalbi}, {Giommi}, {Primavera},
  {Cutini}, \& {Gasparrini}}]{2011arXiv1103.0749S}
{Stratta}, G., {Capalbi}, M., {Giommi}, P., {et~al.} 2011, arXiv e-prints,
  arXiv:1103.0749, \dodoi{10.48550/arXiv.1103.0749}

\bibitem[{{Tavecchio} \& {Ghisellini}(2008)}]{2008MNRAS.386..945T}
{Tavecchio}, F., \& {Ghisellini}, G. 2008, \mnras, 386, 945,
  \dodoi{10.1111/j.1365-2966.2008.13072.x}

\bibitem[{{Tavecchio} \& {Ghisellini}(2015)}]{2015MNRAS.451.1502T}
---. 2015, \mnras, 451, 1502, \dodoi{10.1093/mnras/stv1023}

\bibitem[{{Tavecchio} {et~al.}(2014){Tavecchio}, {Ghisellini}, \&
  {Guetta}}]{2014ApJ...793L..18T}
{Tavecchio}, F., {Ghisellini}, G., \& {Guetta}, D. 2014, \apjl, 793, L18,
  \dodoi{10.1088/2041-8205/793/1/L18}

\bibitem[{{Tramacere} {et~al.}(2022){Tramacere}, {Sliusar}, {Walter},
  {Jurysek}, \& {Balbo}}]{2022A&A...658A.173T}
{Tramacere}, A., {Sliusar}, V., {Walter}, R., {Jurysek}, J., \& {Balbo}, M.
  2022, \aap, 658, A173, \dodoi{10.1051/0004-6361/202142003}

\bibitem[{{Troitsky}(2021)}]{2021PhyU...64.1261T}
{Troitsky}, S.~V. 2021, Physics Uspekhi, 64, 1261,
  \dodoi{10.3367/UFNe.2021.09.039062}

\bibitem[{{{\v{S}}laus} {et~al.}(2020){{\v{S}}laus}, {Smol{\v{c}}i{\'c}},
  {Novak}, {Fotopoulou}, {Ciliegi}, {Jurlin}, {Ceraj}, {Tisani{\'c}},
  {Birkinshaw}, {Bremer}, {Chiappetti}, {Horellou}, {Huynh}, {Intema},
  {Kolokythas}, {Pierre}, {Raychaudhury}, \&
  {Rottgering}}]{2020A&A...638A..46S}
{{\v{S}}laus}, B., {Smol{\v{c}}i{\'c}}, V., {Novak}, M., {et~al.} 2020, \aap,
  638, A46, \dodoi{10.1051/0004-6361/201937258}

\bibitem[{{Wang} {et~al.}(2020){Wang}, {Xi}, {Liu}, {Xue}, \&
  {Wang}}]{2020PhRvD.101h3004W}
{Wang}, Z.-R., {Xi}, S.-Q., {Liu}, R.-Y., {Xue}, R., \& {Wang}, X.-Y. 2020,
  \prd, 101, 083004, \dodoi{10.1103/PhysRevD.101.083004}

\bibitem[{{Wang} {et~al.}(2024){Wang}, {Xue}, {Xiong}, {Wang}, {Sun}, {Peng},
  \& {Mao}}]{2024ApJS..271...10W}
{Wang}, Z.-R., {Xue}, R., {Xiong}, D., {et~al.} 2024, \apjs, 271, 10,
  \dodoi{10.3847/1538-4365/ad168c}

\bibitem[{{Willott} {et~al.}(2001){Willott}, {Rawlings}, {Blundell}, {Lacy}, \&
  {Eales}}]{2001MNRAS.322..536W}
{Willott}, C.~J., {Rawlings}, S., {Blundell}, K.~M., {Lacy}, M., \& {Eales},
  S.~A. 2001, \mnras, 322, 536, \dodoi{10.1046/j.1365-8711.2001.04101.x}

\bibitem[{{Xue} {et~al.}(2019{\natexlab{a}}){Xue}, {Liu}, {Petropoulou},
  {Oikonomou}, {Wang}, {Wang}, \& {Wang}}]{2019ApJ...886...23X}
{Xue}, R., {Liu}, R.-Y., {Petropoulou}, M., {et~al.} 2019{\natexlab{a}}, \apj,
  886, 23, \dodoi{10.3847/1538-4357/ab4b44}

\bibitem[{{Xue} {et~al.}(2019{\natexlab{b}}){Xue}, {Liu}, {Wang}, {Yan}, \&
  {B{\"o}ttcher}}]{2019ApJ...871...81X}
{Xue}, R., {Liu}, R.-Y., {Wang}, X.-Y., {Yan}, H., \& {B{\"o}ttcher}, M.
  2019{\natexlab{b}}, \apj, 871, 81, \dodoi{10.3847/1538-4357/aaf720}

\bibitem[{{Xue} {et~al.}(2021){Xue}, {Liu}, {Wang}, {Ding}, \&
  {Wang}}]{2021ApJ...906...51X}
{Xue}, R., {Liu}, R.-Y., {Wang}, Z.-R., {Ding}, N., \& {Wang}, X.-Y. 2021,
  \apj, 906, 51, \dodoi{10.3847/1538-4357/abc886}

\bibitem[{{Xue} {et~al.}(2022){Xue}, {Wang}, \& {Li}}]{2022PhRvD.106j3021X}
{Xue}, R., {Wang}, Z.-R., \& {Li}, W.-J. 2022, \prd, 106, 103021,
  \dodoi{10.1103/PhysRevD.106.103021}

\bibitem[{{Yao} \& {Komossa}(2023)}]{2023MNRAS.523..441Y}
{Yao}, S., \& {Komossa}, S. 2023, \mnras, 523, 441,
  \dodoi{10.1093/mnras/stad1415}

\bibitem[{{Zdziarski} \& {Bottcher}(2015)}]{2015MNRAS.450L..21Z}
{Zdziarski}, A.~A., \& {Bottcher}, M. 2015, \mnras, 450, L21,
  \dodoi{10.1093/mnrasl/slv039}

\end{thebibliography}
\bibliographystyle{aasjournal}
\end{document}